\documentclass[twocolumn]{aastex6311}
\usepackage{threeparttable}
\usepackage{multirow}
\usepackage{times}
\usepackage{enumitem}
\usepackage{natbib}
\usepackage{url} 
\usepackage{amsmath,bm}
\usepackage{multirow}

\shorttitle{Stellar Parameters and Abundances for J-PLUS DR3 Stars}
\shortauthors{Yang Huang et al.}

\begin{document}

\title{J-PLUS: Beyond Spectroscopy III. Stellar Parameters and Elemental-abundance Ratios for Five Million Stars from DR3}

\author[0000-0001-8424-1079]{Yang Huang}
\affiliation{School of Astronomy and Space Science, University of Chinese Academy of Sciences, Beijing 100049, People's Republic of China}
\affiliation{CAS Key Lab of Optical Astronomy, National Astronomical Observatories, Chinese Academy of Sciences, Beijing 100012, People's Republic of China}
\affiliation{Institute for Frontiers in Astronomy and Astrophysics, Beijing Normal University, Beijing, 102206, People’s Republic of China}

\author[0000-0003-4573-6233]{Timothy C. Beers}
\affiliation{Department of Physics and Astronomy, University of Notre Dame, Notre Dame, IN 46556, USA}
\affiliation{Joint Institute for Nuclear Astrophysics -- Center for the Evolution of the Elements (JINA-CEE), USA}

\author[0000-0001-8424-1079]{Kai Xiao}
\affiliation{School of Astronomy and Space Science, University of Chinese Academy of Sciences, Beijing 100049, People's Republic of China}

\author[0000-0003-2471-2363]{Haibo Yuan}
\affiliation{Institute for Frontiers in Astronomy and Astrophysics, Beijing Normal University, Beijing, 102206, China}
\affiliation{Department of Astronomy, Beijing Normal University, Beijing, 100875, People's Republic of China}

\author[0000-0001-5297-4518]{Young Sun Lee}
\affiliation{Department of Astronomy and Space Science, Chungnam National University, Daejeon 34134, Republic of Korea}
\affiliation{Department of Physics and Astronomy, University of Notre Dame, Notre Dame, IN 46556, USA}

\author[0009-0007-5610-6495]{Hongrui Gu}
\affiliation{CAS Key Lab of Optical Astronomy, National Astronomical Observatories, Chinese Academy of Sciences, Beijing 100012, People's Republic of China}
\affiliation{School of Astronomy and Space Science, University of Chinese Academy of Sciences, Beijing 100049, People's Republic of China}

\author[0000-0002-2453-0853]{Jihye Hong}
\affiliation{Department of Physics and Astronomy, University of Notre Dame, Notre Dame, IN 46556, USA}
\affiliation{Joint Institute for Nuclear Astrophysics -- Center for the Evolution of the Elements (JINA-CEE), USA}

\author[0000-0002-2874-2706]{Jifeng Liu}
\affiliation{CAS Key Lab of Optical Astronomy, National Astronomical Observatories, Chinese Academy of Sciences, Beijing 100012, People's Republic of China}
\affiliation{School of Astronomy and Space Science, University of Chinese Academy of Sciences, Beijing 100049, People's Republic of China}
\affiliation{Institute for Frontiers in Astronomy and Astrophysics, Beijing Normal University, Beijing, 102206, People’s Republic of China}

\author[0000-0002-6790-2397]{Zhou Fan}
\affiliation{CAS Key Lab of Optical Astronomy, National Astronomical Observatories, Chinese Academy of Sciences, Beijing 100012, People's Republic of China}
\affiliation{School of Astronomy and Space Science, University of Chinese Academy of Sciences, Beijing 100049, People's Republic of China}

\author{Paula Coelho}
\affiliation{Instituto de Astronomia, Geofísica e Ci{\^e}ncias Atmosf{\'e}ricas, Universidade de S{\~a}o Paulo, Rua do Mat{\~a}o 1226, S{\~a}o Paulo, 05508-090, SP, Brazil}

\author[0000-0003-1793-200X]{Patricia Cruz}
\affiliation{Centro de Astrobiolog\'{\i}a, CSIC-INTA, Camino bajo del castillo s/n, E-28692, Villanueva de la Can\~ada, Madrid, Spain}

\author[0000-0003-4776-9098]{F. J. Galindo-Guil}
\affiliation{Centro de Estudios de F{\'i}sica del Cosmos de Arag{\'o}n (CEFCA), Unidad Asociada al CSIC, Plaza San Juan 1, 44001, Teruel, Spain}

\author[0000-0001-9205-2307]{Simone Daflon}
\affiliation{Observat{\'o}rio Nacional, Rua Gal. Jos{\'e} Cristino 77, Rio de Janeiro, 20921-400, RJ, Brazil}

\author{Fran Jim\'enez-Esteban}
\affiliation{Centro de Astrobiolog\'{\i}a, CSIC-INTA, Camino bajo del castillo s/n, E-28692, Villanueva de la Can\~ada, Madrid, Spain}

\author{Javier Cenarro}
\affiliation{Centro de Estudios de F{\'i}sica del Cosmos de Arag{\'o}n (CEFCA), Unidad Asociada al CSIC, Plaza San Juan 1, 44001, Teruel, Spain}

\author{David Crist{\'o}bal-Hornillos}
\affiliation{Centro de Estudios de F{\'i}sica del Cosmos de Arag{\'o}n (CEFCA), Unidad Asociada al CSIC, Plaza San Juan 1, 44001, Teruel, Spain}

\author{Carlos Hern{\'a}ndez-Monteagudo}
\affiliation{Instituto de Astrofísica de Canarias (IAC), C/V{\'i}a L{\'a}ctea, S/N, E-38205, La Laguna, Tenerife, Spain}

\author{Carlos L{\'o}pez-Sanjuan}
\affiliation{Centro de Estudios de F{\'i}sica del Cosmos de Arag{\'o}n (CEFCA), Unidad Asociada al CSIC, Plaza San Juan 1, 44001, Teruel, Spain}

\author{Antonio Mar{\'i}n-Franch}
\affiliation{Centro de Estudios de F{\'i}sica del Cosmos de Arag{\'o}n (CEFCA), Unidad Asociada al CSIC, Plaza San Juan 1, 44001, Teruel, Spain}

\author{Mariano Moles}
\affiliation{Centro de Estudios de F{\'i}sica del Cosmos de Arag{\'o}n (CEFCA), Unidad Asociada al CSIC, Plaza San Juan 1, 44001, Teruel, Spain}

\author{Jes{\'u}s Varela}
\affiliation{Centro de Estudios de F{\'i}sica del Cosmos de Arag{\'o}n (CEFCA), Unidad Asociada al CSIC, Plaza San Juan 1, 44001, Teruel, Spain}

\author{H{\'e}ctor V{\'a}zquez Rami{\'o}}
\affiliation{Centro de Estudios de F{\'i}sica del Cosmos de Arag{\'o}n (CEFCA), Unidad Asociada al CSIC, Plaza San Juan 1, 44001, Teruel, Spain}

\author{Jailson Alcaniz}
\affiliation{Observat{\'o}rio Nacional, Rua Gal. Jos{\'e} Cristino 77, Rio de Janeiro, 20921-400, RJ, Brazil}

\author{Renato Dupke}
\affiliation{Observat{\'o}rio Nacional, Rua Gal. Jos{\'e} Cristino 77, Rio de Janeiro, 20921-400, RJ, Brazil}

\author{Alessandro Ederoclite}
\affiliation{Centro de Estudios de F{\'i}sica del Cosmos de Arag{\'o}n (CEFCA), Unidad Asociada al CSIC, Plaza San Juan 1, 44001, Teruel, Spain}

\author{Laerte Sodr{\'e} Jr.}
\affiliation{Instituto de Astronomia, Geofísica e Ci{\^e}ncias Atmosf{\'e}ricas, Universidade de S{\~a}o Paulo, Rua do Mat{\~a}o 1226, S{\~a}o Paulo, 05508-090, SP, Brazil}

\author{Raul E. Angulo}
\affiliation{Donostia International Physics Center, Paseo Manuel de Lardizabal 4, 20018, Donostia-San Sebasti{\,a}n, Spain}

\begin{abstract}

We present a catalog of stellar parameters (effective temperature $T_{\rm eff}$, 
surface gravity $\log g$, age, and metallicity [Fe/H]) and elemental-abundance ratios ([C/Fe], [Mg/Fe], and [$\alpha$/Fe]) for some five million stars (4.5 million dwarfs and 0.5 million giants stars) in the Milky Way, based on stellar colors from the Javalambre Photometric Local Universe Survey (J-PLUS) DR3 and \textit{Gaia} EDR3. These estimates are obtained through the construction of a large spectroscopic training set with parameters and abundances adjusted to uniform scales, and trained with a Kernel Principal Component Analysis.
Owing to the seven narrow/medium-band filters employed by J-PLUS, we obtain precisions in the abundance estimates that are as good or better than derived from medium-resolution spectroscopy for stars covering a wide range of the parameter space: 0.10-0.20\,dex for [Fe/H] and [C/Fe], and 0.05\,dex for [Mg/Fe] and [$\alpha$/Fe].
Moreover, systematic errors due to the influence of molecular carbon bands on previous photometric-metallicity estimates (which only included two narrow/medium-band blue filters) have now been removed, resulting in photometric-metallicity estimates down to [Fe/H] $\sim -4.0$, with typical uncertainties of 0.25\,dex and 0.40\,dex for dwarfs and giants, respectively. 
This large photometric sample should prove useful for the exploration of the assembly and chemical-evolution history of our Galaxy. 
\end{abstract}

\section{Introduction}
Over the past decade, great advances have been achieved in the field of Galactic Archaeology due to the determinations of precise stellar parameters and individual elemental-abundance ratios for large numbers of stars obtained by massive spectroscopic surveys, such as the SDSS survey \citep{York2000}, the RAVE survey \citep{rave}, the SDSS/SEGUE survey \citep{segue1, segue2}, the GALAH survey \citep{galah}, the SDSS/APOGEE survey \citep{apogee}, the LAMOST survey \citep{legue, 2012RAA....12..723Z}, and the Gaia-ESO survey \citep{2022A&A...666A.120G, GaiaESO}.
In the era of {\it Gaia}, accurate 3-D positions and proper motions are now available for billions of stars. However, despite these extensive efforts, the number of stars with spectroscopic information lags far behind 
those with full astrometric information, by at least two orders of magnitude, leading to a limited and potentially biased view of the stellar populations in the Milky Way (MW).

In order to alleviate this mismatch between the numbers of stars with available 
spectroscopic and astrometric information, we have pursued approaches to obtain estimates of stellar parameters (and a limited number of other important elemental-abundance ratios) through the use of ongoing or planned narrow/medium-bandwidth photometric surveys \citep[see the summary in Table\,1 of][]{PaperI}.  In the first two papers of this series \citep[][hereafter Papers I and II]{PaperI, PaperII}, stellar parameters, in particular the metallicity, are derived for nearly 50 million stars covering around $3\pi$ steradians of the sky, using $uv$ narrow-band photometric data obtained from the SAGES DR1 catalog of the Northern sky \citep{SAGES} and the SkyMapper catalog in the Southern sky \citep[SMSS;][]{SMSSDR1, SMSSDR2}, combined with {\it Gaia} EDR3 broad-band photometry and astrometric information \citep{GEDR3}.  The huge numbers of stars and deep limiting magnitudes of the derived parameter catalogs are poised to revolutionize our knowledge of the MW, and also serve to identify stars of particular interest for detailed study at high spectral resolution.  

Building on our efforts with SAGES and SkyMapper, we now take the next step forward, measuring not only the [Fe/H], but also other elemental-abundance ratios (including the most important ratios for analyses of stellar populations -- [C/Fe],  [Mg/Fe], and [$\alpha$/Fe]\footnote{[$\alpha$/Fe] is the total $\alpha$-element abundance relative to iron, and is influenced by a combination of all $\alpha$-elements. In APOGEE, it is mainly determined by O, Mg, S, Si, Ca and Ti. In GALAH, it is mainly determined by O, Ne, Mg, Si, S, Ar, Ca, and Ti.}), based on the filter fluxes from the publicly available third data release of the Javalambre Photometric Local Universe Survey (J-PLUS DR3)\footnote{\url{https://archive.cefca.es/catalogues/jplus-dr3}}. The 12 filters employed by J-PLUS include seven narrow/medium-band filters (with FHWM from $100$ to $400$ \AA; $J0378$, $J0395$, $J0410$, $J0430$, $J0515$, $J0660$, $J0861$), designed to detect prominent stellar absorption features (including the Ca {\sc ii} H \& K lines, the molecular CH $G$-band, H$\delta$, the Mg b triplet, H$\alpha$, and the Ca {\sc I} triplet), along with five SDSS-like broad-band filters ($ugriz$).

The construction of training sets are of crucial importance to calibrate estimates of the stellar parameters and elemental-abundance ratios from photometric colors.  We have thus assembled a large database of several million stars with spectroscopically derived stellar parameters from a number of surveys, carefully calibrated to uniform scales, as described in this paper.

This paper is structured as follows. Section\,2 introduces the main data used in this work.
Section\,3 describes the construction of training sets, including the calibrations of the individual parameter scales. Estimates of stellar parameters and elemental-abundance ratios are presented in Section\,4. Section\,5 describes our estimates of effective temperature, distances, ages, and surface gravities. Section\,6 describes our final sample. Section\,7 provides a summary and future prospects.

\section{Data}

\subsection{J-PLUS DR3}
J-PLUS \citep{jplus} is an ongoing effort aimed at observing about 8500 deg$^2$ of the sky visible from the Observatorio Astrof{\'i}sico
de Javalambre \citep[OAJ;][]{2014SPIE.9149E..1IC}, using the JAST80 telescope equipped with the  panoramic camera T80Cam (2 deg$^2$ field-of-view provided by a  single CCD of 9.2k × 9.2k pixels).  This survey adopted 12 specially designed narrow-, medium- and broad-band optical filters; their properties are summarized in Table\,1. 
J-PLUS observations are mainly made under seeing conditions better than 1.5 arcsec and airmass smaller than 1.5.
Here we use the data from the third public data release, J-PLUS DR3, which covers 3192 deg$^2$ (1642 fields) for all 12 bands, with an $r$-band limiting magnitude down to 21.8 ($5\sigma$, 3 arcsec diameter aperture). 
In total, about 47.4 million sources are released in the J-PLUS DR3 catalog. The photometric observations were initially calibrated using {\it Gaia} BP/RP (XP) ultra low-resolution spectra \citep{jplus_calib1}.  Significant improvements in the zero-points of the filter photometry have been been made through re-calibration using the stellar color regression method and an improved \textit{Gaia} XP synthetic photometry method \citep{Xiao2024}; the final accuracy of the zero-points in the photometric calibrations is 1-5 mmag.

\subsection{Gaia EDR3}

In addition to J-PLUS DR3, {\it Gaia} EDR3 broad-band photometry ($G$, $G_{\rm BP}$, and $G_{\rm RP}$) are also used in this study. In {\it Gaia} EDR3 \citep{GEDR3}, the broad-band photometry, as well as astrometric information (parallaxes and proper motions), are provided for about 1.5 billion sources with magnitudes down to $G \sim 21$, although the completeness is quite complicated at the faint end \citep[see details in][]{2021A&A...649A...3R}. The photometric uncertainty is only a few mmag for the $G$-band photometry even at $G = 20$, around 10 mmag for $G_{\rm BP}$ and $G_{\rm RP}$ at $G = 17$, and no worse than 100 mmag for $G_{\rm BP}$ and $G_{\rm RP}$ at $G = 20$.

By cross-matching J-PLUS DR3 with {\it Gaia} EDR3 and requiring $r \le 21.0$ and {\it class\_star\,$\ge 0.6$} yielded by {\tt SExtractor} \citep{1996A&AS..117..393B}, over 16.5 million stars are left from the original J-PLUS DR3 sample.  
This sample is adopted in the following analysis.
In this study, the extinction map of \citet[][hereafter the SFD map\footnote{Note that the over-estimated 14\% systematic offset in extinction is corrected for.}]{SFD98} is adopted for reddening corrections, since over 90\% of the J-PLUS DR3 stars are located at higher Galactic latitudes ($|b| \ge 15^{\circ}$).  The reddening coefficients ($k_{\chi} = \frac{A_{\chi}}{E (B-V)}$, see Table\,1) for the J-PLUS and {\it Gaia} EDR3 photometric passbands are  empirically estimated in \citet{2021A&A...654A..61L} and \citet{2021ApJ...907...68H},  respectively, using the star-pair technique described by \citet{2013MNRAS.430.2188Y}.

\subsection{Spectroscopic Surveys}
External estimates of the stellar parameters ($T_{\rm eff}$, log\,$g$, and [Fe/H]) are adopted from a master catalog assembled from completed/ongoing large-scale spectroscopic surveys, including the SDSS/SEGUE, LAMOST,  SDSS/APOGEE. and GALAH surveys. 

The SDSS/SEGUE stellar parameters for hundreds of thousands of stars are those released in SDSS DR12 \citep{SDSSDR12}, based on low-resolution ($R \sim 2000$) optical spectra collected by the 2.5 meter Sloan Foundation Telescope at Apache Point Observatory (APO) \citep{2006AJ....131.2332G}.
The stellar parameters are derived using the  SEGUE Stellar Parameter Pipeline \citep[SSPP;][]{2008AJ....136.2022L}.
The typical metallicity uncertainty is around 0.15\,dex.

Nearly five million stars with precise metallicity estimates, as well as other stellar parameters, are adopted from DR9\footnote{\url{http://www.lamost.org/dr9/v1.1/}} of LAMOST, a 4 meter quasi-meridian reflecting Schmidt telescope equipped with 4000 fibers distributed over a
field of view of 5$^{\circ}$ in diameter \citep{lamost}; these parameters are estimated from SDSS-like low-resolution ($R \sim 2000$) optical spectra by using the LAMOST Stellar Parameter Pipeline \citep[LASP;][]{2015RAA....15.1095L}.  The metallicity uncertainty is 0.10--0.15\,dex.

The latest SDSS DR17 \citep{SDSSDR17} has released atmospheric parameters and over 20 elemental-abundance ratios for over six hundred thousand stars based on near-infrared ({\it H} band; 1.51--1.70\,$\mu$m) high-resolution ($R \sim 22,500$) spectra collected by the APOGEE-1 and APOGEE-2 surveys \citep{apogee}. These spectra are obtained by the 2.5 meter Sloan Foundation Telescope \citep{2006AJ....131.2332G} 
%and the 1m New Mexico State University (NMSU) Telescope \citep{2018AJ....156..125H}
at APO in the Northern Hemisphere,
and the 2.5 m Irénée du Pont Telescope \citep{1973ApOpt..12.1430B} at Las Campanas Observatory (LCO) in the Southern Hemisphere.
The uncertainties in [Fe/H] and for the 
elemental-abundance ratios we employ in this study ([C/Fe], [Mg/Fe] and [$\alpha$/Fe]) are  
0.10\,dex, 0.02\,dex, and 0.02\,dex, respectively.  

The GALAH survey is a large optical high-resolution ($R \sim 28,000$) spectroscopic survey using the HERMES spectrograph installed on the 3.9 m Anglo-Australian Telescope \citep{galah}. GALAH DR3 has released metallicity estimates and up to 30 elemental-abundance ratios for over 0.5 million stars \citep{GALAHDR3}; the typical uncertainties of the [Fe/H] estimates are better than 0.10\,dex, and the pertinent elemental-abundance ratios are precise to better than 0.02--0.03\,dex.

 \begin{table}
\centering
\caption{Summary of the J-PLUS Filter System}
\begin{tabular}{ccccc}
\hline
Filter& $\lambda_{\rm eff}$ &FWHM&$k_{\chi} = \frac{A_{\chi}}{E (B-V)}$&Comments\\
    &(\AA)&(\AA)&&\\
\hline
\hline
$u$&3485&508&4.479&Balmer-break region\\
$J0378$&3785&168&4.294& O II\\
$J0395$&3950&100&4.226& Ca H+K\\
$J0410$&4100&200&4.023& H$\delta$\\
$J0430$&4300&200&3.859&CH $G$-band\\
$g$&4803&1409&3.398&SDSS\\
$J0515$&5150&200&3.148&Mg $b$ triplet\\
$r$&6254&1388&2.383&SDSS\\
$J0660$&6600&138&2.161&H$\alpha$\\
$i$&7668&1535&1.743&SDSS\\
$J0861$&8610&400&1.381& Ca triplet\\
$z$&9114&1409&1.289& SDSS\\
\hline 
\hline
\end{tabular}
\end{table}

%\subsection{Pan-STARRS DR1}

%\vskip 1cm
\section{Training sets and derivation of uniform parameter scales}

In this section, we describe the training sets we employ for obtaining estimates of the metallicity ([Fe/H]) and elemental abundances ([C/Fe], [Mg/Fe], [$\alpha$/Fe]) measured by previous spectroscopic surveys.  We emphasize that this effort is not only important for the current study, but also can be adopted by future work on estimating stellar parameters either from multiple colors or spectroscopy.

\begin{figure}
\begin{center}
\includegraphics[scale=0.325,angle=0]{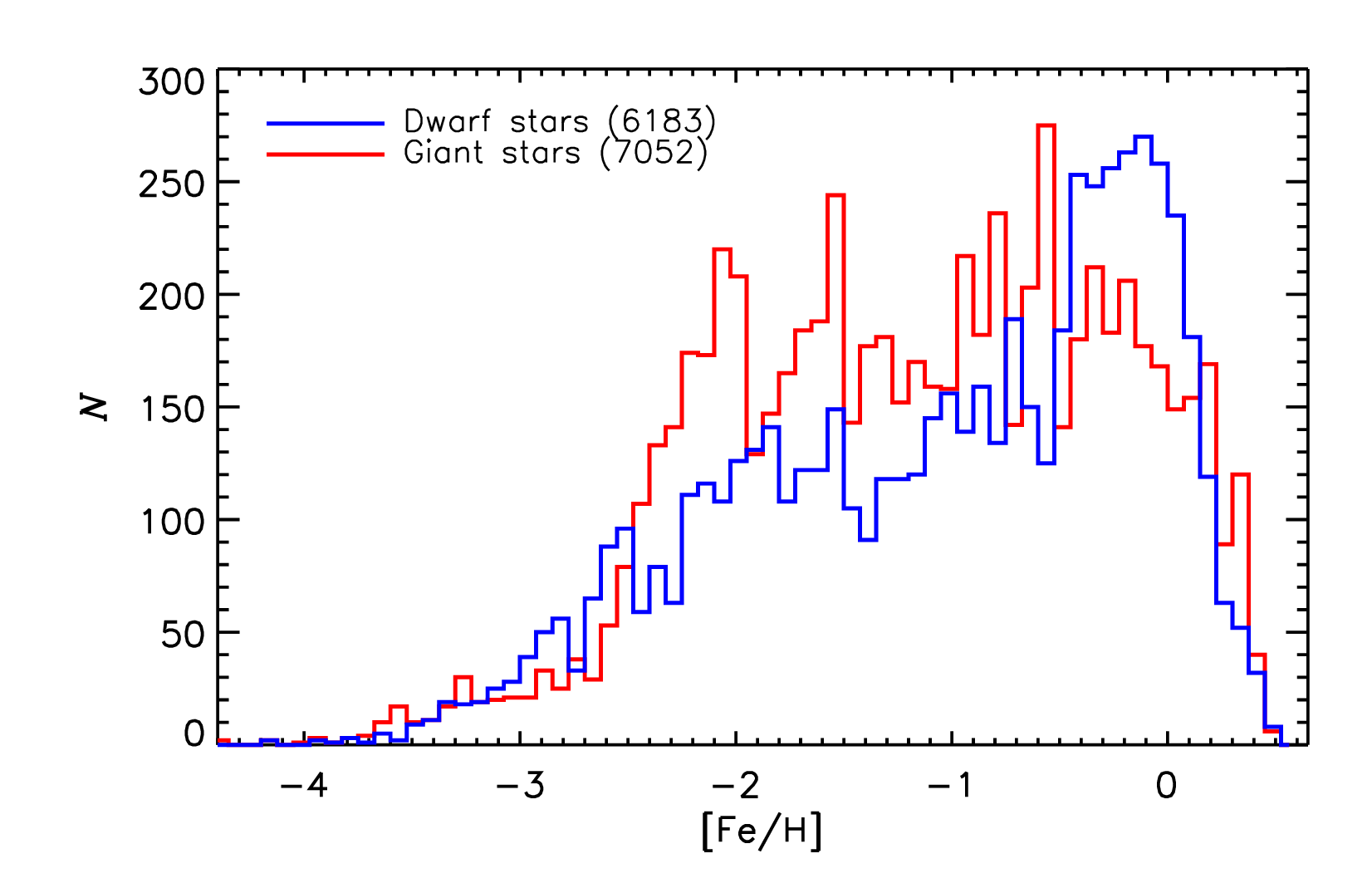}
\caption{Histograms of the training sets for metallicity ([Fe/H]) for dwarfs (blue line) and giants (red line).}
\end{center}
\end{figure}

\begin{figure*}
\begin{center}
\includegraphics[scale=0.425,angle=0]{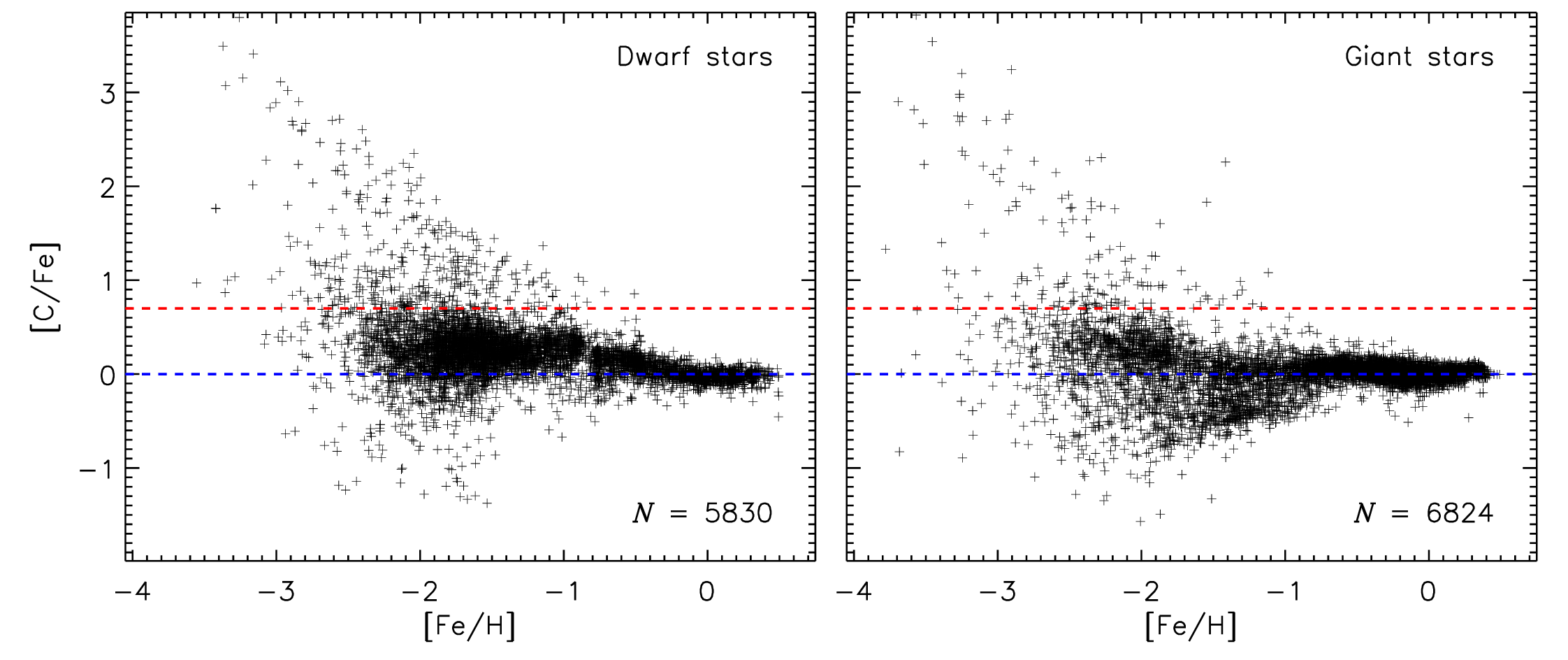}
\includegraphics[scale=0.425,angle=0]{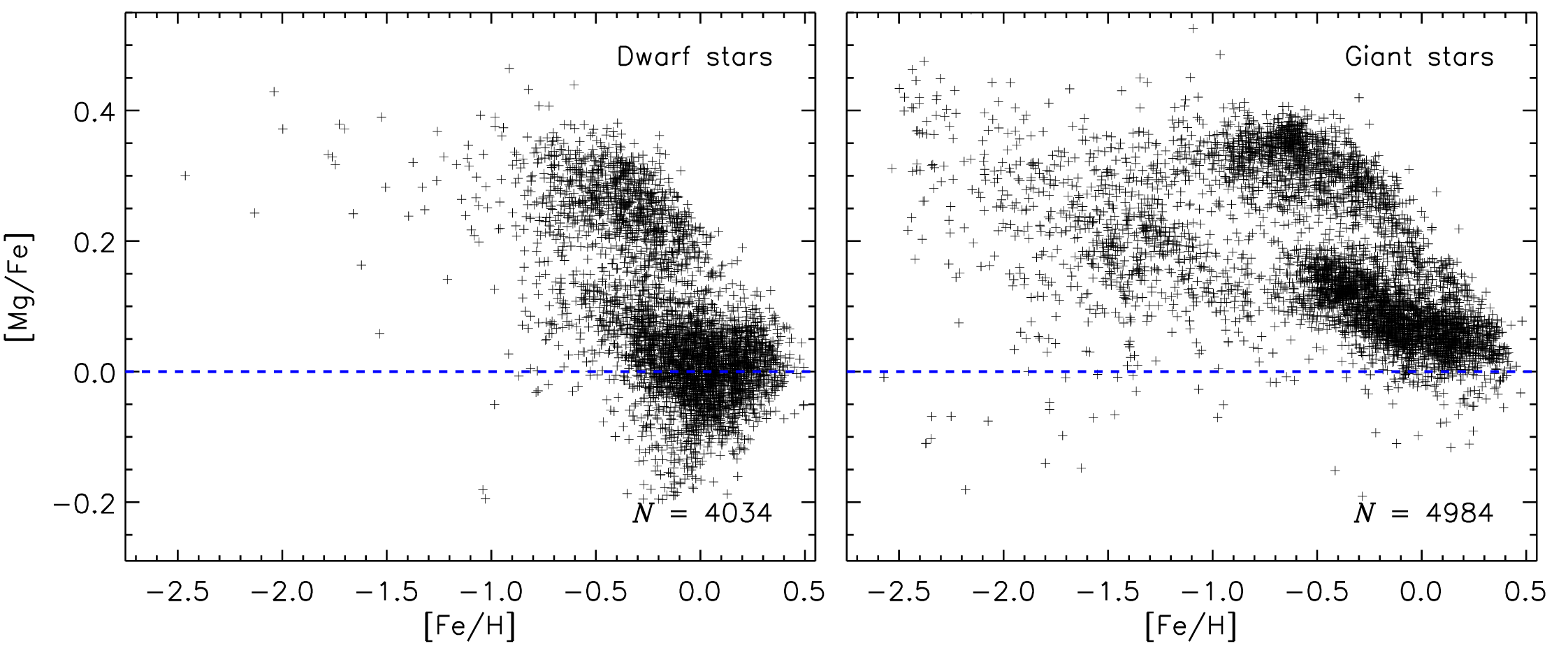}
\includegraphics[scale=0.425,angle=0]{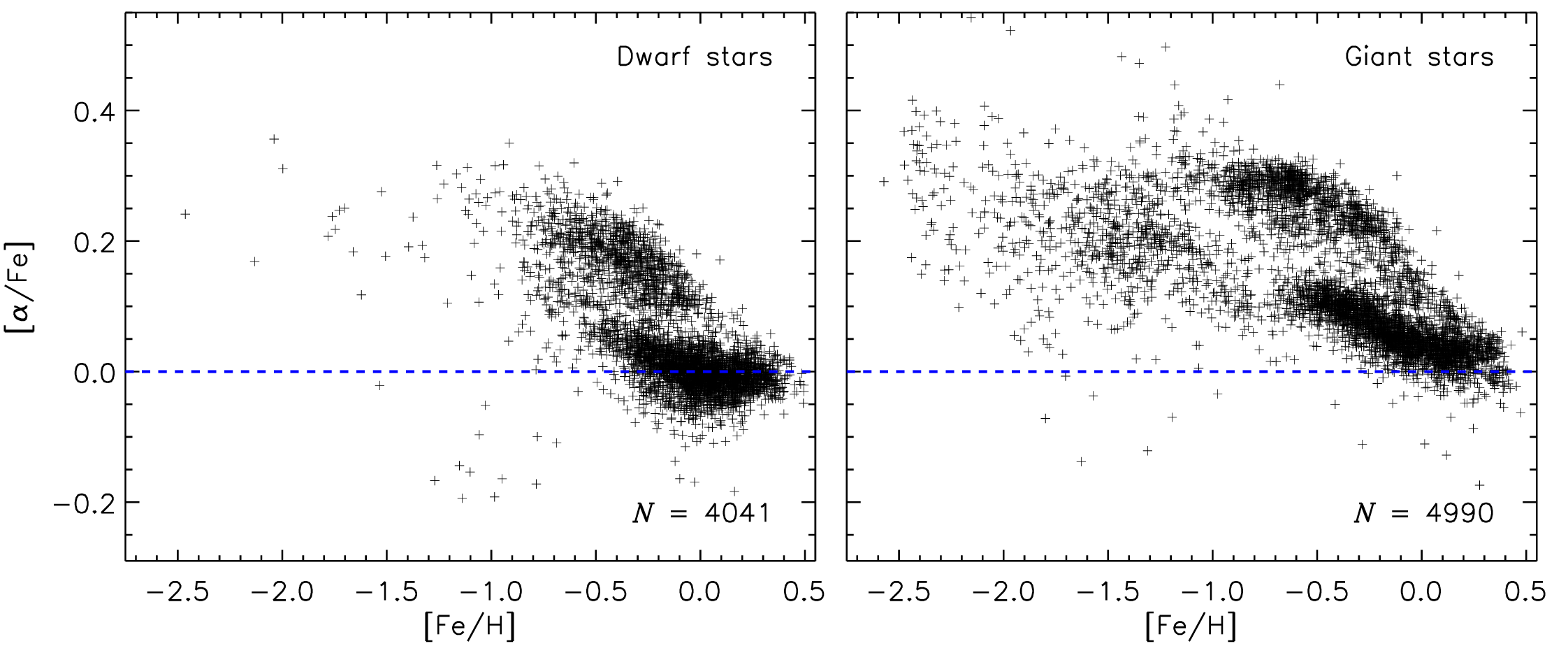}
\caption{Distributions of training-sample stars for [C/Fe] in the [C/Fe] vs. [Fe/H] space (top panels), for [Mg/Fe] in the [Mg/Fe] vs. [Fe/H] space (middle panels), and for [$\alpha$/Fe] in the [$\alpha$/Fe] vs. [Fe/H] space (bottom panels). The left column of panels is for dwarf stars and the right column of panels is for giant stars. The red-dashed lines in the top panels mark the [C/Fe] value of +0.7, the criterion often used to define carbon-enhanced metal-poor (CEMP) stars.  Note that for our present purpose, we report the measured estimate of [C/Fe], without applying evolutionary corrections. The blue-dashed lines in each panel indicate the Solar ratios.}
\end{center}
\end{figure*}

\begin{figure*}
\begin{center}
\includegraphics[scale=0.355,angle=0]{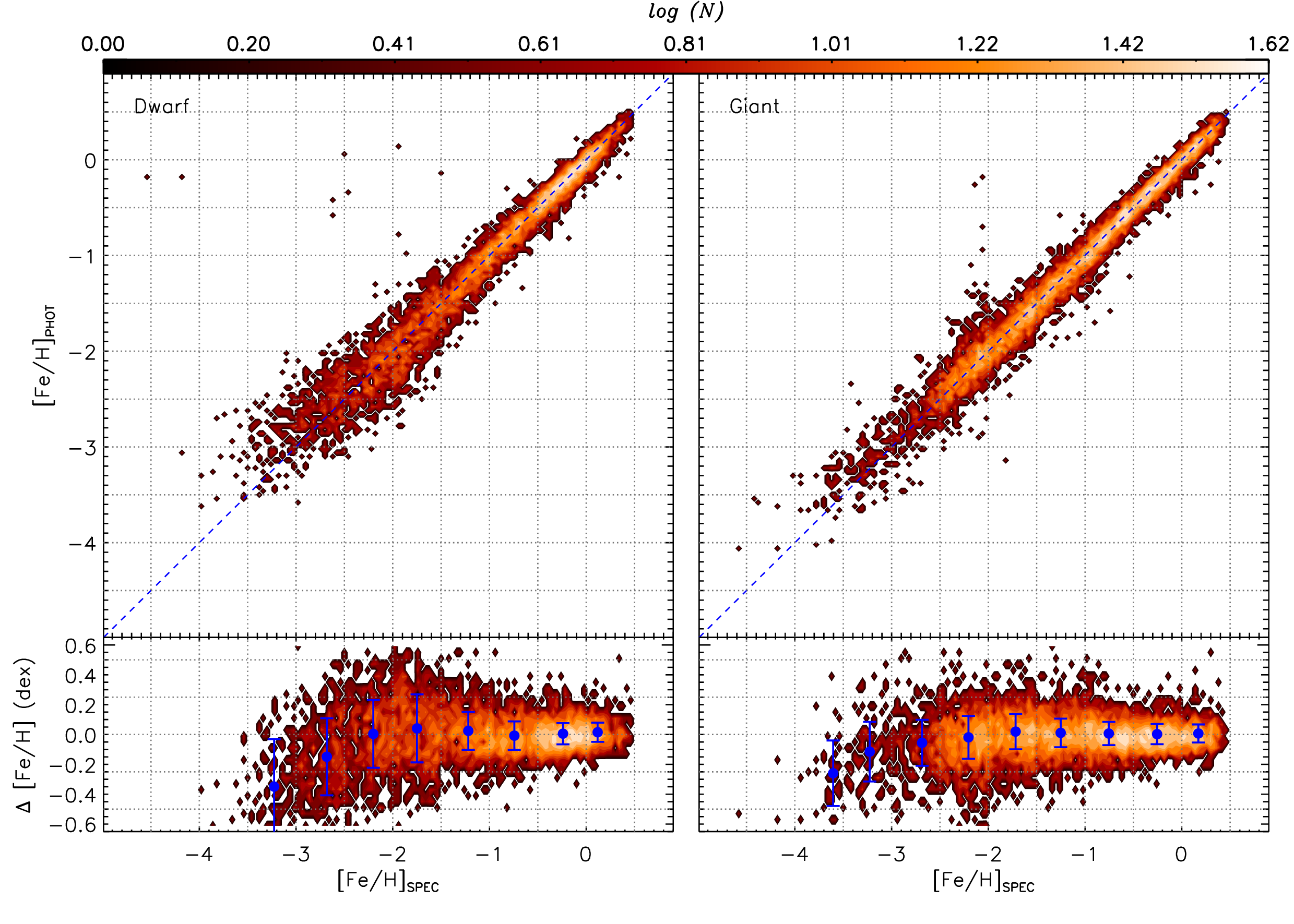}
\includegraphics[scale=0.355,angle=0]{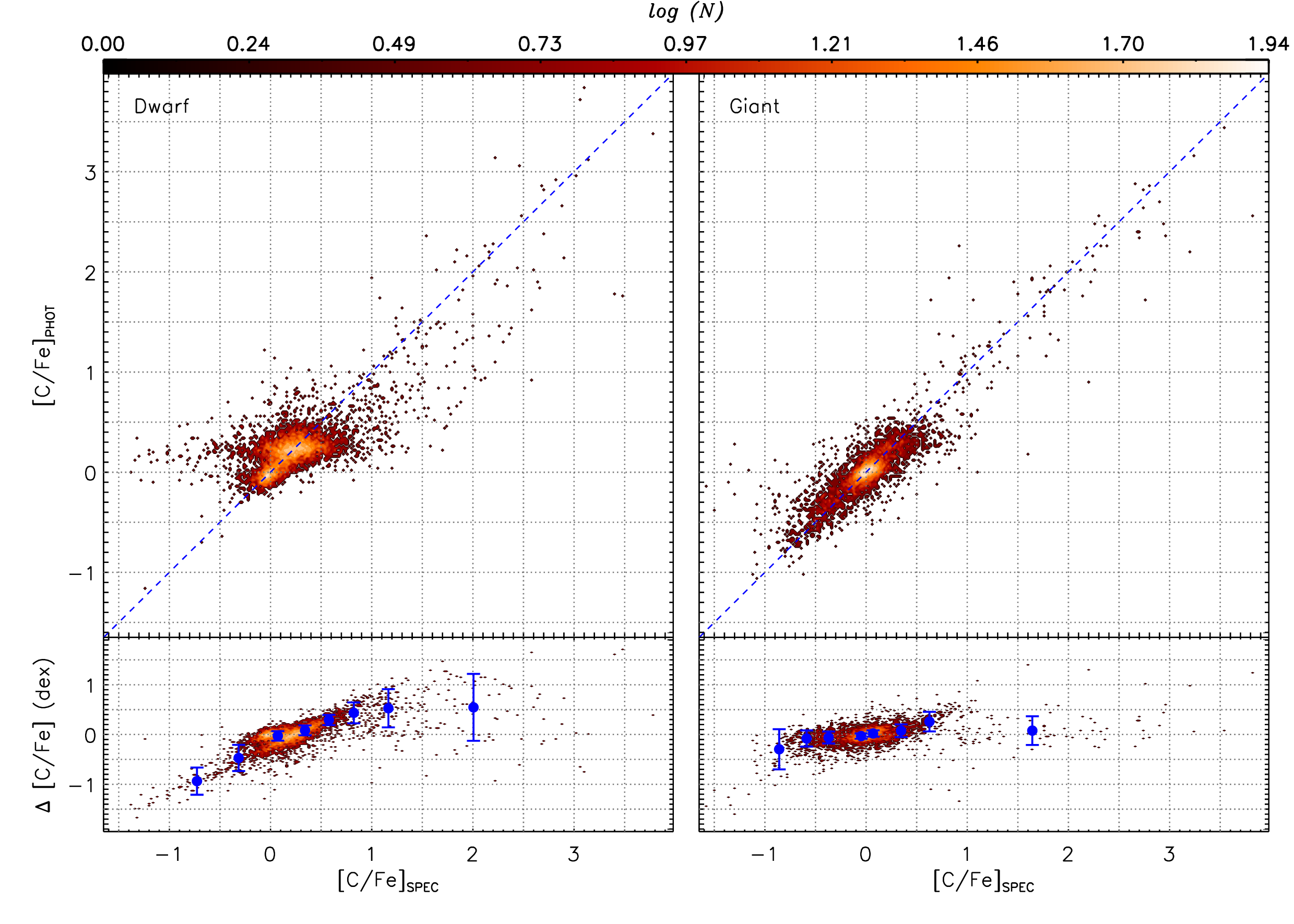}
\caption{Comparisons between the photometric and spectroscopic metallicities (top panels) and the photometric and spectroscopic [C/Fe] (bottom panels) for the training-sample dwarf stars (left column of panels) and giant stars (right column of panels). The photometric results are estimated from the multiple colors formed with the combination of the J-PLUS DR3 and {\it Gaia} EDR3 magnitudes using a Kernel Principal Component Analysis (KPCA) technique (see text). The lower part of each panel shows the parameter differences (photometric minus spectroscopic), as a function of the spectroscopic determinations. The blue dots and error bars in each panel represent the median and dispersion of the parameter differences in the individual parameter bins.  The blue-dashed lines are the one-to-one lines.  A color bar representing the numbers of stars is provided at the top of each set of panels.}
\end{center}
\end{figure*} 

\begin{figure*}
\begin{center}
\includegraphics[scale=0.355,angle=0]{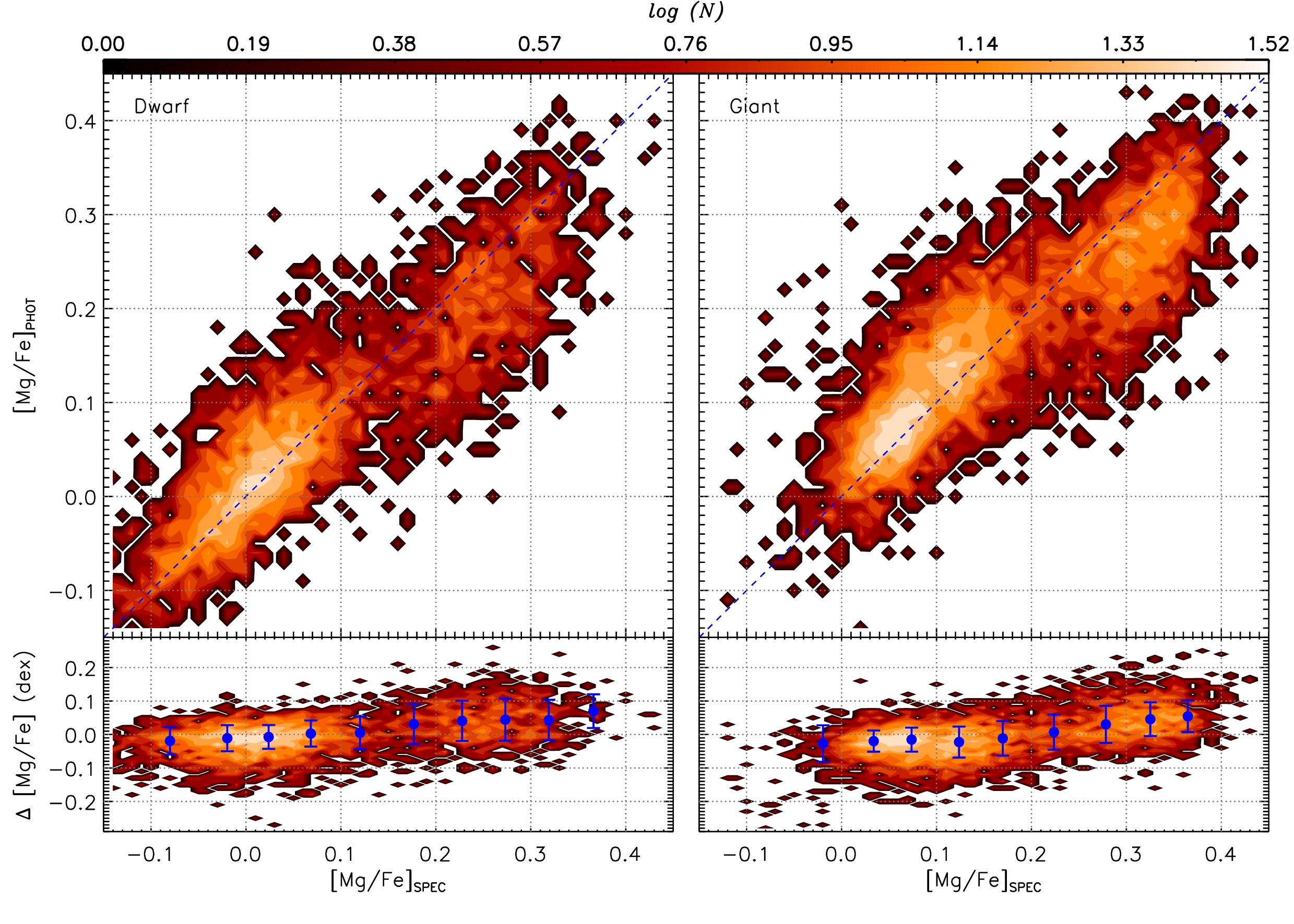}
\includegraphics[scale=0.355,angle=0]{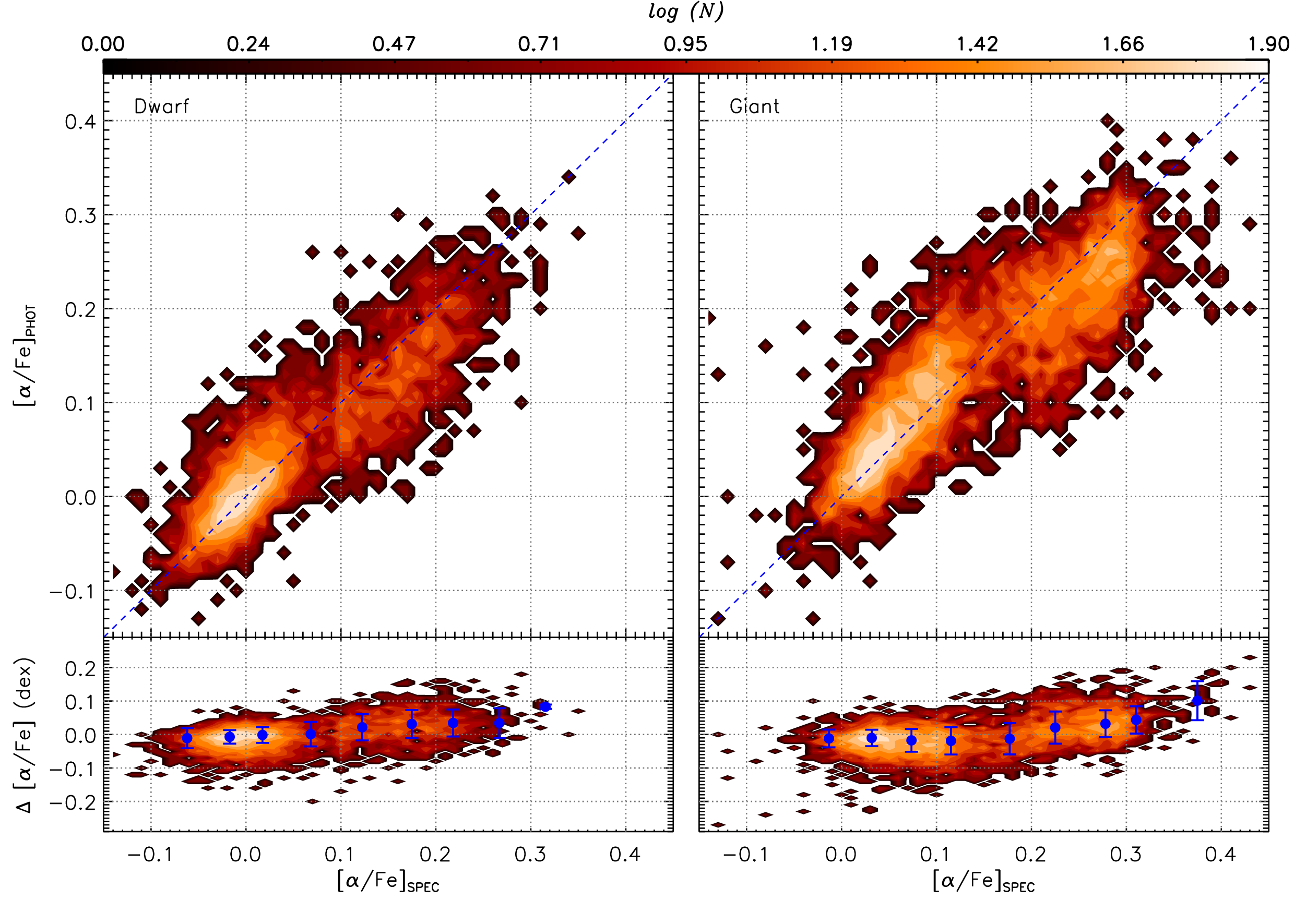}
\caption{Similar to Fig.\,3, but for [Mg/Fe] (top panels) and [$\alpha$/Fe] (bottom panels).}
\end{center}
\end{figure*} 

\begin{figure*}
\begin{center}
\includegraphics[scale=0.315,angle=0]{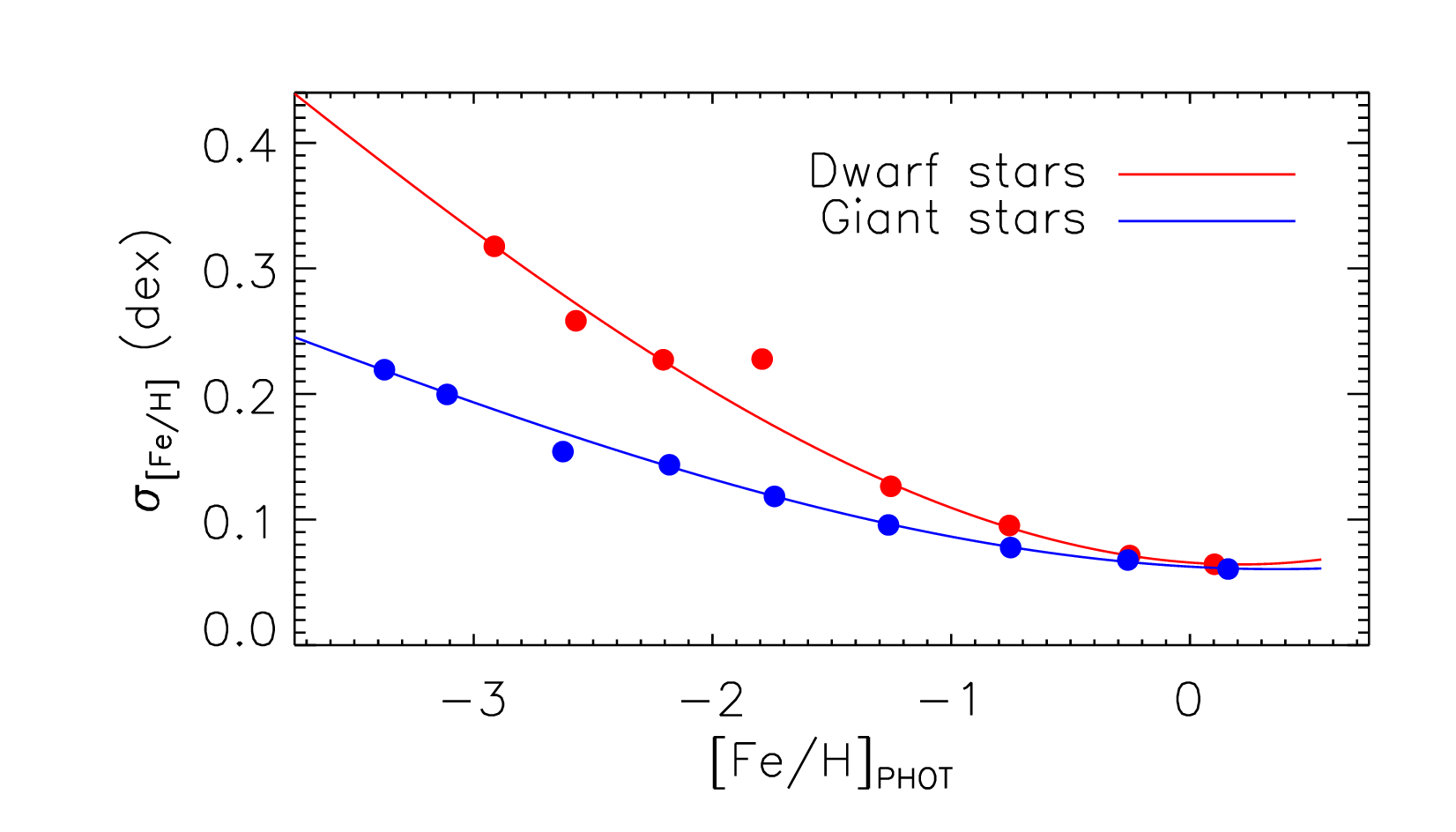}
\includegraphics[scale=0.315,angle=0]{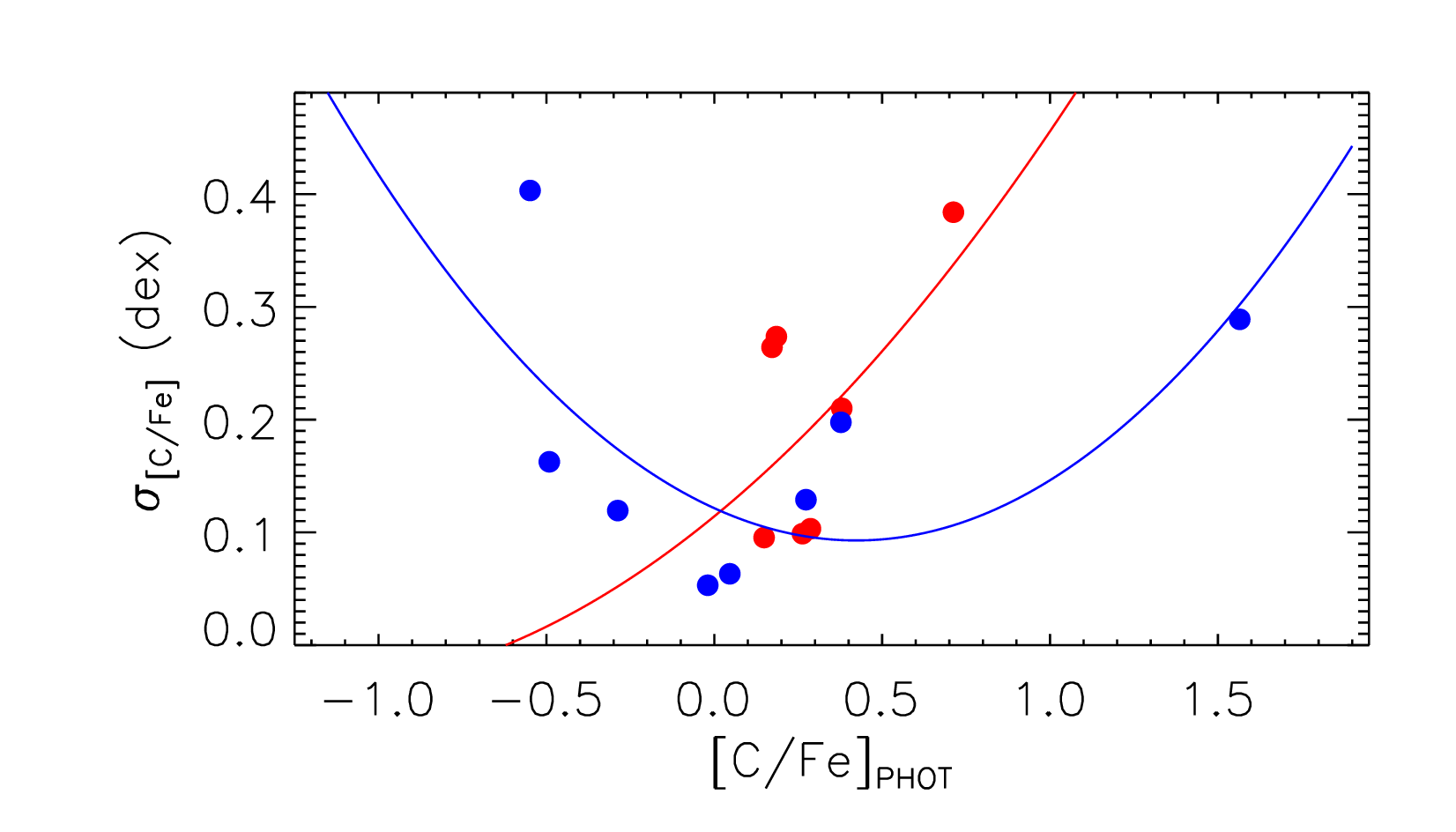}
\includegraphics[scale=0.315,angle=0]{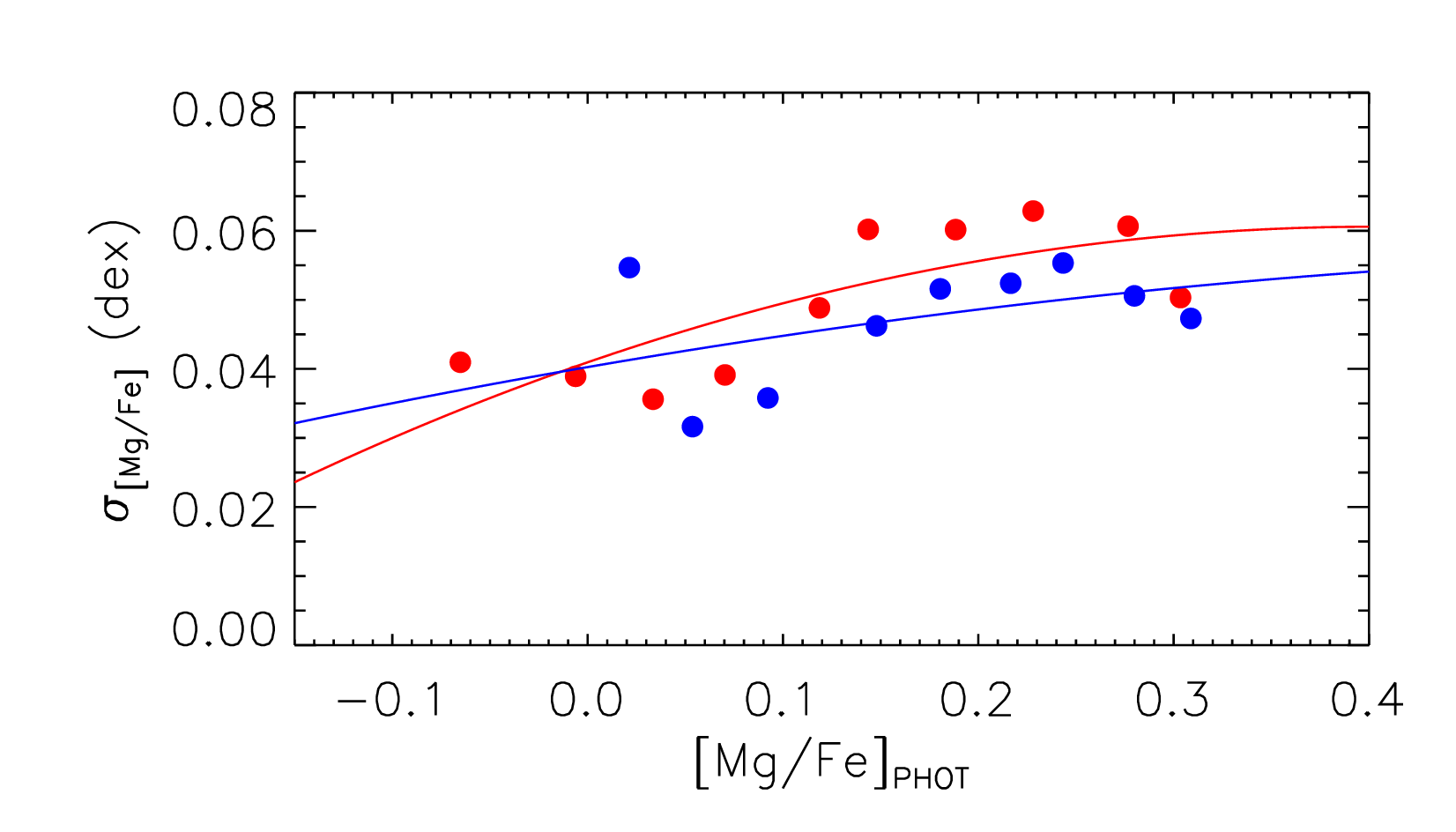}
\includegraphics[scale=0.315,angle=0]{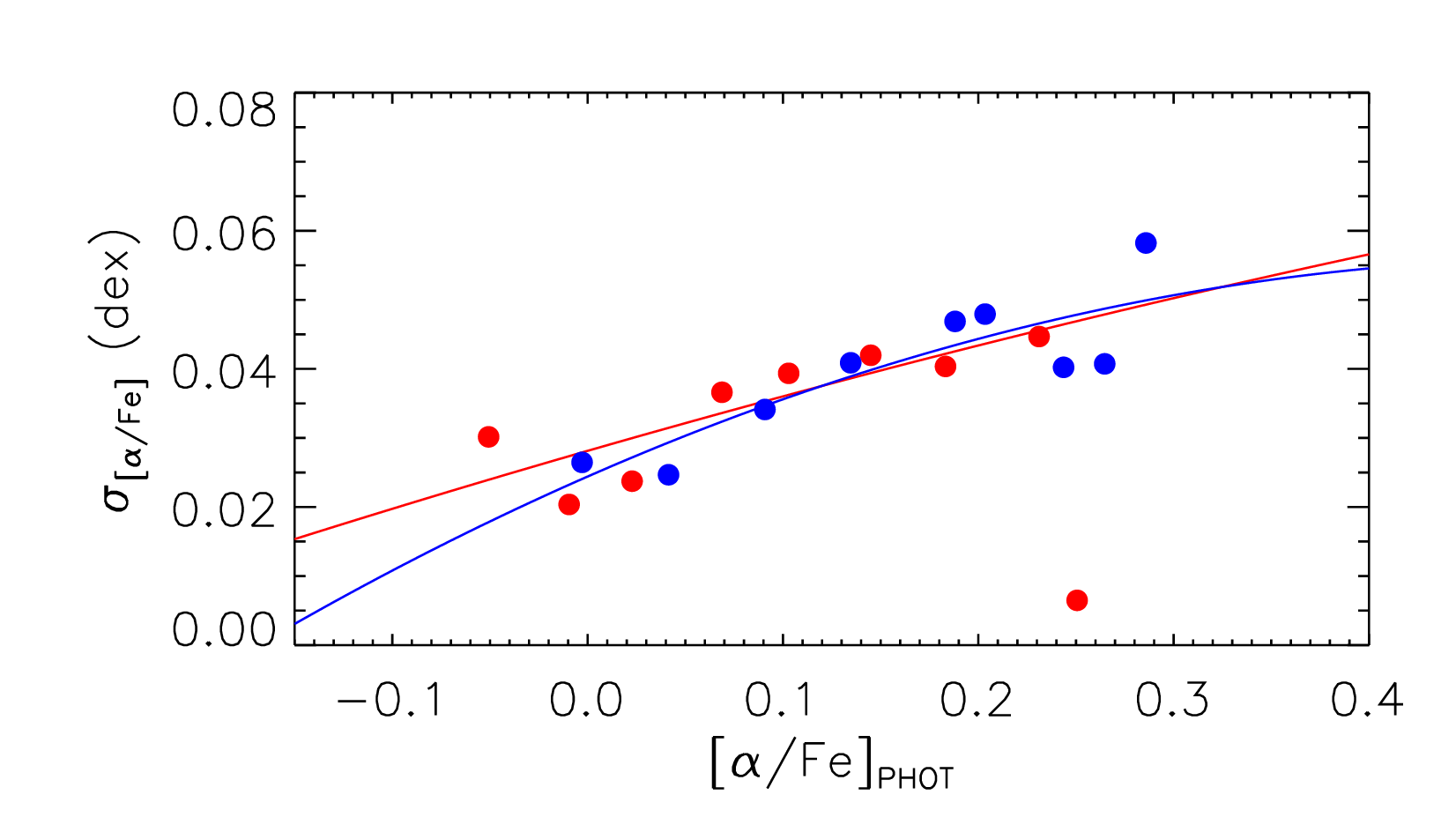}
\caption{Uncertainties in the photometric estimates for [Fe/H] (top-left), [C/Fe] (top-right), [Mg/Fe] (bottom-left) and [$\alpha$/Fe] (bottom-right), as functions of 
the photometric estimates, obtained using the training sets shown in Figs. 3 and 4. The red and blue dots represent the results for dwarf and giant stars, respectively. 
The red and blue lines are second- and third-order polynomial fits to these data 
points, respectively}
\end{center}
\end{figure*} 

\begin{figure*}
\begin{center}
\includegraphics[scale=0.245,angle=0]{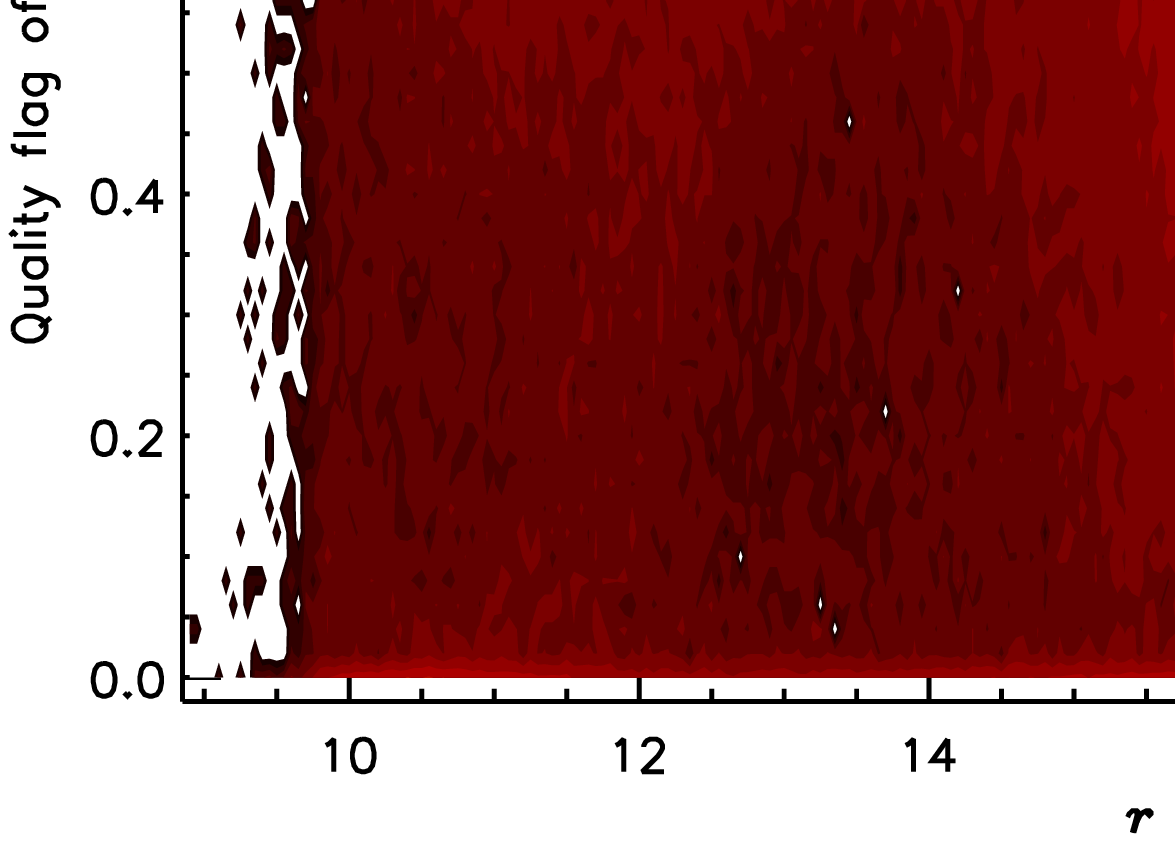}
\includegraphics[scale=0.245,angle=0]{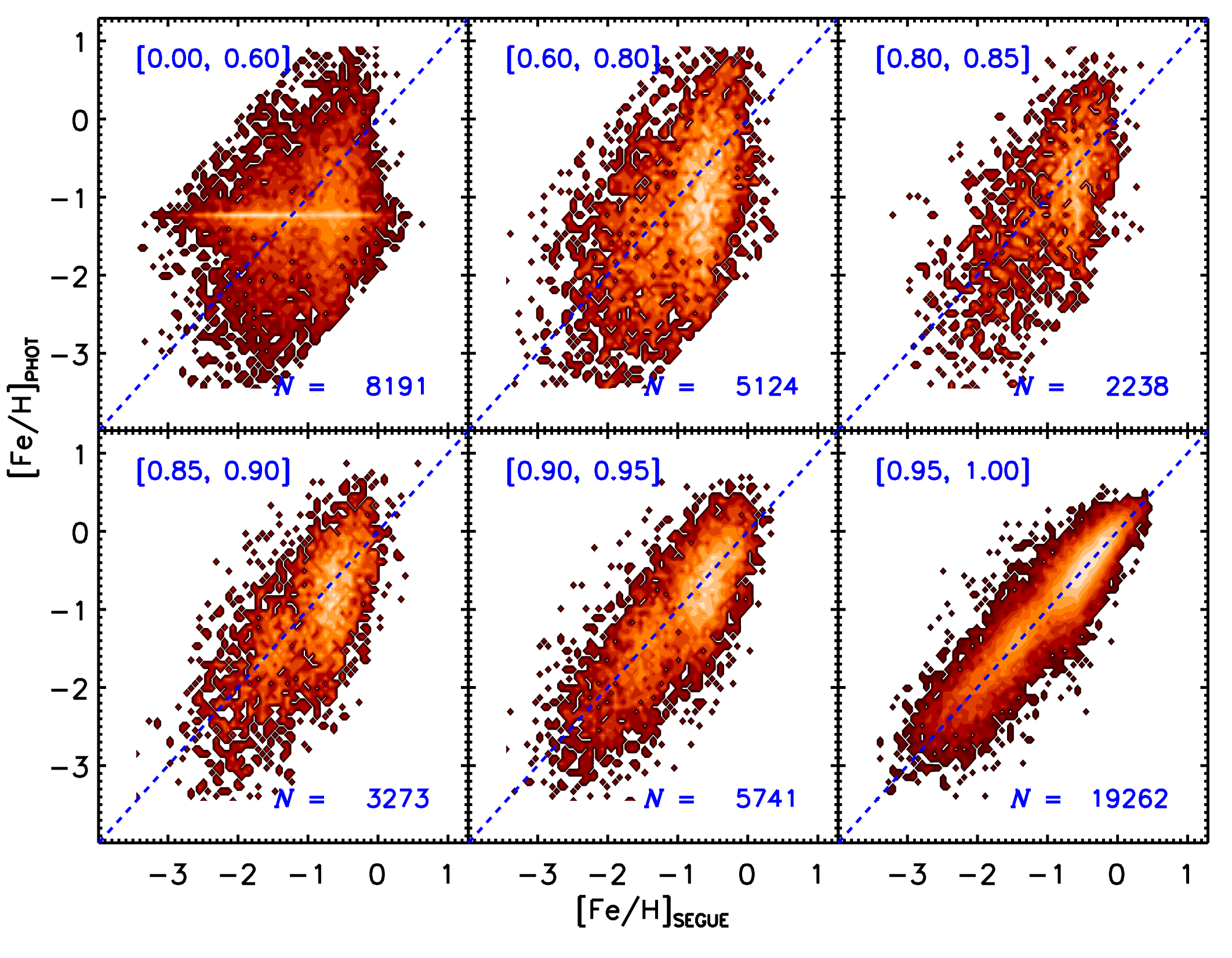}
\caption{Left panel: Density distribution of the [Fe/H] quality flag {\tt flg}$_{\rm [Fe/H]}$, as a function of $r$-band magnitude.  A color bar representing the numbers of stars is provided at the top of the panel.  Right panel: Comparisons of the photometric metallicity and the SDSS/SEGUE spectroscopic metallicity for different ranges of the [Fe/H] quality flag (as marked in the top-left corner of each sub-panel).The SDSS/SEGUE spectroscopic metallicity is corrected for the systematic offsets described in Appendix\,A. The blue-dashed lines are the one-to-one lines. The total number of stars used in the comparison is marked in the bottom-right corner of each sub-panel.}
\end{center}
\end{figure*} 

\begin{figure*}
\begin{center}
\includegraphics[scale=0.35,angle=0]{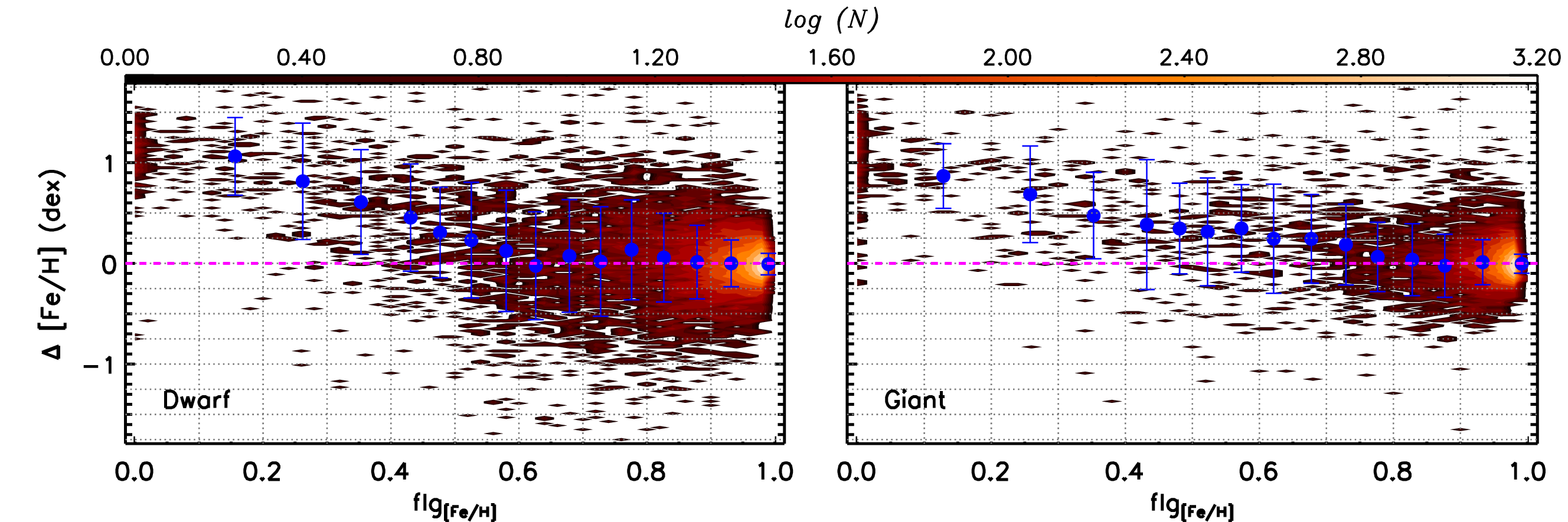}
\includegraphics[scale=0.35,angle=0]{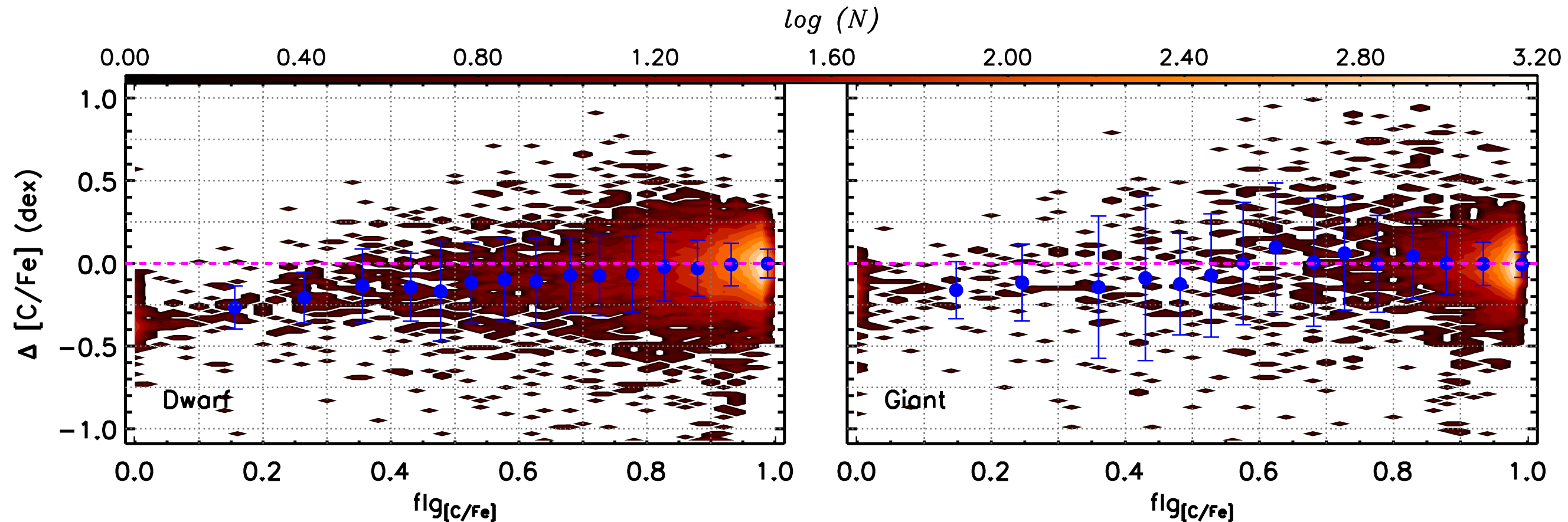}
\includegraphics[scale=0.35,angle=0]{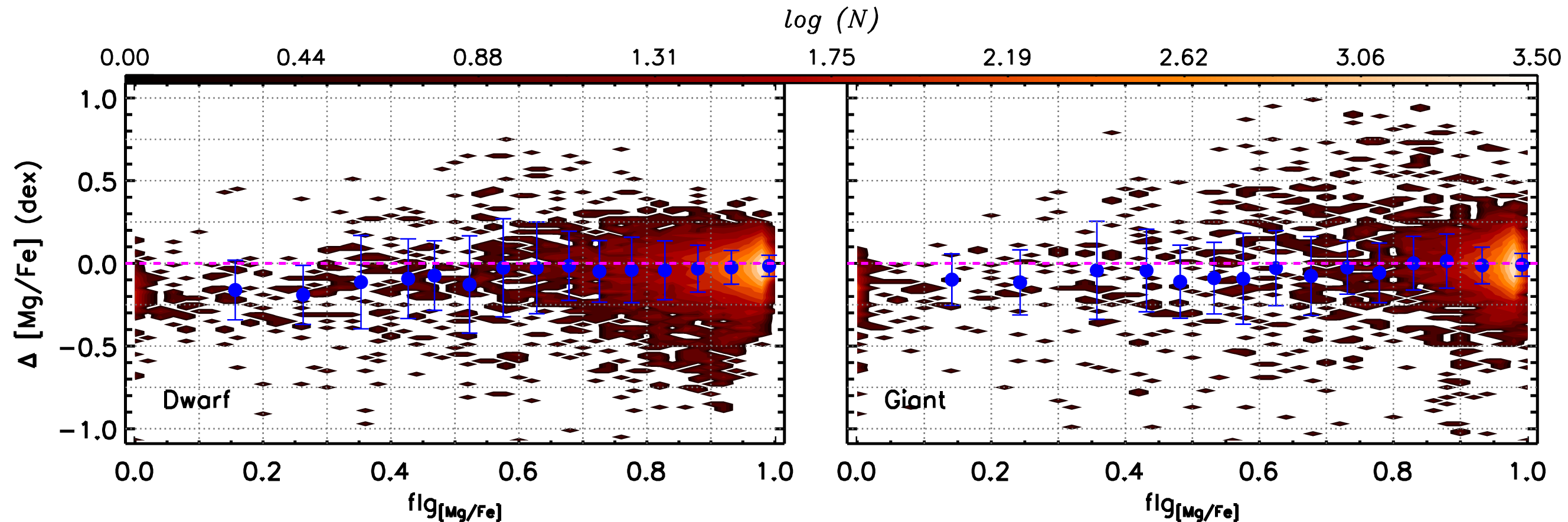}
\includegraphics[scale=0.35,angle=0]{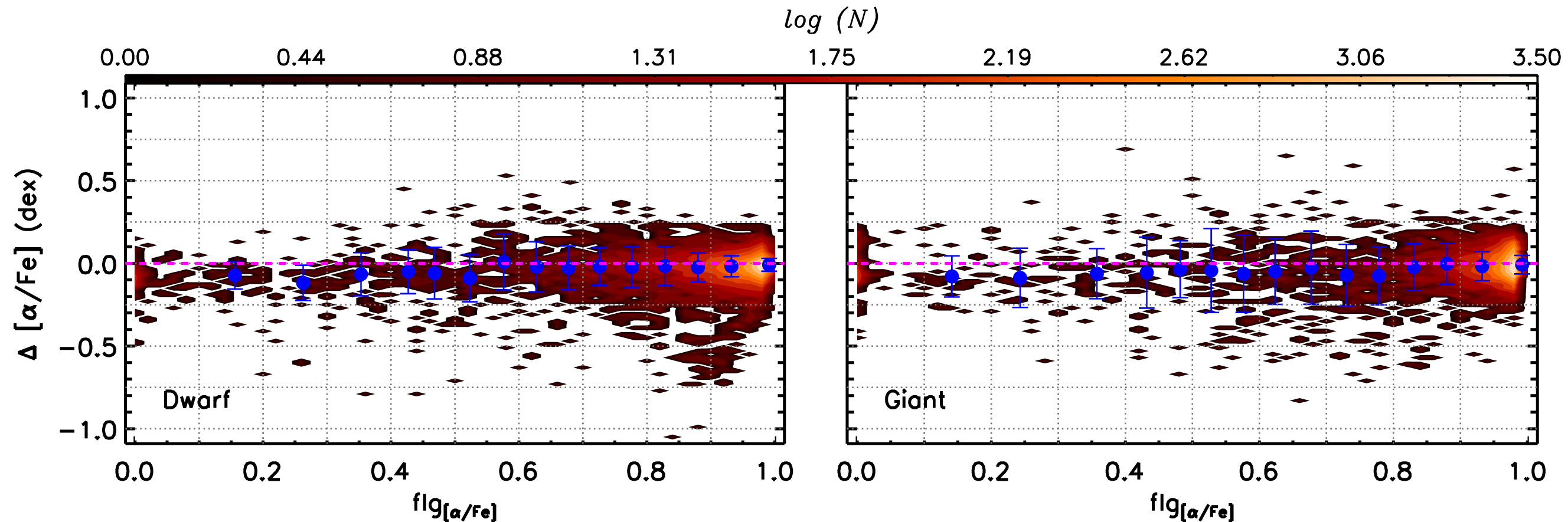}
\caption{Density distributions of the metallicity and elemental-abundance differences (APOGEE estimates minus photometric estimates), as a function of the quality flag for [Fe/H] (first row), [C/Fe] (second row), [Mg/Fe] (third row), and [$\alpha$/Fe] (bottom row). The left column of panels is for dwarf stars and the right column is for giant stars. The blue dots and error bars in each panel represent the median and dispersion of the parameter differences in the individual  bins of the parameter quality flag. Magenta-dashed lines indicate zero residuals in each panel. A color bar representing the numbers of stars is provided at the top of each row of panels. }
\end{center}
\end{figure*} 

\vskip 2cm
\subsection{Metallicity ([Fe/H])}

We adopted metallicity measurements ([Fe/H]) based on high-resolution spectroscopy (HRS) as the reference scale, merging measurements from the Stellar Abundances for Galactic Archaeology (SAGA) Database \citep{2008PASJ...60.1159S} and the PASTEL catalog \citep{2016A&A...591A.118S}.

For stars with multiple measurements, the average values are adopted.
As evaluated in Fig.\,1 of  Paper I, the accuracy is better than 0.05\,dex for stars with [Fe/H]\,$> -2.0$ and 0.05--0.10\,dex for stars with [Fe/H]\,$\leq-2.0$.
The metallicity scale of our compiled sample is compared to the largest homogeneous sample of four hundred very metal-poor stars collected by Subaru Telescope \citep[][see Fig.\,A3]{2022ApJ...931..147L}. From inspection, they are very consistent with each other, with small mean residuals, and scatter of 0.17\,dex.
In total, 24,160 stars with [Fe/H] between $-5.7$ and +1.0 form the HRS database as the reference scale (see Table\,A1).

The metallicity estimated from the completed/ongoing spectroscopic surveys are then compared to the HRS sample (Fig.\,A1). Generally, the metal-rich stars ([Fe/H]\,$>-1.5$) of SDSS/APOGEE, GALAH, and SDSS/SEGUE surveys are consistent with those of the HRS sample, while the metal-poor regions ([Fe/H]\,$\le-1.5$) deviate from the HRS metallicity scale, and exhibit a larger scatter.

To correct for the deviations, second- to third-order polynomials are adopted with the coefficients marked in Fig.\,A1. 
The revised APOGEE metallicity is then adopted to calibrate the stellar parameters from LAMOST in the $T_{\rm eff}$, log\,$g$, and [Fe/H] spaces (Fig.\,A2), given the large number of stars in 
common. As shown in Figs.\,A1 and A2, most of the usual survey pipelines\footnote{For example, the LASP by \citet{2015RAA....15.1095L} and the APOGEE Stellar Parameter and Chemical Abundances Pipeline (ASPCAP) by \citet{2016AJ....151..144G}.} are only able to derive metallicity down to [Fe/H] = $-2.5$, and not much more metal-poor than [Fe/H] = $-3.0$, from the existing spectroscopic surveys.
To improve the calibration on the metal-poor end, we adopted a custom version of the SSPP \citep[the LSSPP;][]{2015AJ....150..187L} for the LAMOST spectra, and the latest version of the SSPP for the SDSS/SEGUE spectra, in order to derive metallicities for hundreds of thousands of metal-poor stars (down to [Fe/H] = $-4.0$) from both surveys.

All of the spectra for those metal-poor stars were visually inspected (by Beers), and any problematic spectra (e.g., defects that might compromise their determinations, as well as contamination from cool white dwarfs, hot B-type subdwarf stars, and emission line objects) are excluded.  The comparisons show that the LSSPP/SSPP metallicity is consistent with that of the HRS scale (see Fig.\,A3), even at the metal-poor end ([Fe/H]$ \sim -4.0$), with no significant offset, and a typical scatter of 0.20--0.30\,dex.

The HRS, SDSS/APOGEE, LAMOST, and LAMOST+SEGUE very metal-poor (VMP; [Fe/H] $\leq -2.0$) samples are then cross-matched with the J-PLUS parent sample to build up the training sample at low metallicity.
The SDSS/SEGUE and GALAH sample stars are then used for validation purposes.

When constructing the training set, the stars are required to satisfy the following  criteria: 1) photometric uncertainties smaller than 0.04\,mag in the 12 J-PLUS bands, and 0.02\,mag in the 3 {\it Gaia} bands; 2) higher Galactic latitudes ($|b| \ge 15^{\circ}$) and low-extinction values ($E (B-V) \le 0.04$); and 3) metallicities estimated from high-quality spectra with signal-to-noise ratio (SNR) at least 20 per pixel.
The training-sample stars are further divided into dwarfs and giants according to an empirical cut in the $(G_{\rm BP} - G_{\rm RP})_0$--$M_{G}$ diagram (see Paper I); the $G$-band absolute magnitudes are derived from {\it Gaia} $G$ (extinction corrected), and geometric distances are estimated from Bayesian-based {\it Gaia} parallax measurements \citep{BJ21}.
All of the metal-poor stars ([Fe/H]\,$\le -2.0$) that satisfy the above criteria are selected.
To achieve good training results for the entire metallicity range, a similar number of relatively more 
metal-rich ([Fe/H]\,$> -2.0$) stars are randomly selected from the millions of stars passing the above cuts.
In total, 6183 and 7052 dwarf and giant stars are selected as the training sets for [Fe/H]; their metallicity distributions are shown in Fig.\,1.
Clearly, the metallicity of our training sample can be extended to [Fe/H]\,$\sim -4.0$. 

 \subsection{Carbon-to-Iron Abundance Ratios} 
 
Unlike for [Fe/H], there are no biographically compiled large catalog of [C/Fe] measurements from HRS observations covering the full range of [Fe/H] and [C/Fe] we consider here. We therefore adopt [C/Fe] measured from the SDSS/APOGEE survey as the reference scale. The comparison shown in Fig.\,A5 indicates that the [C/Fe] measurements of the LAMOST/SEGUE VMP samples are consistent with those of SDSS/APOGEE, with negligible offsets, and scatters of around 0.1\,dex\footnote{The scatter is slightly larger, around 0.3\,dex, for the LAMOST VMP samples.}.
Again, the  SDSS/APOGEE and LAMOST/SEGUE VMP samples are cross-matched with the J-PLUS parent sample to construct the training sets for [C/Fe]; the GALAH sample stars are used for test 
purposes. 
Here we note that evolution-dependent corrections \citep[e.g.,][]{Placco14} are not made for the adopted [C/Fe] estimates.
The strategy for defining training stars is the same as we have used for [Fe/H]. 
In total, the [C/Fe] training set contains 5830 and 6824 dwarf and giant stars, respectively; their distributions in the [C/Fe] vs. [Fe/H] space are shown in the upper panels of Fig.\,2.
As found by many previous studies, the fraction of CEMP stars ([C/Fe]\,$> +0.7$) increases rapidly with decreasing [Fe/H].

\vskip 1cm
 \subsection{Magnesium-to-Iron and $\alpha$-to-Iron Abundance Ratios}

Similar to [C/Fe], [Mg/Fe] and [$\alpha$/Fe] from the SDSS/APOGEE survey are adopted for the reference scales. 
The results from GALAH are quite consistent with those from SDSS/APOGEE, as shown in Fig.\,A5.
The uncertainties of [Mg/Fe] and [$\alpha$/Fe] in the LAMOST/SEGUE VMP samples are quite large; thus they are not adopted in the training sets. 
The lack of VMP sample stars in our training sets may introduce issues for estimating [Mg/Fe] and [$\alpha$/Fe] for VMP stars, which are discussed in Section\,4.4.
Finally, 4043 dwarf and 4984 giant stars, respectively, are chosen for the training sets of [Mg/Fe] (middle panels in Fig.\,2); 4041 dwarf and 4990 giant stars, respectively, are selected for the training sets of [$\alpha$/Fe] (bottom panels in Fig.\,2).
The GALAH sample stars are again adopted for checking the estimation precisions of [Mg/Fe] and  [$\alpha$/Fe].

\section{Estimates of [F\MakeLowercase{e}/H], [C/F\MakeLowercase{e}], [M\MakeLowercase{g}/F\MakeLowercase{e}], and [$\alpha$/F\MakeLowercase{e}] from the J-PLUS and {\it G\MakeLowercase{aia}} colors}

\subsection{Kernel Principal Component Analysis}

%In this study, a Kernel Principal Component Analysis \citep[KPCA;][]{kpca} is adopted  to construct empirical relations between the stellar labels ([Fe/H], [C/Fe], [Mg/Fe], and [$\alpha$/Fe]) and the colors from J-PLUS and {\it Gaia}. 
Principal Component Analysis (PCA) is widely used in astronomy for transforming observational features (e.g., spectra, multiple colors) to a set of uncorrelated orthogonal principal components.
Kernel Principal Component Analysis \citep[KPCA;][]{kpca} is an extension of PCA, which adds a kernel technique for nonlinear feature extraction. This approach has been applied to estimate stellar parameters, including atmospheric parameters, mass, and age, from stellar spectra \citep[e.g.,][]{Huang15, Huang20, Xiang17, Wu19}. 
Here we adopt the KPCA  technique\footnote{Here we set the kernel to be a Gaussian radial basis function.} to derive metallicity and elemental-abundance ratios (i.e., [C/Fe], [Mg/Fe] and [$\alpha$/Fe]) from the multiple colors formed with the combination of J-PLUS DR3 and {\it Gaia} EDR3 \citep{GEDR3} magnitudes.
The number of principal components and the radius of the Gaussian radial basis are chosen by a series of tests to achieve a tradeoff between reducing the training residuals and avoiding over-fitting. 

\subsection{Training Results}

To derive photometric estimates of metallicity and elemental-abundance ratios, 13 stellar colors: $G_{\rm BP} - G_{\rm RP}$, $G_{\rm BP} - u$, $G_{\rm BP} - g$, $G_{\rm RP} - r$,  $G_{\rm RP} - i$, $G_{\rm RP} - z$,  $G_{\rm BP} - J0378$, $G_{\rm BP} - J0395$, $G_{\rm BP} - J0410$, $G_{\rm BP} - J0430$, $G_{\rm BP} - J0515$, $G_{\rm RP} - J0660$, and $G_{\rm RP} - J0861$ are used as observational inputs.
All the colors are de-reddened using the SFD extinction map, as well as the extinction coefficients described in Section\,2.2.
With the KPCA technique, we construct the relations between the stellar labels from the aforementioned training sets and these 13 colors.

The training results for [Fe/H] and [C/Fe] are shown in Fig.\,3; those for [Mg/Fe] and [$\alpha$/Fe] are shown in Fig.\,4.

Generally, the trained photometric metallicity estimate is consistent with that of the spectroscopic estimate, even at the metal-poor end, down to [Fe/H]\,$\sim -4.0$. 
No significant offsets are detected down to [Fe/H] = $-2.5$, while small deviations ($-0.10$ to $-0.25$\,dex) are found at the more metal-poor end.
The dispersion is tiny ($< 0.10$\,dex) for the metal-rich region ([Fe/H]\,$> -1.0$), and 0.10--0.30\,dex for the metal-poor region ([Fe/H]\,$< -1.0$).
We note that, overall,  the [Fe/H] precision for giant stars is better than that for the dwarf stars, especially at the metal-poor end, as is expected due to the weaker absorption lines for the warmer dwarfs. 

The photometric [C/Fe] estimates for dwarf stars have moderate offsets in both the 
carbon-rich region ([C/Fe]\,$ >  +1.0$; the offset is around $+0.5$\,dex in the sense spectroscopic minus photometric) and the carbon-poor region ([C/Fe]\,$< -0.5$; the offset is around $-0.5$\,dex). The precision is also a function of [C/Fe]; about 0.05--0.10\,dex in the middle region ($-0.5 <$\,[C/Fe]\,$<+0.5$), and up to 0.2--0.4\,dex in the carbon-rich/poor ends. For giant stars, the photometric [C/Fe] estimate is quite good compared to the spectroscopic one; no significant offset is detected.
The precision is about 0.05--0.10\,dex for the middle region ($-0.5 <$\,[C/Fe]\,$<+0.5$) and 0.15--0.25\,dex for the carbon-rich and carbon-poor ends.

For [Mg/Fe] and [$\alpha$/Fe], the photometric estimates agree with those of the spectroscopic results, with precisions better than 0.05\,dex for both dwarf and giant stars, although slight offsets are found in the high [Mg/Fe] and [$\alpha$/Fe] ranges.
The overall precision of [$\alpha$/Fe] is slightly better than that of [Mg/Fe].
The scatter of the comparisons between the photometric and spectroscopic estimates for these parameters, as a function of the photometric estimates, are shown  in Fig.\,5. 

 Finally, at the suggestion of the referee, we clarify the error levels associated with our determinations of [Fe/H], [Mg/Fe], and [$\alpha$/Fe].
Our estimates are based on the labels from spectroscopy, in particular from the HRS; thus the errors are ``inherited" from the labels used for the calibration stars. The best precision we can achieve is thus close to that of the HRS.   
For the [Fe/H] HRS measurements, the error comes from two sources: 1) a statistical error from the scatter of measurements for different iron lines; and 2) a systematic error from the uncertainties in the atmospheric parameters (which can differ somewhat between different sub-samples in the HRS). In most cases, the latter one dominates the total error of the [Fe/H] determinations (e.g., a typical 100~K $T_{\rm eff}$ error will result in about a 0.1 dex error in [Fe/H] for Solar-abundance stars). For [Mg/Fe] or [$\alpha$/Fe], the error origin is the same as that due to [Fe/H]. The statistical error is similar to that for [Fe/H]; the systematic error is largely reduced since, using Fe as the reference element, as Mg shares similar systematics. 
This accounts for why the uncertainty of [Fe/H] is larger than that of [Mg/Fe] and [$\alpha$/Fe].

\begin{figure*}
\begin{center}
\includegraphics[scale=0.315,angle=0]{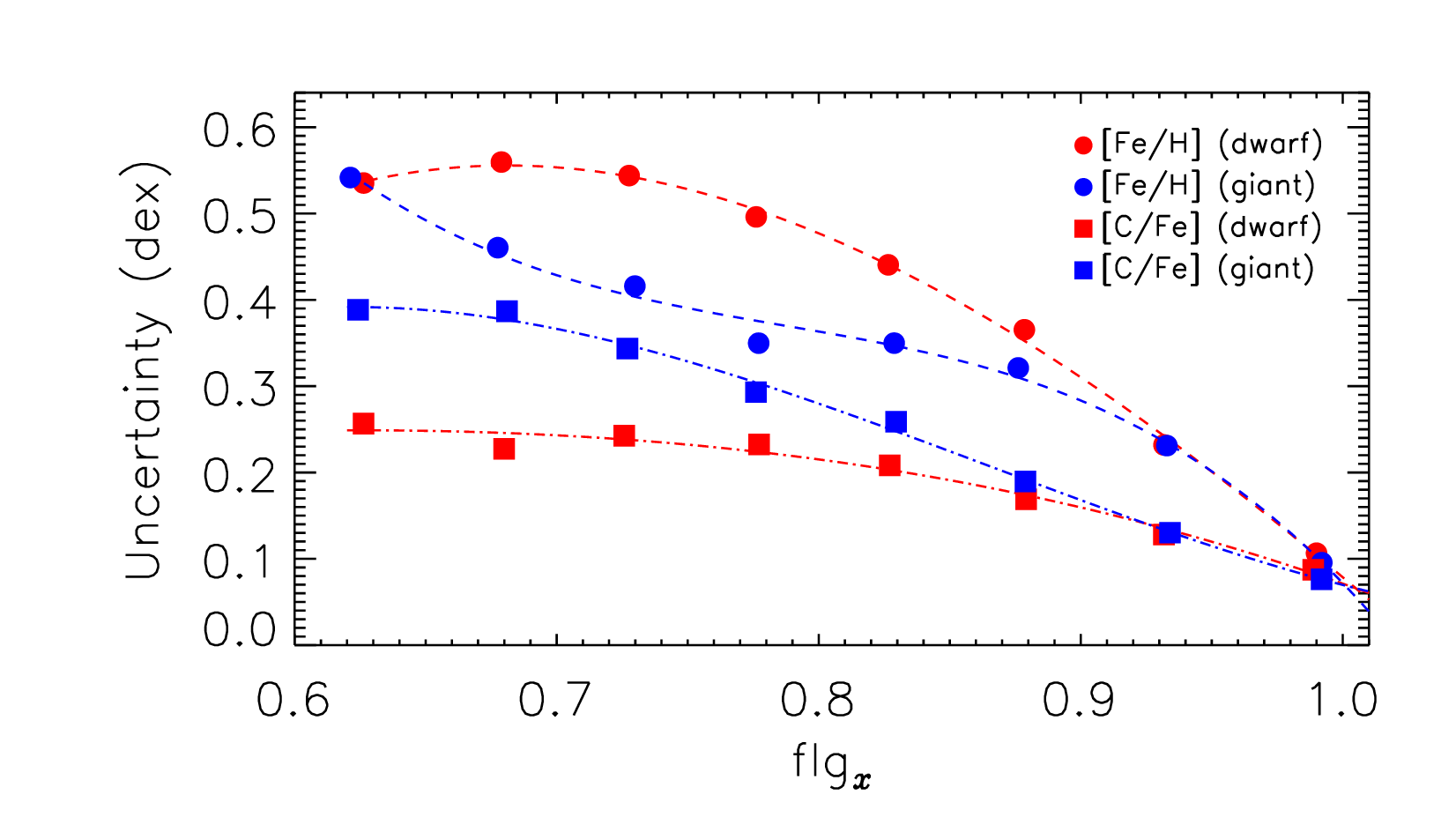}
\includegraphics[scale=0.315,angle=0]{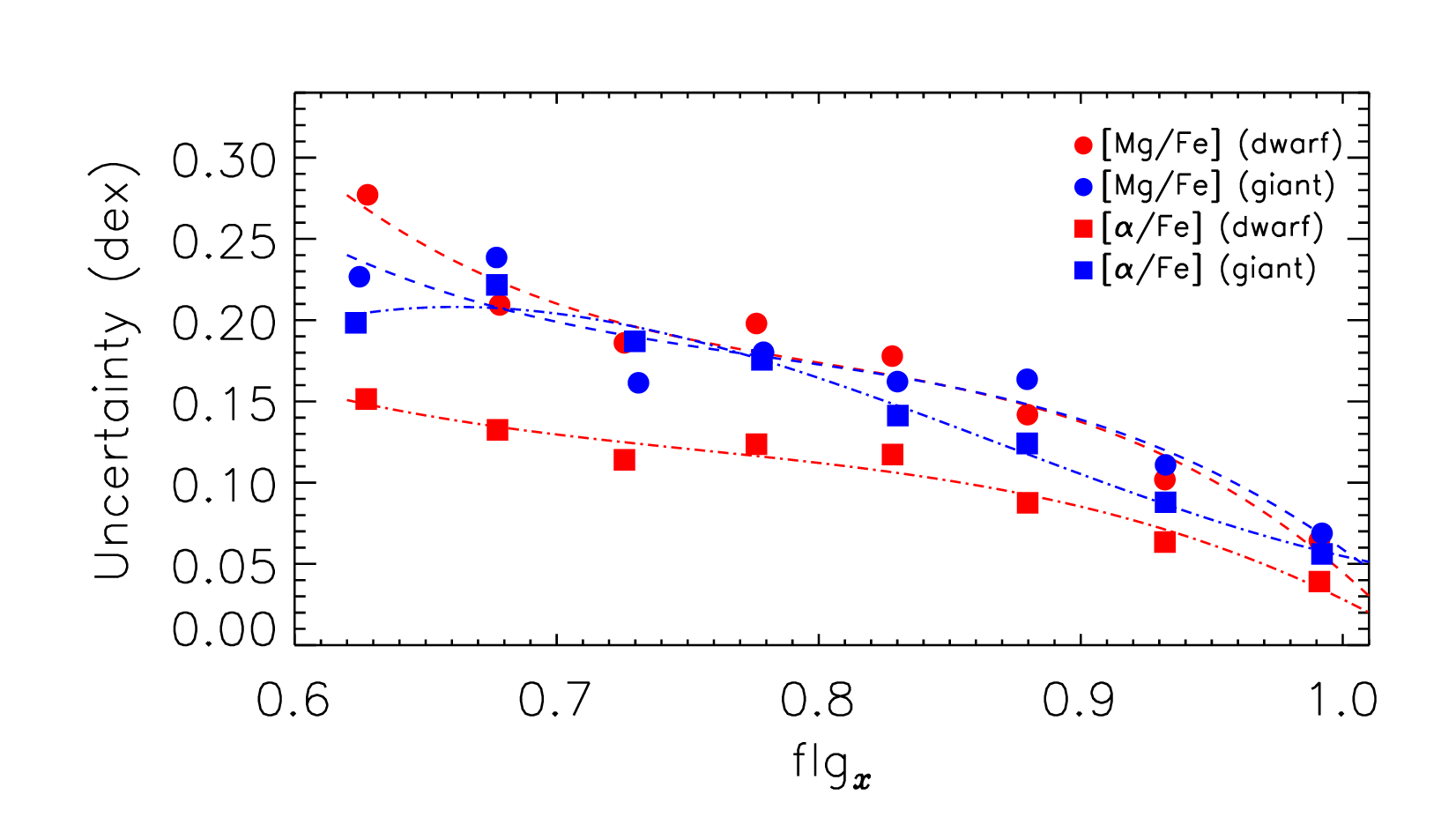}
\caption{Uncertainties of photometric estimates for [Fe/H] and [C/Fe] (left panel), and [Mg/Fe], [$\alpha$/Fe] (right panel), derived from comparison with the APOGEE--J-PLUS stars in common, as function of the photometric quality flags, {\tt flg$_x$}. Red and blue symbols represent the results for dwarf and giant stars, respectively. The dashed lines represent third-order polynomial fits to these data points.}
\end{center}
\end{figure*}

\begin{figure*}
\begin{center}
\includegraphics[scale=0.35,angle=0]{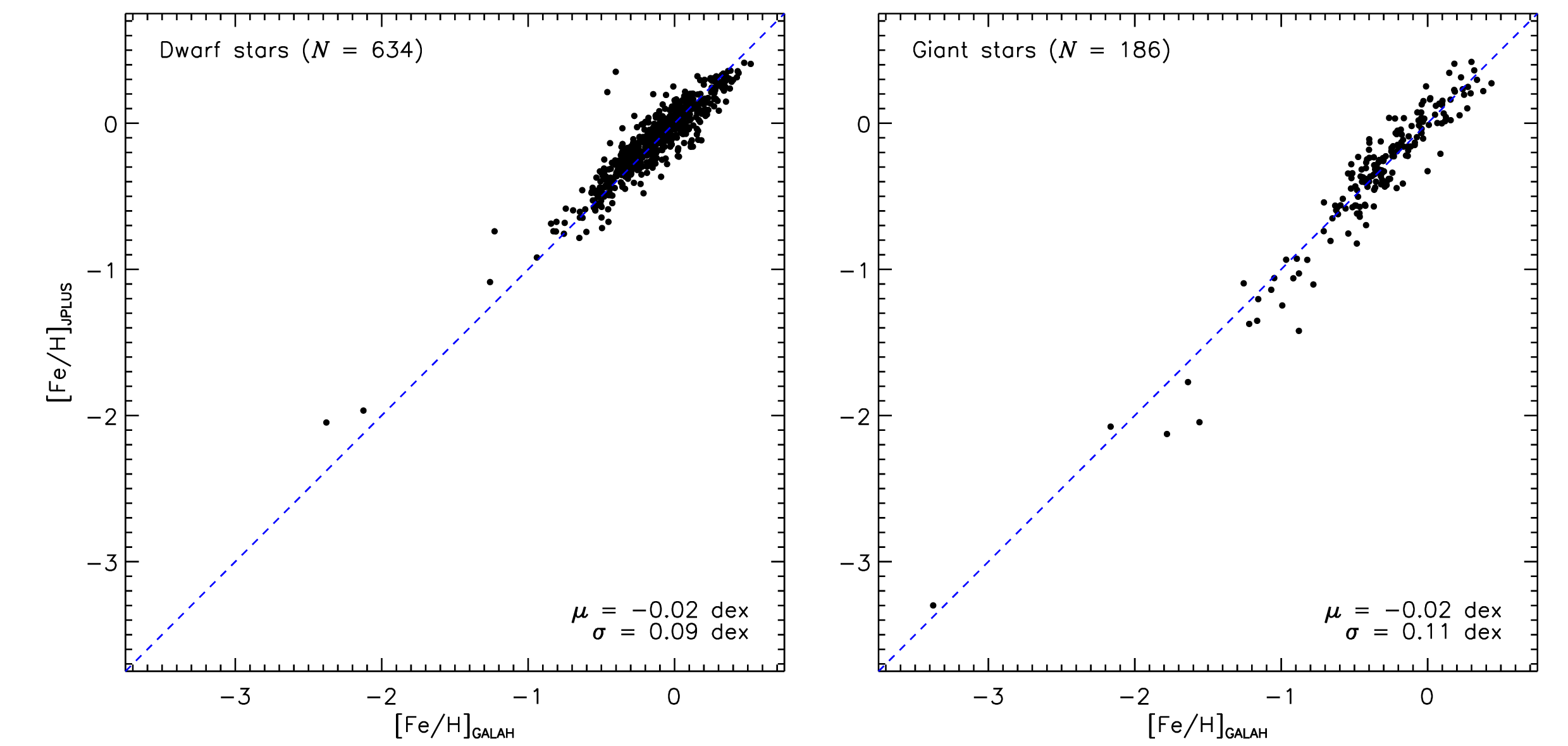}
\includegraphics[scale=0.35,angle=0]{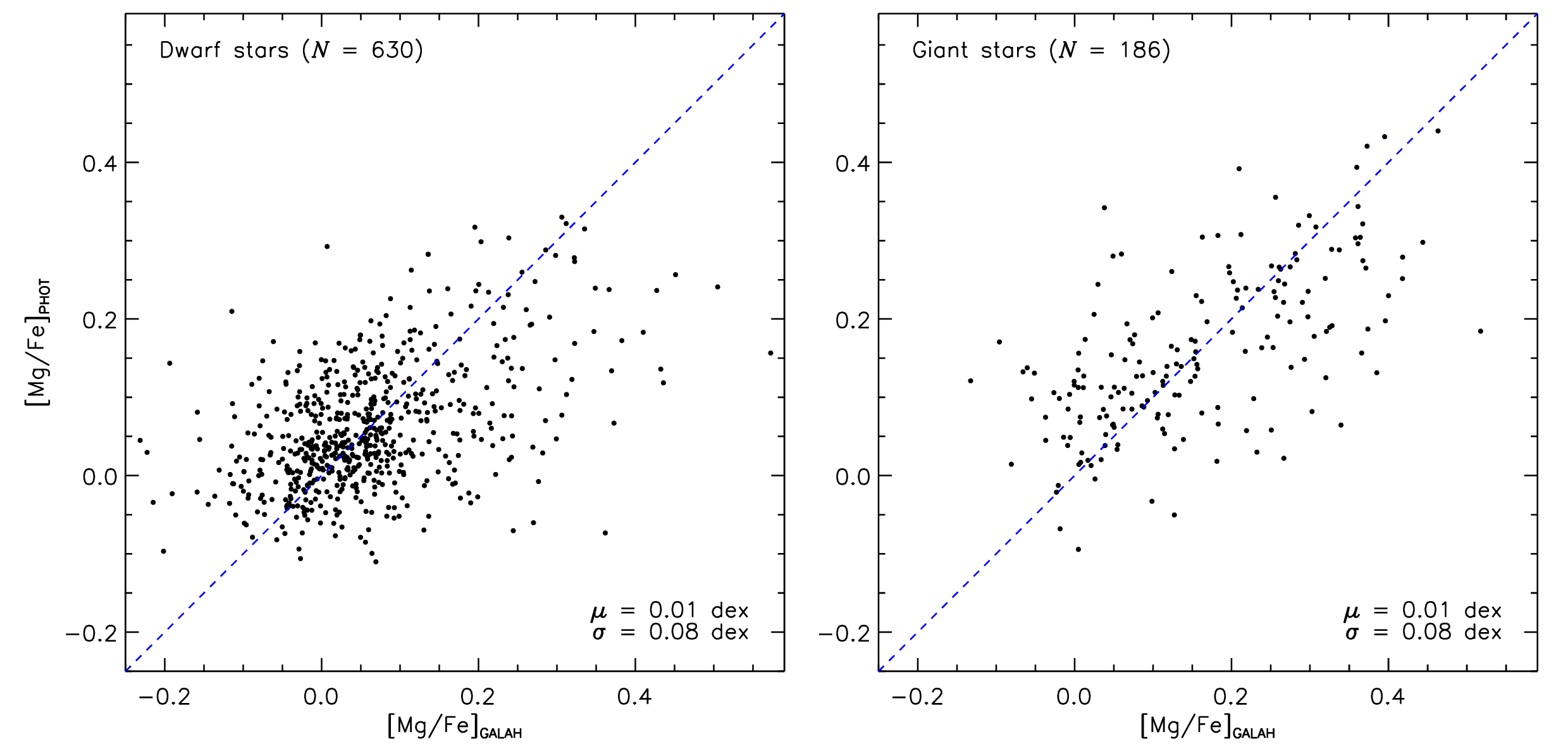}
\includegraphics[scale=0.35,angle=0]{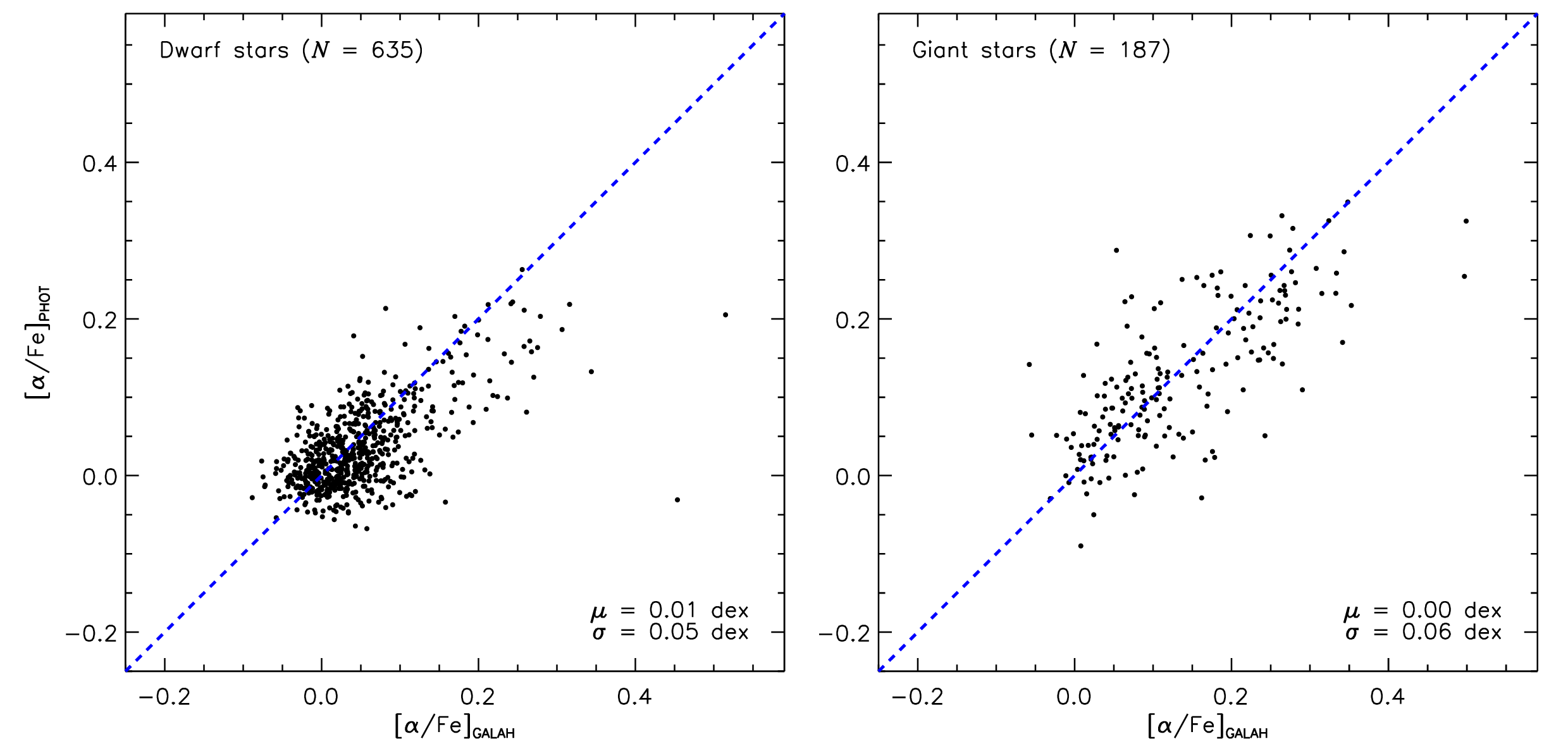}
\caption{Comparisons of the photometric estimates of [Fe/H] (top panels), [Mg/Fe] 
(middle panels), and [$\alpha$/Fe] (bottom panels) with spectroscopic estimates from GALAH DR3. The left column of plots apply to dwarf stars and the right column applies to giant stars. The blue-dashed lines are the one-to-one lines. The overall median offset and standard deviation are marked in the bottom-right corner of each panel.}
\end{center}
\end{figure*} 

\begin{figure*}
\begin{center}
\includegraphics[scale=0.35,angle=0]{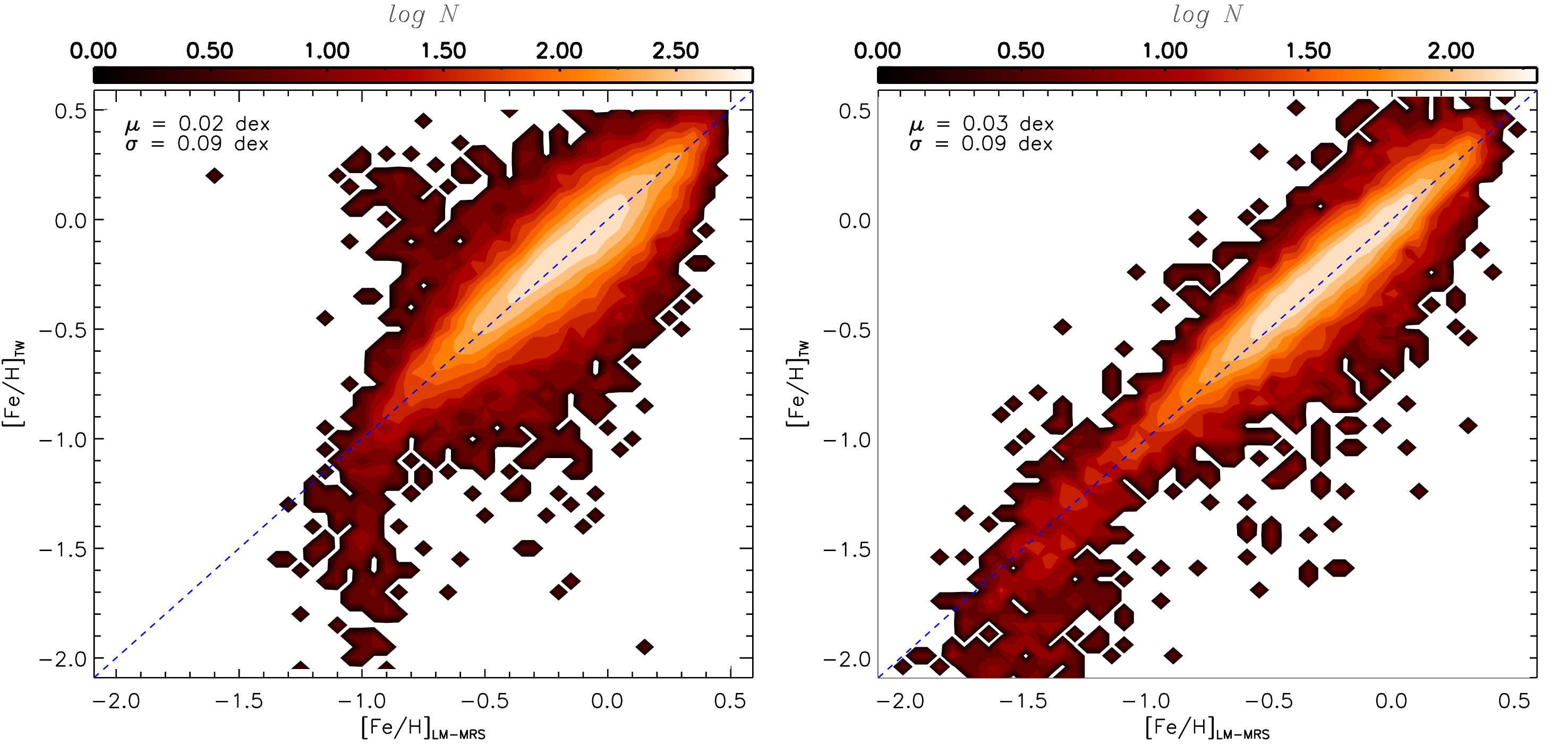}
\caption{Comparisons of the photometric estimates with those from \cite{2023MNRAS.523.5230L} for dwarf (left panel) and giant (right) stars.
The blue-dashed lines are the one-to-one lines. 
The overall median offset and standard deviation are marked in the top-left corner of each panel. Color bars representing the numbers of stars are provided at the top of each
panel.}
\end{center}
\end{figure*} 

\begin{figure*}
\begin{center}
\includegraphics[scale=0.335,angle=0]{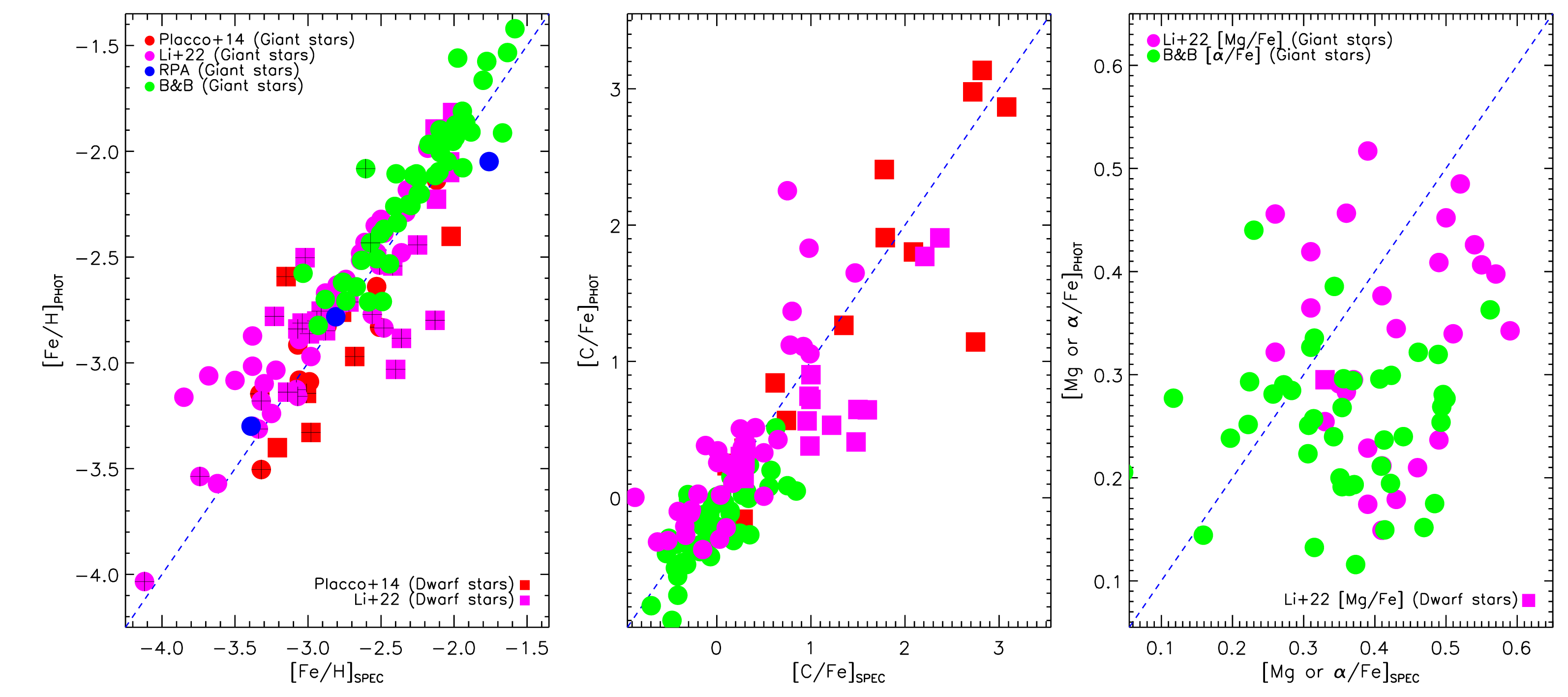}
\caption{Comparisons of the photometric estimates of [Fe/H] (left panel), [C/Fe] (middle panel), and [Mg/Fe] or [$\alpha$/Fe] (right panel) with those from the Best \& Brightest Survey \citep[green dots;][]{BNB} and the HRS samples, the CEMP stars are from \citet[][red dots]{Placco14}, the $R$-Process Alliance
sample \citep[blue dots;][]{2018ApJ...858...92H, 2018ApJ...868..110S, 2020ApJ...898..150E, 2020ApJS..249...30H}, and the Subaru follow-up observations of LAMOST VMP candidates \citep[magenta dots;][]{2022ApJ...931..146A, 2022ApJ...931..147L}. In the left panel, the plus symbols mark the CEMP stars with [C/Fe]\,$> +0.7$.  The blue-dashed lines are the one-to-one lines. }
\end{center}
\end{figure*} 

\begin{figure}
\begin{center}
\includegraphics[scale=0.235,angle=0]{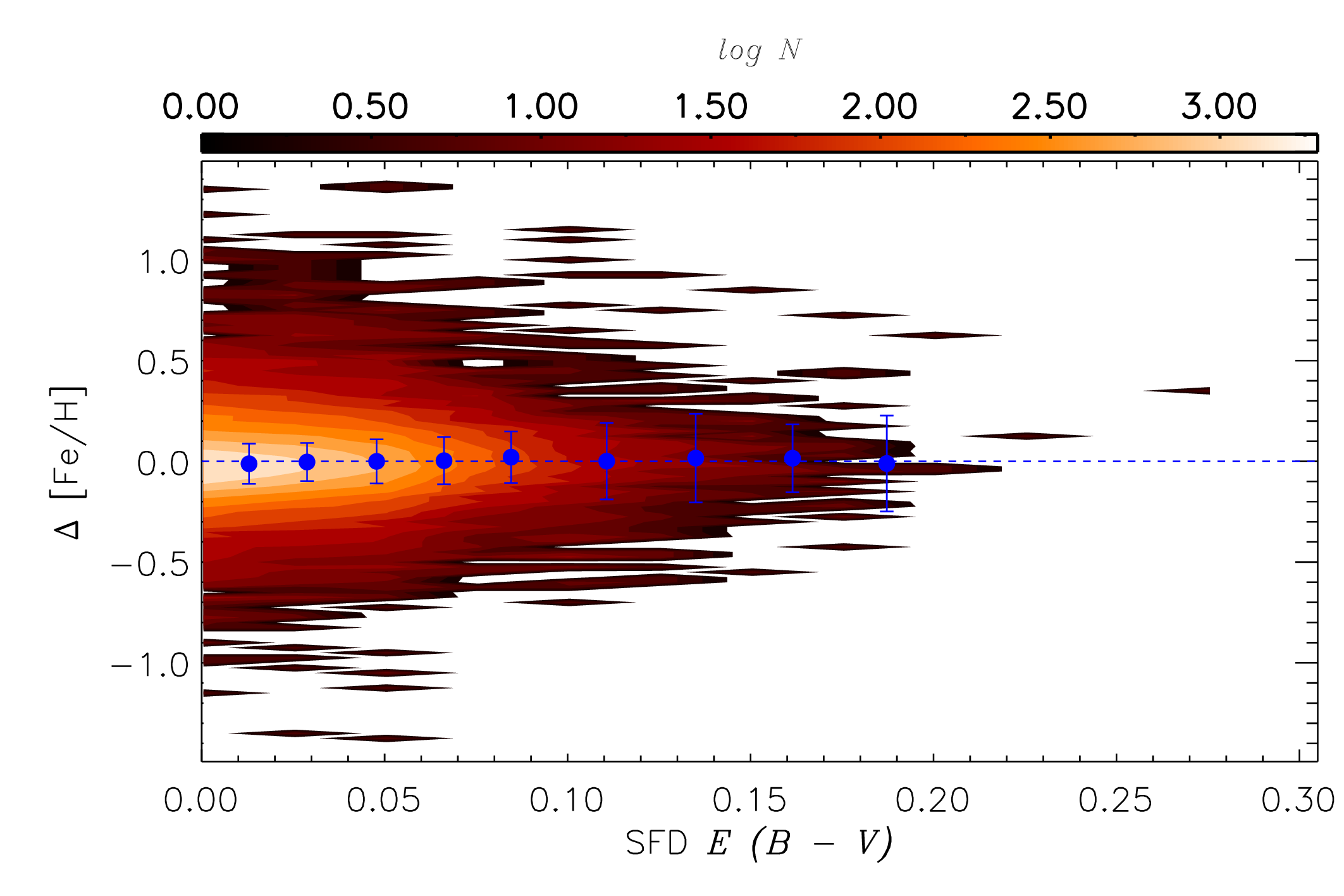}
\caption{Density distributions of the metallicity difference for APOGEE--J-PLUS, as a function of SFD $E (B - V)$. The blue-dashed line indicates zero residuals. The blue dots and error bars represent the median and dispersion
of the metallicity difference in the individual bins of SFD $E (B - V)$ reddening estimate. 
No trend of the metallicity difference with SFD $E (B - V)$ is detected.
The color bar at the top of represents the numbers of stars.}
\end{center}
\end{figure} 

\begin{figure*}
\begin{center}
\includegraphics[scale=0.375,angle=0]{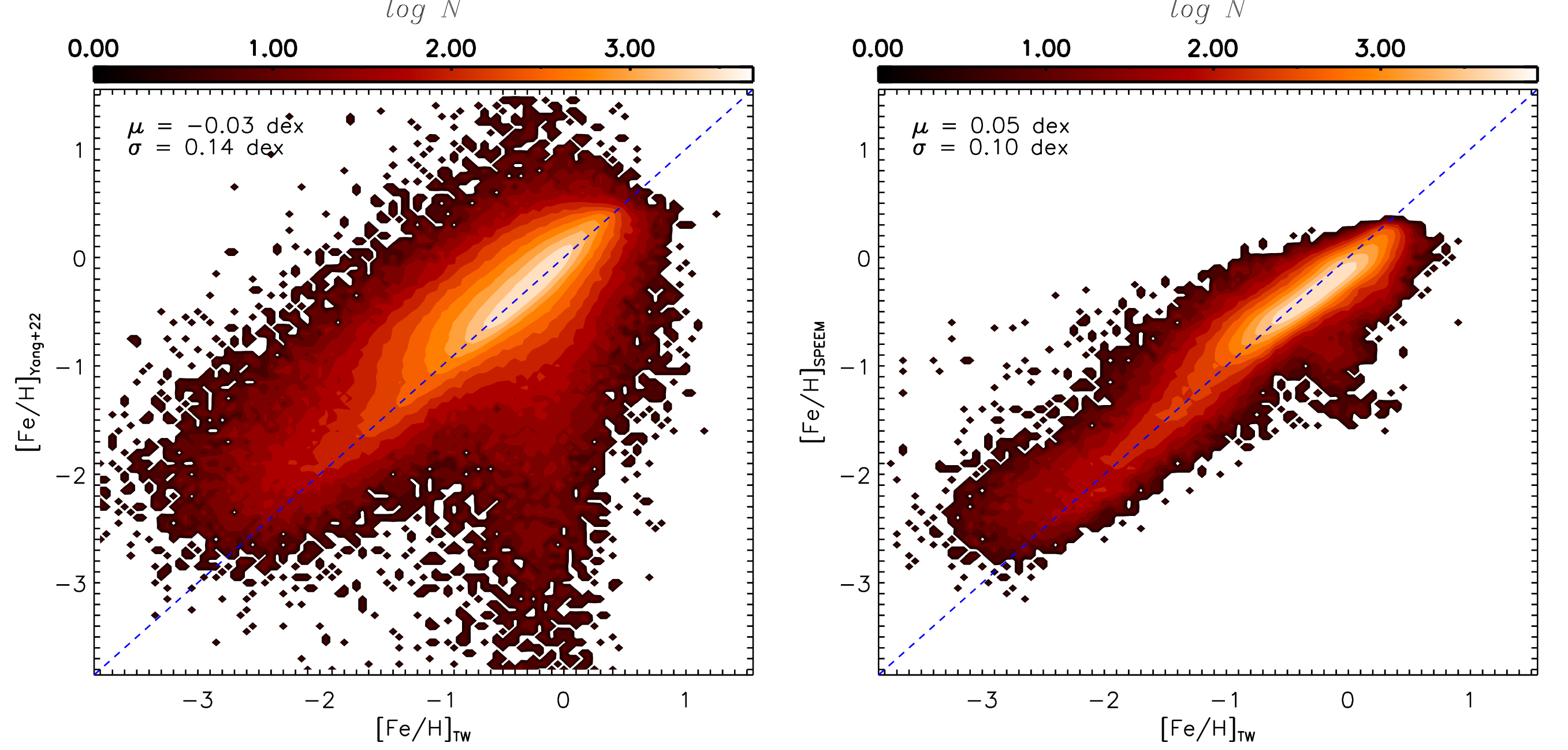}
\caption{Comparisons of the photometric estimates of [Fe/H] with those from \citet[][left panel]{2022A&A...659A.181Y} and those from \citet[][right panel]{2022A&A...657A..35G} using around 0.5 million stars in common. The blue-dashed lines are the one-to-one lines. The overall median offset and standard deviation are marked in the top-left corner of each panel. Color bars representing the numbers of stars are provided at the top of each panel.}
\end{center}
\end{figure*} 

\begin{figure*}
\begin{center}
\includegraphics[scale=0.325,angle=0]{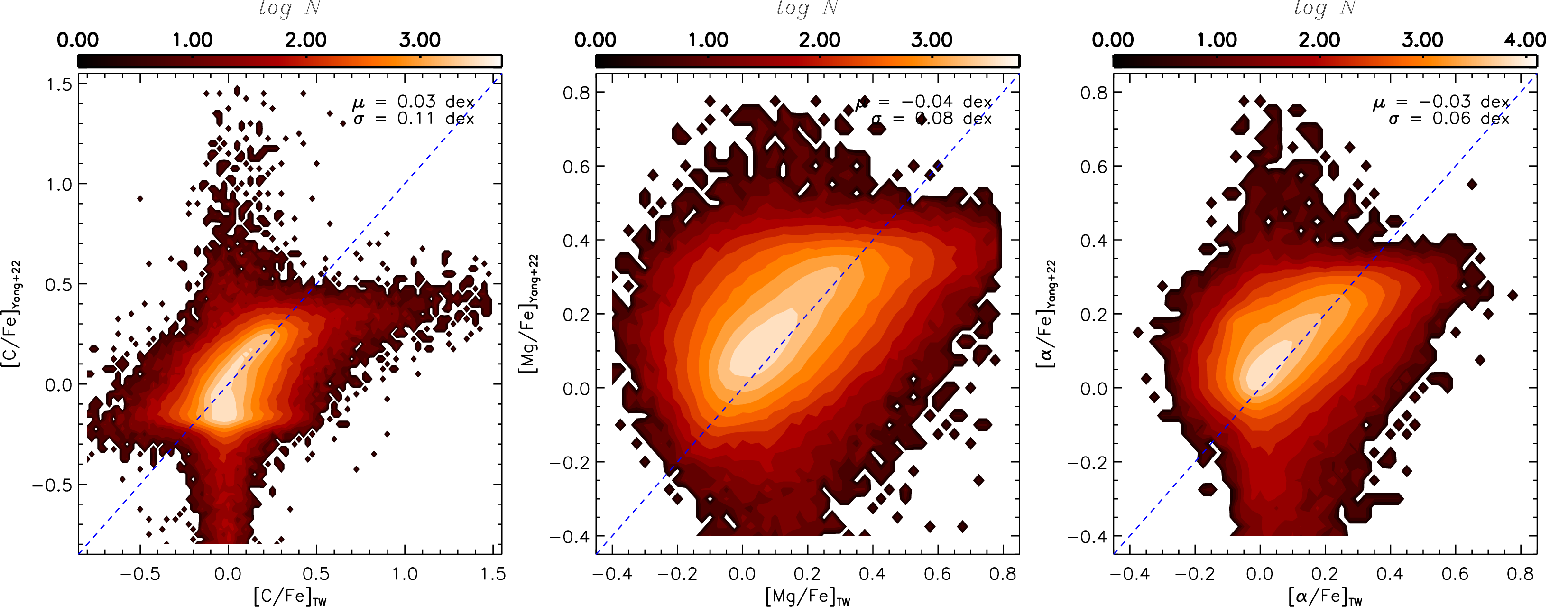}
\caption{Comparisons of the photometric estimates of [C/Fe] (left panel), [Mg/Fe] (middle panel), and [$\alpha$/Fe] (right panel) to those from \citet{2022A&A...659A.181Y}, using the approximately 0.5 million stars in common. The blue-dashed lines are the one-to-one lines. The overall median offset and standard deviation are marked in the top-right corner of each panel. Color bars representing the numbers of stars are provided at the top of each panel.}
\end{center}
\end{figure*} 

\begin{figure*}
\begin{center}
\includegraphics[scale=0.295,angle=0]{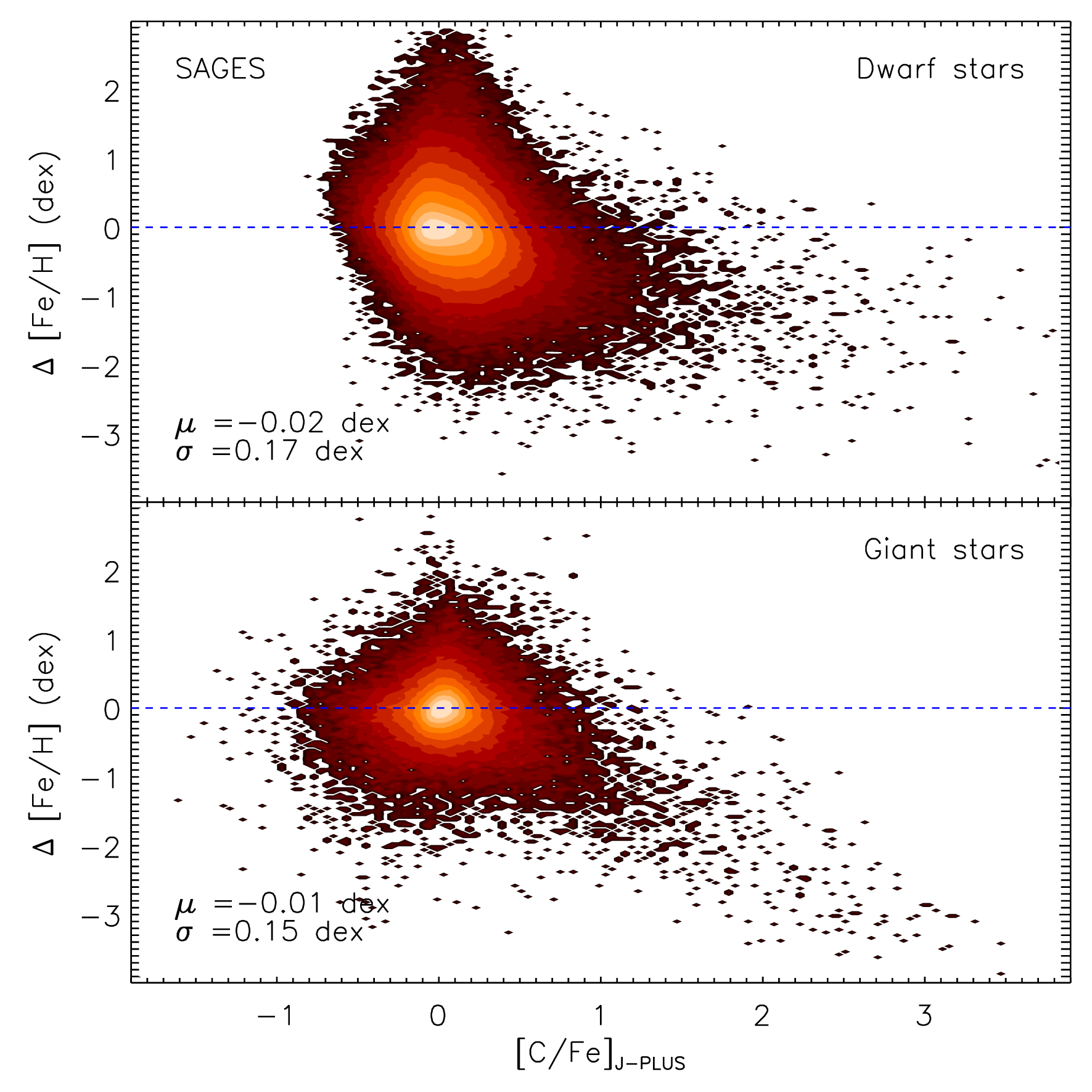}
\includegraphics[scale=0.295,angle=0]{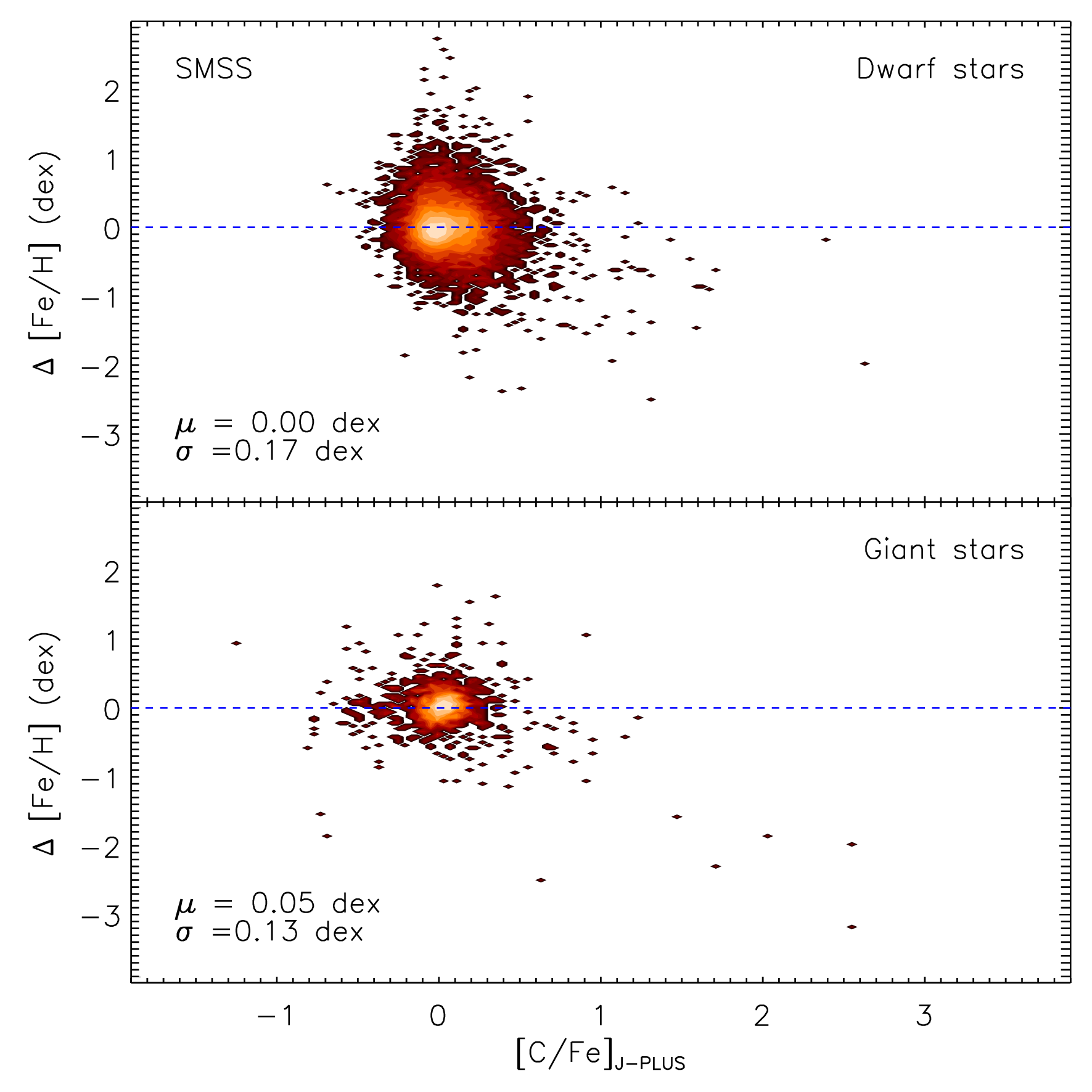}
\includegraphics[scale=0.295,angle=0]{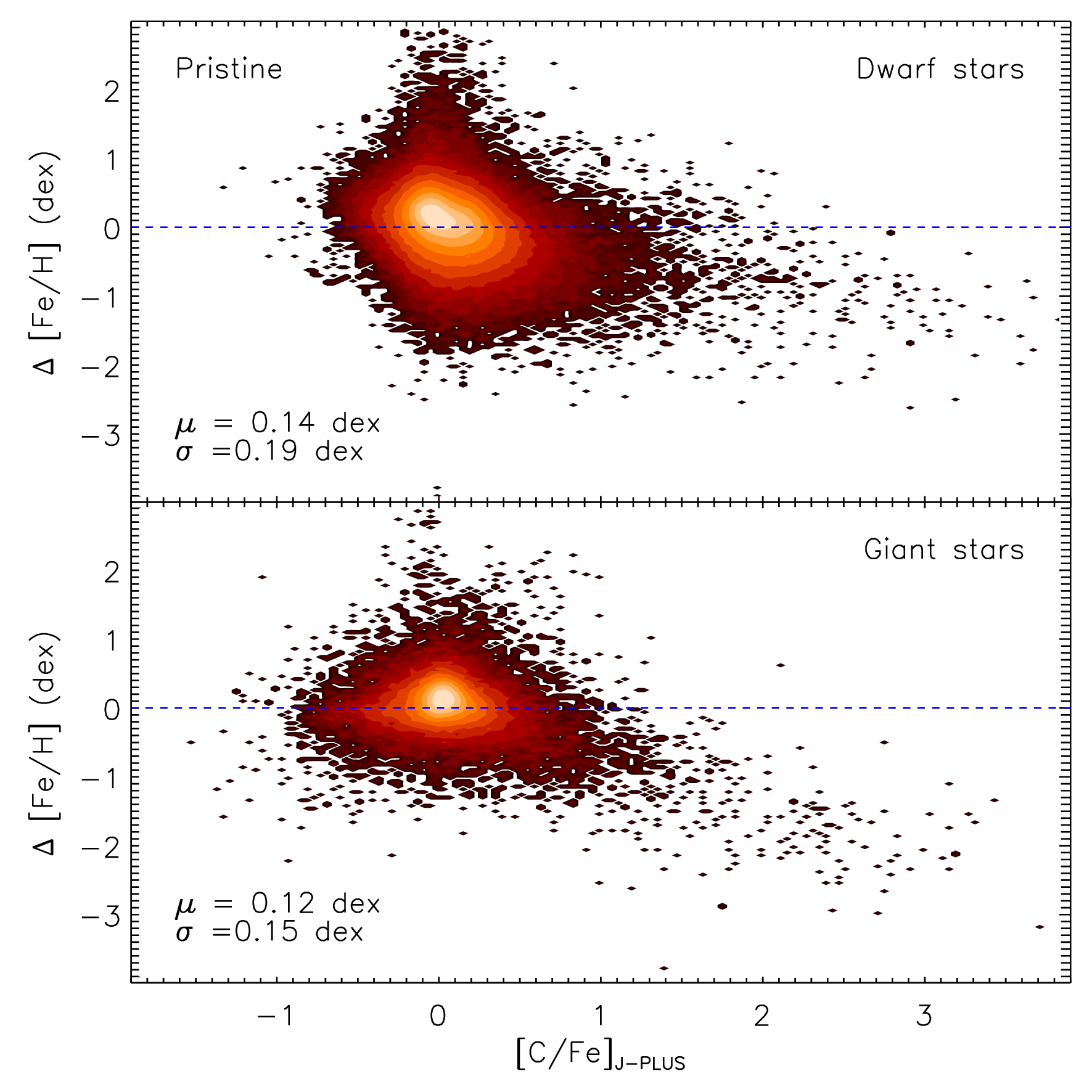}
\includegraphics[scale=0.295,angle=0]{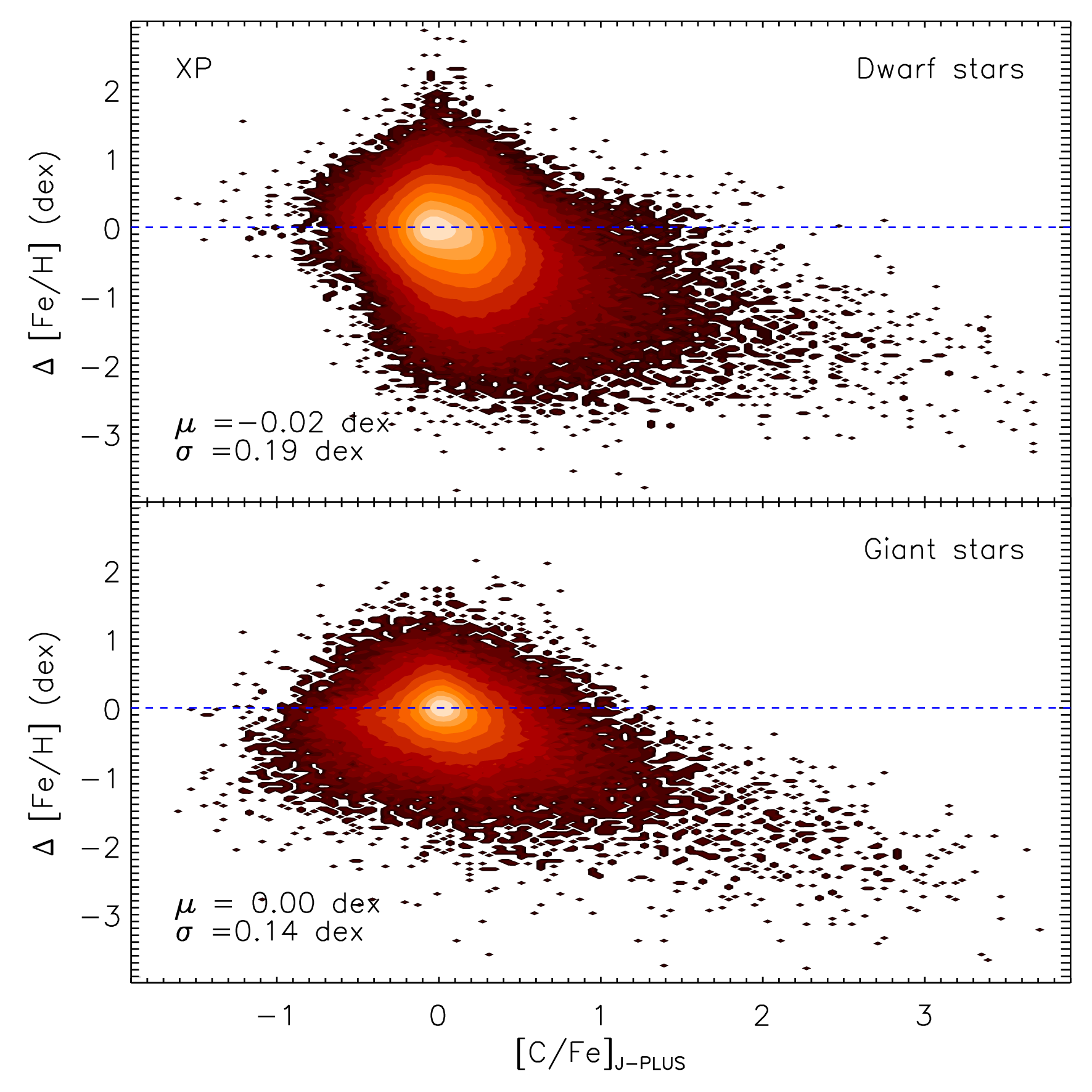}
\caption{Density distributions of the metallicity differences for J-PLUS--SAGES (top-left), J-PLUS--SMSS (top-right), J-PLUS--Pristine (bottom-left), and J-PLUS--{\it Gaia} XP 
(bottom-right), as a function of J-PLUS photometric [C/Fe]. Blue-dashed lines indicate the zero residuals in each panel.  The overall median offset and standard deviation are marked in the bottom-left corner of each panel.}
\end{center}
\end{figure*} 

\begin{figure*}
\begin{center}
\includegraphics[scale=0.325,angle=0]{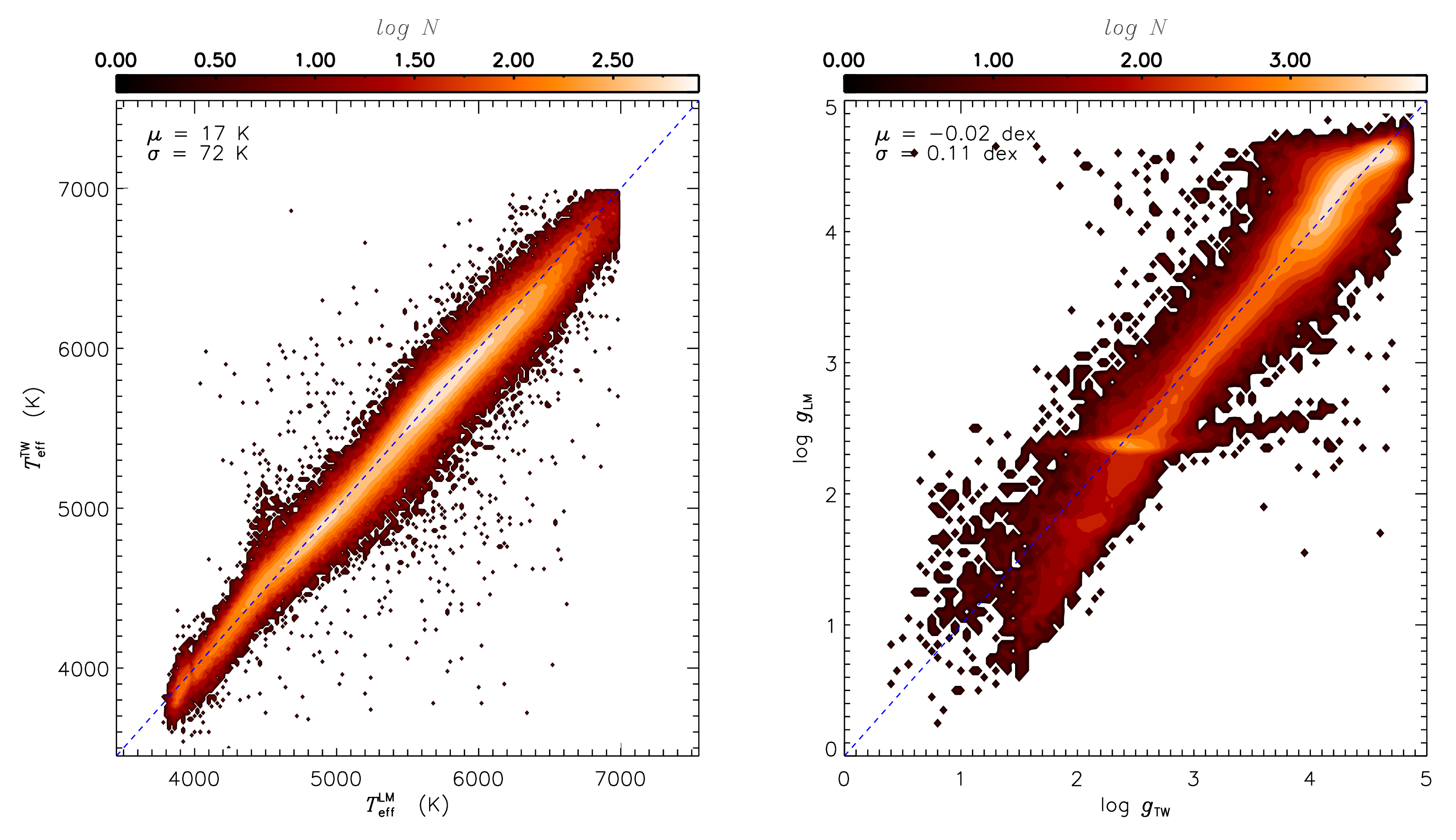}
\caption{Comparisons of our estimates of $T_{\rm eff}$ (left panel) and log\,$g$ (right panel) to those from DR9 of the LAMOST low-resolution survey. The blue-dashed lines are the one-to-one lines. The overall median
offset and standard deviation are marked in the top-left corner of each panel. Color bars representing the numbers of stars are provided at the top of each panel.}
\end{center}
\end{figure*} 

\begin{figure*}
\begin{center}
\includegraphics[scale=0.355,angle=0]{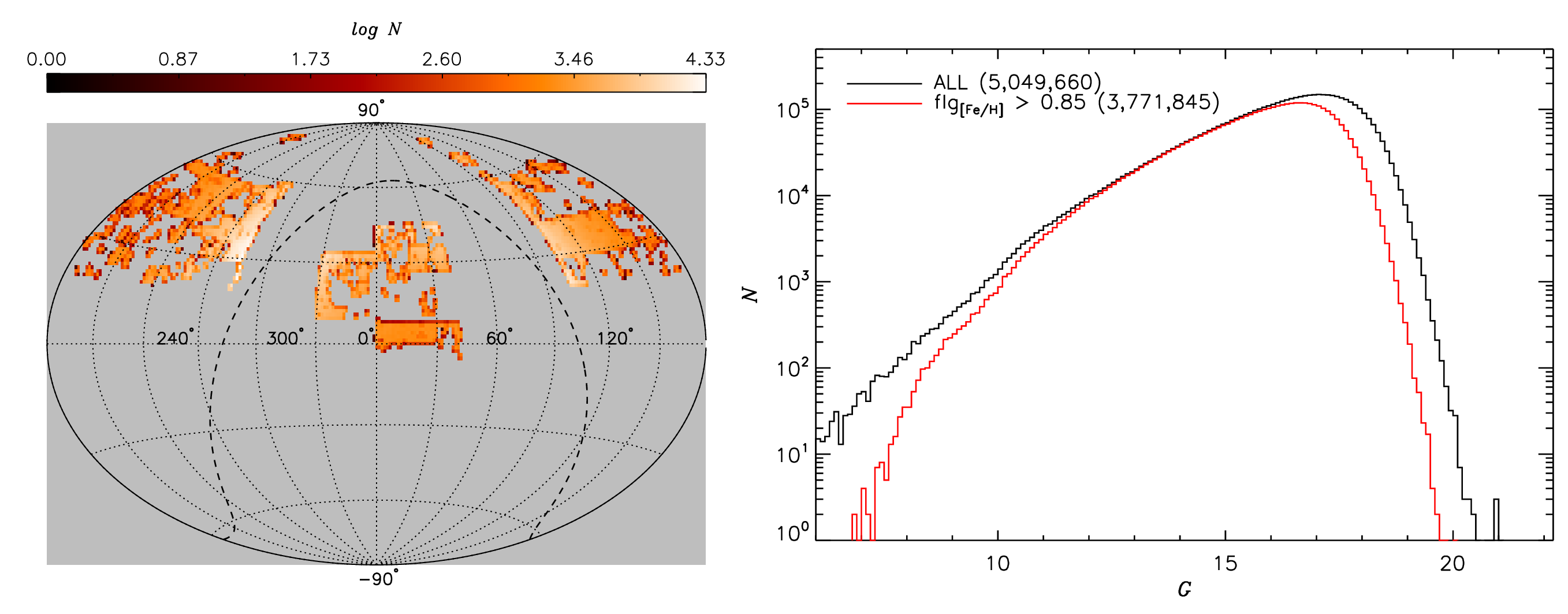}
\caption{Left panel: Stellar density distribution of the final J-PLUS parameter sample stars across the sky in equatorial coordinates. The black-dashed line marks the Galactic plane. Right panel: Magnitude distribution of the final J-PLUS parameter sample stars in the 
{\it Gaia G}- band. The red histogram represents the magnitude distribution for sample stars with {\tt flg$_{\rm [Fe/H]}$} $\ge 0.85$.}
\end{center}
\end{figure*} 

\begin{figure*}
\begin{center}
\includegraphics[scale=0.295,angle=0]{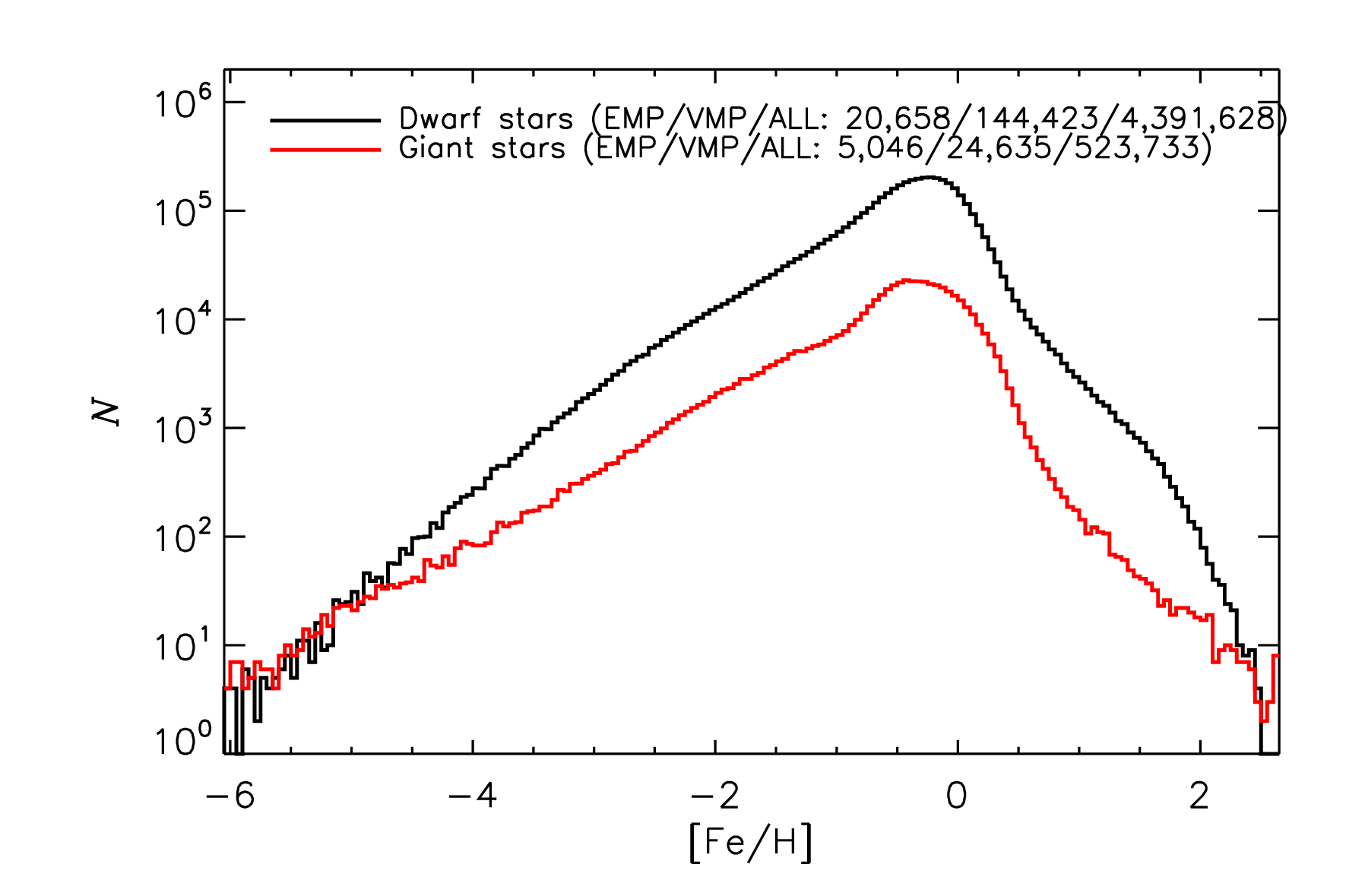}
\includegraphics[scale=0.295,angle=0]{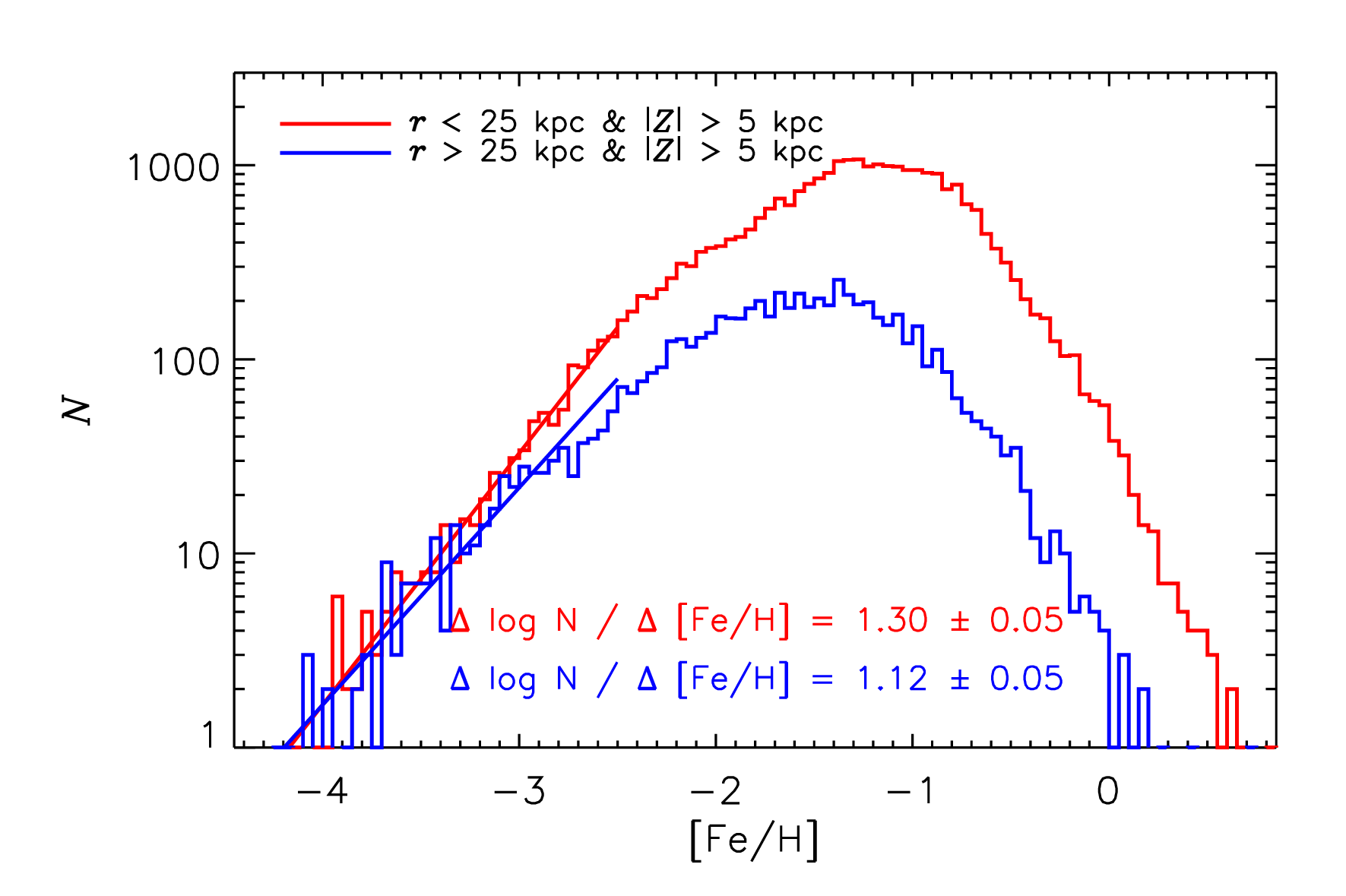}
\caption{Left panel: The J-PLUS photometric-metallicity distributions for dwarf 
(black histogram) and giant (red histogram) stars. Right panel: The metallicity distribution functions of inner-halo (red histogram; $r < 25$\,kpc and $|Z| > 5$\,kpc) and outer-halo (blue histogram; $r > 25$\,kpc and $|Z| > 5$\,kpc)
stars in the J-PLUS giant sample. The red and blue lines represent the best fits for their corresponding MDFs between [Fe/H] = $-2.75$ and $-4.0$, with the slopes listed at the bottom of the panel.}
\end{center}
\end{figure*} 

\begin{figure*}
\begin{center}
\includegraphics[scale=0.275,angle=0]{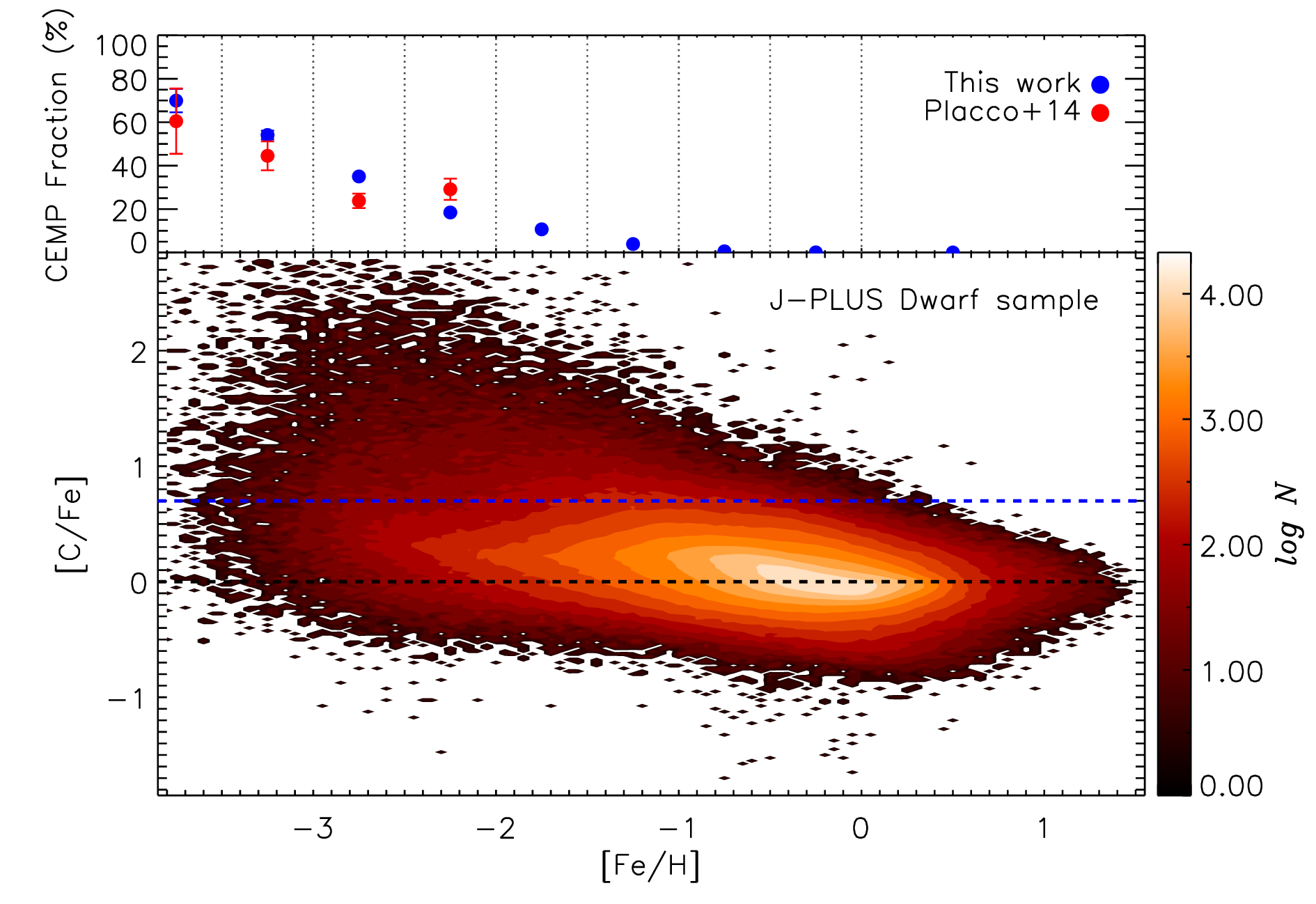}
\includegraphics[scale=0.275,angle=0]{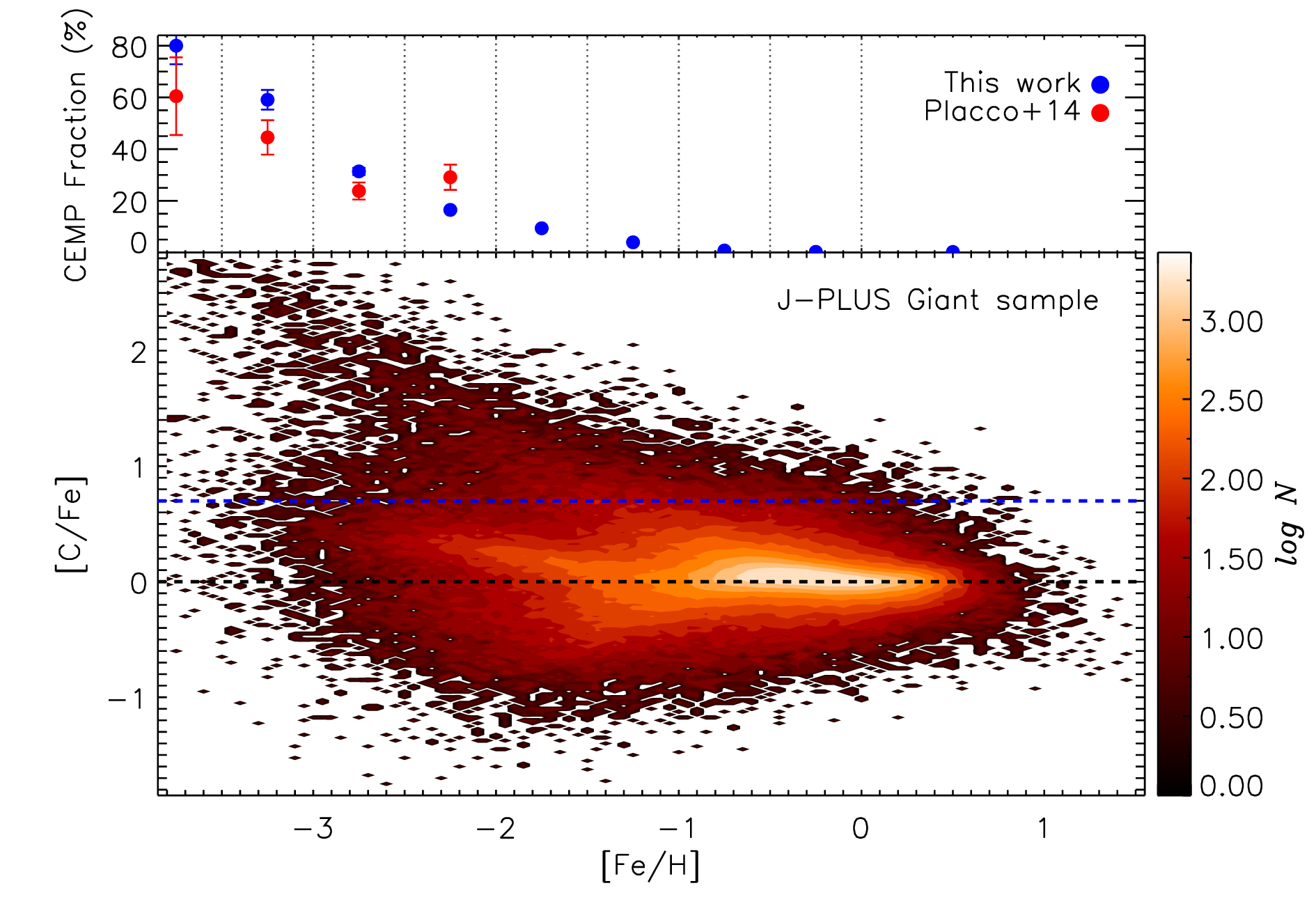}
\caption{Left panel: Density distribution of [C/Fe] vs. [Fe/H] for dwarf stars with flg$_{\rm [Fe/H]} \ge 0.85$, flg$_{\rm [C/Fe]} \ge 0.85$, and RUWE\,$\le 1.4$, with a color bar shown on the right side. The black-dashed and blue-dashed lines represent [C/Fe] = 0 and 
[C/Fe] = +0.7, respectively. Stars with [Fe/H] $\leq -1.0$ and [C/Fe] $ > +0.7$ are CEMP stars. The top sub-panel plots the fraction of CEMPs as a function of [Fe/H]. The red dots are the results taken from \citet{Placco14}. Right panel: Similar to the left panel, but for giant stars.  Note that, at present, we have not applied evolutionary corrections to the measured [C/Fe] in our sample. }
\end{center}
\end{figure*} 

\begin{figure*}
\begin{center}
\includegraphics[scale=0.385,angle=0]{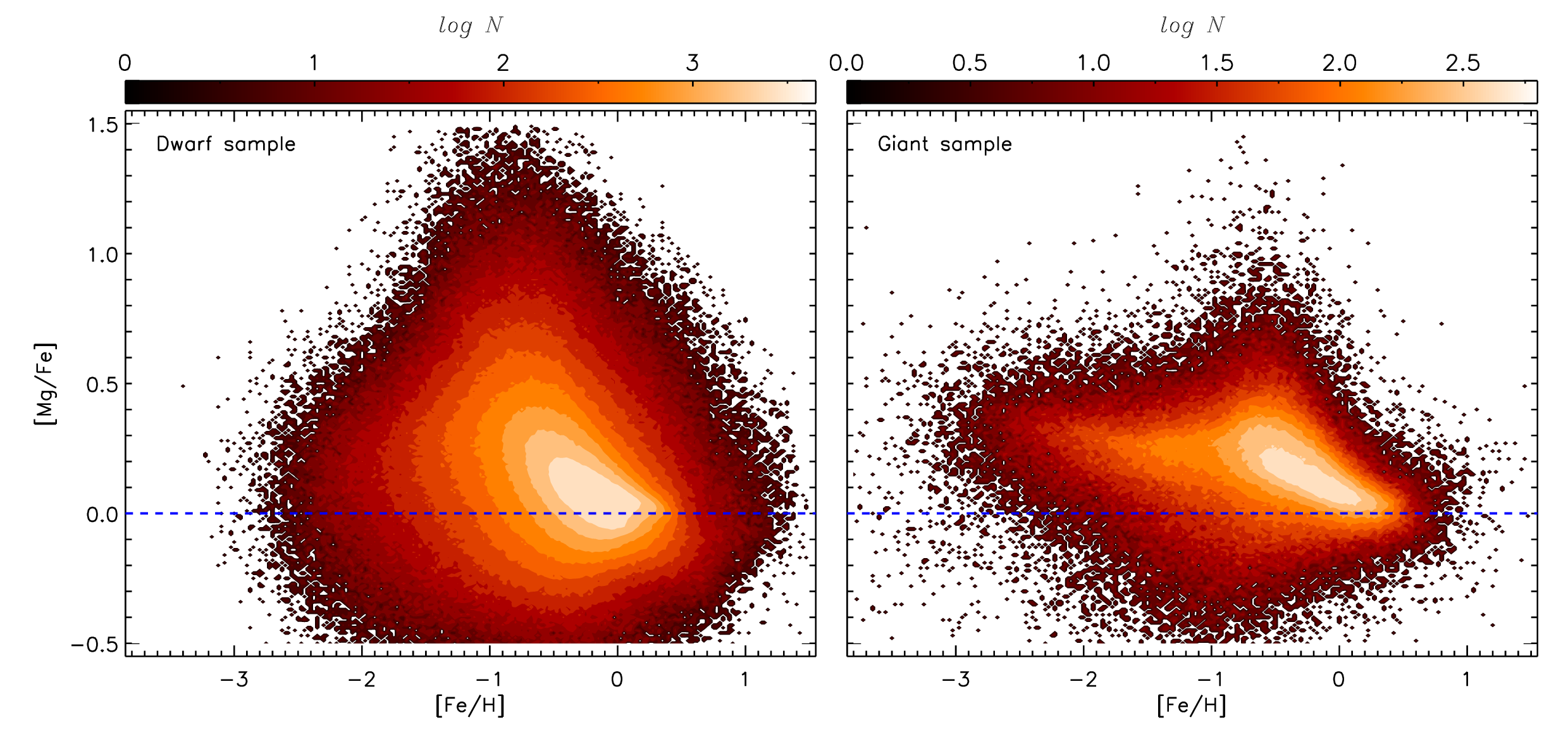}
\includegraphics[scale=0.385,angle=0]{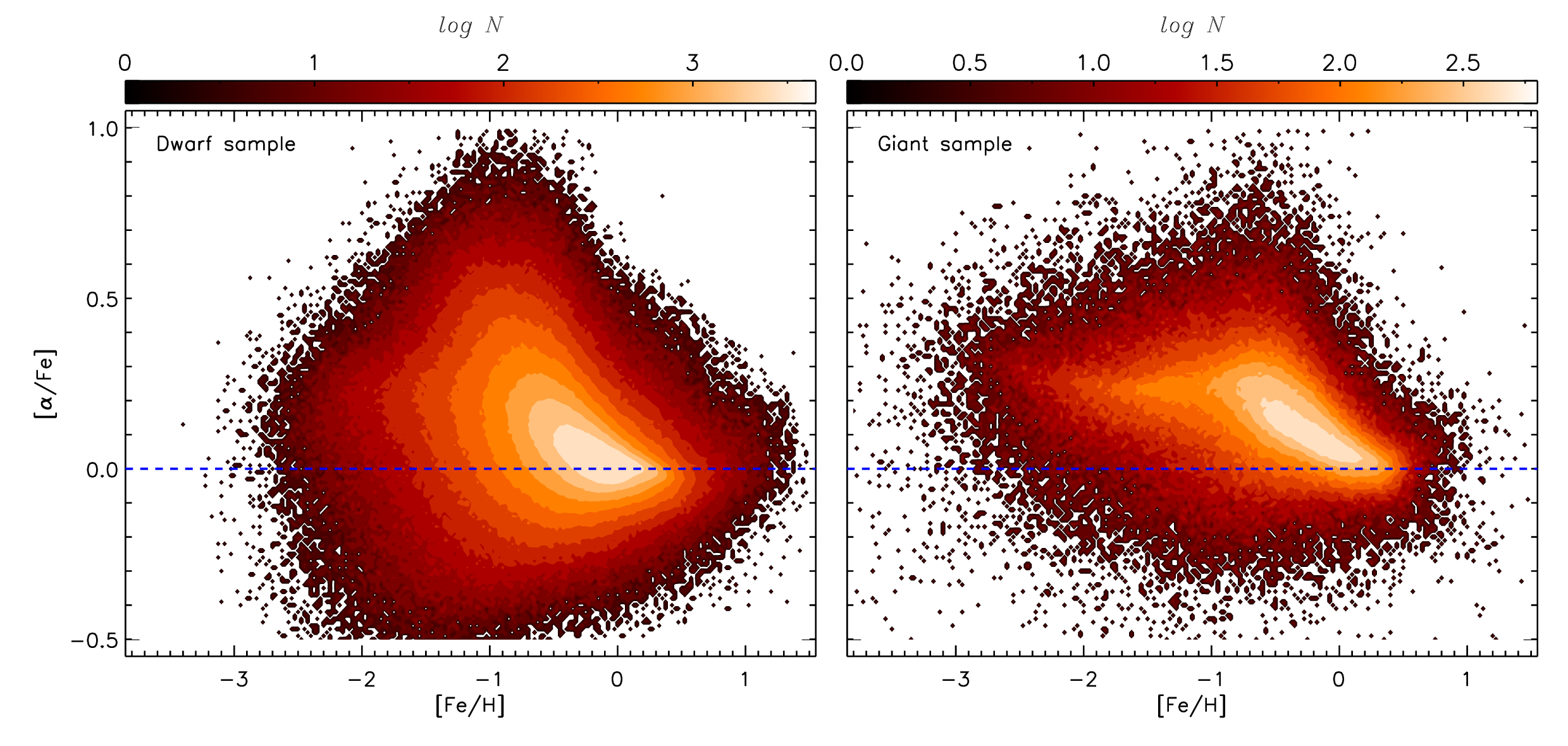}
\caption{Upper panel: Density distributions of [Mg/Fe] vs. [Fe/H] for dwarf (left) and giant (right) stars with flg$_{\rm [Fe/H]} \ge 0.85$, flg$_{\rm [Mg/Fe]} \ge 0.85$, and RUWE $\le 1.4$, with a color bar shown on the top side.
Lower panel: Similar to upper panel, but for [$\alpha$/Fe] vs. [Fe/H]. The blue-dashed lines in each panel indicate the Solar ratios.}
\end{center}
\end{figure*} 

\subsection{Application to the J-PLUS Parent Sample}

The above trained relationships are applied to the J-PLUS parent sample of $\sim 16.5$ million stars defined in Section\,2.2.
To evaluate the quality of the estimated abundances, we define the quality parameter {\tt flg$_x$} (here $x$ represents [Fe/H], [C/Fe], [Mg/Fe], or [$\alpha$/Fe]), which is given by the maximal kernel value between the target stellar colors and those in the training set.
The value of {\tt flg$_x$} can vary from 0 to 1, with unity representing exact agreement between a target stellar colors and these in the training sets.
As an example, Fig.\,6 shows the distribution of  {\tt flg$_{\rm [Fe/H]}$} as a function of $r$-band magnitude.
As expected,  {\tt flg$_{\rm [Fe/H]}$} is close to 1 for stars brighter than $r$ = 18, and quickly drops to 0 at the faint end due to the larger photometric errors.
We note that a small fraction of stars may have low values of  {\tt flg$_{\rm [Fe/H]}$} in the bright range, likely due to stellar variability, binarity, emission-line objects, etc.

We also compare the J-PLUS photometric metallicity estimates to those from SDSS/SEGUE medium-resolution spectroscopy in different bins of  {\tt flg$_{\rm [Fe/H]}$} (see right panel of Fig.\,6).
Generally, the dispersion of the metallicity difference between the photometric and spectroscopic estimates increases with decreasing  {\tt flg$_{\rm [Fe/H]}$}.
For {\tt flg$_{\rm [Fe/H]}$}\,$< 0.60$, the dispersion is larger than 0.8\,dex, and obvious artificial features are seen; thus, the photometric estimates of metallicity  are not recommended for any stars with {\tt flg$_{\rm [Fe/H]}$}\,$< 0.60$.

After removing stars used in the training sets, the remaining APOGEE--J-PLUS stars in common are used to consider the abundance differences (spectroscopic minus photometric), as a function of   {\tt flg$_x$}, in Fig.\,7.
Clearly, the scatter increases with decreasing {\tt flg$_x$} for all estimated abundances (see Fig.\,8), i.e., [Fe/H], [C/Fe], [Mg/Fe], and [$\alpha$/Fe].
The trends fitted by third-order polynomials can be taken as the random error of the abundance estimate $\sigma_{x}$.
By considering the fitting uncertainty (hereafter, the method error $\sigma_{m}$) as functions of the abundances themselves in Fig.\,5, the final uncertainty of the abundance measurement for a specific star is given by $\sqrt{\sigma_x^2 + \sigma_m^2}$. 
In total, over five million stars have determinations of stellar abundances with at least one parameter quality flag ({\tt flg$_x$} ) greater than 0.6 (hereafter referred to as the J-PLUS parameter sample); the total number is about 4.8 million if all quality flags are required to satisfy {\tt flg$_x$}\,$\geq 0.6$. 

\vskip 1.5cm
\subsection{Validation}
\subsubsection{Comparison with GALAH DR3}

As mentioned in Section\,3, stars from the GALAH survey are not used in the training sets, and thus can be adopted to examine our photometric estimates.
The GALAH DR3 is cross-matched with the J-PLUS parameter sample, by requiring GALAH {\tt SNR\_C2\_IRAF}\,$\ge 30$ per pixel and J-PLUS  {\tt flg$_x$}\,$\ge 0.9$.
The comparison results are shown in Fig.\,9.
For [Fe/H], our photometric results are in excellent agreement with the HRS estimates from GALAH, with negligible offsets and scatter of around 0.10\,dex for both dwarf and giant stars.
Generally, the photometric [Mg/Fe] estimates agree with those from GALAH, with a moderate scatter of 0.08\,dex.
The comparisons show that the photometric estimate  of [$\alpha$/Fe] is quite precise, with scatter of only 0.05\,dex for dwarf stars and 0.06\,dex for giant stars.
This is also in-line with our training results, as shown in Fig.\,4.

\subsubsection{Comparison with LAMOST Medium-Resolution Survey}
Most recently, \cite{2023MNRAS.523.5230L} derived stellar parameters from over 4 million LAMOST medium-reslution ($R \sim 7500$) spectra using the RRNet technique and the APOGEE DR17 stellar parameters as training labels.
This sample is cross-matched with the J-PLUS parameter sample, by requiring {\tt SNR\_BLUE}\,$\ge 20$ per pixel, J-PLUS  {\tt flg$_{\rm [Fe/H]}$}\,$\ge 0.9$, and Gaia {\tt RUWE}\,$< 1.4$.
In total, 53,676 stars in common are found.
Generally, our photometric estimates agree very well with those of \cite{2023MNRAS.523.5230L}, with negligible offsets and a small scatter around 0.09\,dex for both dwarf and giant stars (see Figure\,10).
However, we note that their metallicity estimate are truncated at [Fe/H]\,$= -1.0$ for dwarf stars and [Fe/H]\,$= -1.5$ for giant stars (see Fig.\,10).

\subsubsection{Comparison with Metal-poor Samples from the Literature}

The identification of large numbers of metal-poor stars is one of the main goals of this project. We therefore want to compare our photometric abundance estimates to those derived from spectroscopy in the literature.

First, our photometric abundances are compared to those for low-metallicity candidates from the Best \& Brightest (B\&B) Survey with medium-resolution ($R \sim 2000$) follow-up spectroscopic observations \citep{BNB}. More recently, \citet{2019ApJ...870..122P} and \citet{Limberg21} present metallicity estimates, along with other elemental-abundance ratios ([C/Fe] and [Mg/Fe]), for nearly 1900 stars using the medium-resolution spectra from follow-up of stars in the B\&B survey.\footnote{Note that the spectroscopic abundance ratios for [C/Fe] and [Mg/Fe] or [$\alpha$/Fe] reported by \citet{2019ApJ...870..122P} and \citet{Limberg21} have been superseded by the values listed in \citet{2022ApJ...926...26S}, who corrected an error that existed in their calculation in the previous works.}
In total, 37 metal-poor ([Fe/H] \,$\le -1.5$) giant stars in common (with {\tt flg$_{\rm [Fe/H]}$}\,$\ge 0.9$, {\tt flg$_{\rm [C/Fe]}$}\,$\ge 0.9$, or {\tt flg$_{\rm [Mg/Fe]}$}\,$\ge 0.9$ ) are found, with the lowest metallicity approaching [Fe/H] $\sim -3.0$.
Moderate offsets of $-0.11$\,dex (B\&B minus photometric estimates) are found in the metallicity differences, as shown in Fig.\,11.
The scatter is only 0.15\,dex, consistent with the expectation from our internal check shown in Fig.\,5 (see section\,4.2).
Generally, the photometric estimate of [C/Fe] \footnote{The [C/Fe] estimates refer to results without evolution-dependent corrections, since no such corrections were made for the training sets (see Section\,3.2).} agrees with that of the B\&B follow-up, with a moderate offset of $+0.13$\,dex and scatter of 0.19\,dex.
The scatter of the [$\alpha$/Fe] difference is large; this is as expected, due to the lack of metal-poor stars with precise estimates of [Mg/Fe] in our training sets (see Section\,3.3). %Hence, adoption of the photometric estimates of [$\alpha$/Fe] and [Mg/Fe] for metal-poor stars are not recommended at present. 

Secondly, photometric abundances are compared to the HRS sample stars, only a few if which are included in our training sets.
The HRS sample stars are collected from the CEMP sample for over 600 stars \citep{Placco14}, the $R$-Process Alliance project \citep{2018ApJ...858...92H, 2018ApJ...868..110S, 2020ApJ...898..150E, 2020ApJS..249...30H} for over 600 VMP stars, and the 400 LAMOST-selected VMP candidates with chemical abundances determined from Subaru high-resolution spectroscopic follow-up \citep{2022ApJ...931..146A, 2022ApJ...931..147L}. 
In total, there are 79 stars in common (28 dwarf and 51 giant stars) with {\tt flg$_{\rm [Fe/H]}$} greater than 0.9, covering a metallicity range of [Fe/H] = [$-4.0$, $-2.0$].
For [Fe/H], the median difference is minor for both dwarf and giant stars; the scatter is 0.32\,dex for dwarf stars and about 0.17\,dex for giant stars, which are consistent with the performance of the training exercise (see Fig.\,5).

Of central importance, unlike the case for SMSS and SAGES (which only employ two narrow/medium-band filters, the $u$- and $v$-bands), our photometric metallicities for CEMP stars (those with pluses in Fig.\,11) are no longer over-estimated, since the $J0395$ filter responds to metallicity, while the $J0430$
filter independently measures the molecular carbon centered on the CH $G$-band.
Among the 73 stars in common (25 dwarf and 48 giant stars) with {\tt flg$_{\rm [C/Fe]}$}\,$\ge 0.9$, almost all the CEMP stars with [C/Fe]\,$> +0.7$ in the HRS sample are recovered by our photometric measurements. 
The median offset (HRS minus photometric) of [C/Fe] is minor for giant stars and 0.44\,dex for dwarf stars, similar to that found for our training sets (bottom panels of Fig.\,3).
The overall scatter of the [C/Fe] difference is 0.52\,dex and 0.27\,dex for dwarf and giant stars, respectively.
Again, as expected, the scatter of the difference between the photometric and HRS estimates of [Mg/Fe] or [$\alpha$/Fe] is quite large.

\subsubsection{Examination of Metallicity Systematics in Relation to Reddening}
As mentioned in Section\,2.2, we adopted the SFD map to correct the extinction since most of sample stars (92.7\%) are located in low extinction regions with $E (B -V) < 0.1$.
However, the SFD map is a two-dimensional map, and the reddening of nearby stars may be over-estimated.
To examine this potential systematic, the J-PLUS sample is cross-matched with the DR17 of APOGEE.\footnote{We note that only a few thousand stars with extremely low extinction values of $E(B - V) < 0.04$ were used as the training sample. Therefore, it is appropriate to re-examine the metallicity difference with $E(B - V)$ using APOGEE stars.}
By requiring APOGEE spectral {\tt SNR $\ge 50$}, J-PLUS  {\tt flg$_{\rm [Fe/H]}$}\,$\ge 0.9$, and Gaia {\tt RUWE } $< 1.4$, a total of 32,088 stars in common are found.  Fig.\,12 shows the metallicity difference (APOGEE minus J-PLUS), as a function of SFD $E (B -V)$.
The median offsets are almost zero for different bins of $E (B - V)$, ranging from 0 to 0.20. 
This result confirms that the SFD map is sufficiently accurate for photometric estimates of stellar parameters, as expected given the low extinction values for most of the sample stars.

\subsection{Comparisons with Previous Estimates from J-PLUS DR1 and DR2}

Prior to this work, several attempts have provided estimates of stellar parameters, as well as elemental abundances, from J-PLUS DR1 \citep{2019A&A...622A.182W, 2022A&A...659A.181Y} and DR2 \citep{2022A&A...657A..35G}.
Based on J-PLUS DR1, stellar parameters and [C/Fe], [N/Fe], [Mg/Fe], [Ca/Fe], and [$\alpha$/Fe] are derived for two million stars \citep[][hereafter Yang+22]{2022A&A...659A.181Y}.
As shown in the left panel of Fig.\,13, our metallicity is consistent with that of Yang+22 with a negligible offset of $-0.03$\,dex (this work minus Yang+22) and a scatter of 0.14\,dex.
At the metal-poor end, the metallicity of Yang+22 is over-estimated and truncated at [Fe/H]$\sim -2.5$ due to their training labels from LAMOST DR5 (see Fig.\,A2).
The estimates of [C/Fe], [Mg/Fe], and [$\alpha$/Fe] are also compared to those from Yang+22 in Fig.\,14. Generally, they agree with other within the typical errors.
By using the Sellar Parameters Estimation based on Ensemble Methods (SPEEM) pipeline, \citet{2022A&A...657A..35G} have derived stellar atmospheric parameters ($T_{\rm eff}$, 
log\,$g$, and [Fe/H]) for a ``gold sample" of 746,531 stars from J-PLUS DR2.
The comparison of our photometric estimates of [Fe/H] to those from \citet{2022A&A...657A..35G} is shown in the right panel of Figure\,13.
The median difference is 0.05 dex (this work minus SPEEM) and the scatter is only 0.10\,dex.
Simlarly, the metallicity estimated by SPEEM is over-estimated at the metal-poor end ([Fe/H]$<-2.0$), which has been already mentioned in \citet{2022A&A...657A..35G}.

\subsection{Comparisons with Other (Spectro)photometric Estimates}

In this subsection, the J-PLUS metallicity estimates are compared to other photometric results from surveys that employ one/two blue narrow-band filters, those from the SAGES DR1 (Paper II), the SMSS DR2 (Paper I) and the Pristine Survey DR1 \citep{2023arXiv230801344M}.

Due to contamination of the blue narrow/medium-band filters by molecular carbon bands such as CN, the photometric estimates of [Fe/H] are often over-estimated for VMP stars -- the most interesting targets for understanding the early chemical evolution of the universe.  This is due to the large fraction of VMP stars that are carbon enhanced, causing the colors $(u/v - G_{\rm BP})$ to appear redder than for a VMP star with normal carbon abundance \citep[][Papers I and II]{2017MNRAS.471.2587S}.
As discussed earlier, owing to use of the seven narrow/medium-band filters employed by J-PLUS, the above degeneracy can be broken; the metallicity and [C/Fe] can be both photometrically measured.

Generally, the SAGES, SMSS, and Pristine [Fe/H] estimates are quite consistent with those from J-PLUS (see Fig.\,15), with scatters of $\sim$\,0.20\,dex for dwarf stars and $\sim$\,0.15\,dex for giant stars.
There is an offset of 0.10--0.15\,dex between J-PLUS and Pristine (former minus latter).
As compared to the J-PLUS estimates, the SAGES, SMSS, and Pristine [Fe/H] estimates exhibit strong biases for carbon-enhanced stars; their [Fe/H] estimates are over-estimated by 1--2\,dex at [C/Fe]\,$\ge +1.0$ for both dwarf and giant stars (see Fig.\,15).

Unexpectedly, the XP metallicity estimates \citep{2023ApJS..267....8A} exhibit similar trends along with [C/Fe], when compared to the J-PLUS metallicity estimates.
In principle, the XP spectra contain sufficient information to determine [Fe/H] and [C/Fe] simultaneously \citep[see][]{2023MNRAS.523.4049L}.
One possible explanation is that there are few carbon-enhanced stars in their training sets. 

\section{Effective temperatures, distances, ages, and surface gravities}
Through use of the metallicity-dependent $T_{\rm eff}$--color relations constructed in Paper\,I of this series, the effective temperatures for J-PLUS dwarf and giant stars are derived from $(G_{\rm BP} - G_{\rm RP})_0$ and the photometric [Fe/H] estimated above. 
As shown in Paper II, the typical uncertainty of the derived effective temperature is within 100\,K, when compared to the spectroscopic uncertainty.
For instance, as examined with over 400,000 common stars, the effective temperature estimated in this work is quite consistent with that from LAMOST, with a tiny offset around 14\,K (LAMOST minus this work) and a scatter of only
72\,K (see Fig. 16).

The strategy of distance determinations is again similar to that described in Papers I and II.
For stars with reliable parallax measurements from {\it Gaia} EDR3 (precision better than 30\%, parallax greater than 0.15\,mas, and renormalized unit weight error (RUWE) smaller than 1.4), the distances are directly adopted from \citet{BJ21}.
The further classifications (turn-off, main-sequence, and binary; see subtype in Table\,2), based on positions of the stars on the Hertzsprung-Russell (H-R) Diagram, are obtained by comparison with the PARSEC isochrones \citep{2012MNRAS.427..127B, 2017ApJ...835...77M}.

Using a Bayesian method similar to that in Paper I, the stellar ages for stars are determined with the constraints from $(G_{\rm BP} - G_{\rm RP})_0$, $G$-band absolute magnitude, and photometric metallicity. 
In this work, the surface gravity is also estimated using the isochrone-fitting technique described above.
In this manner, nearly 3.8 million stars have their distances, ages, and surface gravities, as well as luminosity classifications assigned.

As mentioned in Paper I, the isochrone-fitting method mainly works for turn-off stars with a typical uncertainty of 20\%.
For distant dwarf and giant stars, the empirical metallicity-dependent color-absolute magnitude relations/fiducials from Paper I are adopted.
Interested readers are referred to Paper I for additional details.
By combining with {\it Gaia} and Pan-STARRS photometry, and the SFD reddening map, distances for around  0.5 million distant stars are estimated. 
This is important for our giant sample stars, since 40\% of their distances are derived in this way.
As examined from nearly 700 stars in common with the SDSS/SEGUE K-giant sample \citep{2014ApJ...784..170X}, the precision of our distances from the color-absolute magnitude fiducials is better than 16\%, without significant offsets.
Again using over 400,000 stars in common, the surface gravity estimated in this work is also consistent with that of the LAMOST spectroscopic survey, with a tiny offset of $-0.02$\,dex (LAMOST minus this work) and a scatter of only 0.11\,dex (see Fig. 16).
  
\section{The J-PLUS Parameter Sample}
Using data from J-PLUS DR3 and Gaia EDR3, the photometric metallicity, carbon-to-iron abundance ratio, magnesium-to-iron abundance ratio, and alpha-to-iron abundance ratio are estimated for about 4.5 million dwarf and 0.5 million giant stars with quality flags {\tt flg}$_x >$ 0.6. Their spatial coverage and magnitude distributions are shown in Figure\,17.

The metallicity distribution functions (MDFs) for dwarf and giant stars are shown in Fig.\,18. 
In total, over 160,000 VMP stars are found.
As an example of the utility of these VMP stars, we investigate the MDFs of the metal-poor halo stars for the J-PLUS giant star sample. 
The slope is found to be $\frac{\Delta\,N}{\Delta\,{\rm [Fe/H]}}\,= 1.30 \pm 0.05$ and $\frac{\Delta\,N}{\Delta\,{\rm [Fe/H]}}\,= 1.12 \pm 0.05$, respectively, for the inner stellar halo ($r < 25$\,kpc and $|Z| > 5$\,kpc) and the outer stellar halo ($r > 25$\,kpc and $|Z| > 5$\,kpc) with [Fe/H] between $-2.75$ and $-4.0$.
The result found for the inner halo is commensurate with other recent determinations \citep[e.g.,][]{2020MNRAS.492.4986Y, 2021MNRAS.507.4102Y,2021ApJ...912..147W}. 
This first application shows that the slope of the MDF may evolve with $r$, which is worth exploring further with the (presumably minimal) selection effects properly considered.

We also show the [C/Fe] vs. [Fe/H] distributions of our sample in Fig.\,19.
As found by previous studies, the carbon-enhanced ([C/Fe]\,$> +0.7$) are mostly found in the metal-poor regime ([Fe/H] $\le -1.0$); they are therefore referred to as  CEMP stars.  
The fraction of CEMP stars is a strong increasing function of declining [Fe/H], with a value of a few per cent at [Fe/H]\,$\sim -1.0$ to values as high as 70\% at [Fe/H]\,$< -3.0$.  Recall that, at present, we have not applied corrections to the photometric [C/Fe] estimates arising from evolutionary effects. Even so, the observed trend is consistent with that found from high-resolution spectroscopy \citep{Placco14}. It is notable that our sample contains over 120,000 CEMP stars (100,800 dwarfs and 15,000 giants), which is a lower limit due to the lack of evolutionary corrections.
Finally, the distributions of [Mg/Fe]--[Fe/H] and [$\alpha$/Fe]--[Fe/H] are shown in Fig.\,20.

Table\,2 summarizes the contents of our final parameter sample.
From a series of well-established techniques in our previous studies (Paper I and II), the effective temperature, distances, surface gravities, and ages are derived for all 4.3 million and 3.8 million stars in the parameter sample. The astrometric information (i.e., parallaxes, proper motions, and their uncertainties), taken from Gaia EDR3 \citep{GEDR3}, as well as the available radial velocities, from a number of sources, is also included.
The sample will be made publicly available at \url{https://zenodo.org/records/13160149}.
The applicable range and typical uncertainty of the derived parameters are summarized in Table\,3.  Note that, although we report elemental-abundance estimates in our J-PLUS parameter sample over a wide range, the quoted uncertainties only apply to the listed range.  Outside of these ranges, the typical errors increase.

\begin{table*}
\centering
\caption{Description of the Final Sample}
\begin{tabular}{lll}
\hline
\hline
Field&Description&Unit\\
\hline
Sourceid&Gaia EDR3 source ID&--\\
ra&Right Ascension from J-PLUS DR3 (J2000)&degrees\\
dec&Declination from J-PLUS DR3 (J2000)&degrees\\
gl&Galactic longitude derived from ICRS coordinates&degrees\\
gb&Galactic latitude derived from ICRS coordinates&degrees\\
\text{mag}$_{1 ... 12}$&Magnitudes of J-PLUS twelve bands&--\\
err\_\text{mag}$_{1 ... 12}$&Uncertainties of magnitudes of J-PLUS twelve bands&mag\\
g/r/i&Magnitudes from Pan-STARRS1&--\\
err\_g/r/i&Uncertainties of magnitudes from Pan-STARRS1&mag\\
G/BP/RP&Magnitudes for the {\it Gaia} three bands from EDR3; note G represents a calibration-corrected G magnitude&--\\
err\_G/BP/RP&Uncertainties of magnitudes for the three {\it Gaia} bands from EDR3&mag\\
ebv\_sfd&Value of $E (B - V)$ from the extinction map of SFD98, corrected for a 14\% systematic&--\\
\text{[Fe/H]}&Photometric metallicity&--\\
\text{err\_[Fe/H]}&Uncertainty of photometric metallicity&dex\\
\text{flg\_[Fe/H]}&Quality flag of [Fe/H]&--\\
\text{[C/Fe]}&Photometric carbon-to-iron abundance ratio&--\\
\text{err\_[C/Fe]}&Uncertainty of photometric carbon-to-iron abundance ratio&dex\\
\text{flg\_[C/Fe]}&Quality flag of [C/Fe]\\
\text{[Mg/Fe]}&Photometric magnesium-to-iron abundance ratio&--\\
\text{err\_[Mg/Fe]}&Uncertainty of photometric magnesium-to-iron abundance ratio&dex\\
\text{flg\_[Mg/Fe]}&Quality flag of [Mg/Fe]&-- \\
\text{[$\alpha$/Fe]}&Photometric alpha-to-iron abundance ratio&--\\
\text{err\_[$\alpha$/Fe]}&Uncertainty of photometric alpha-to-iron abundance ratio &dex\\
\text{flg\_[$\alpha$/Fe]}&Quality flag of [$\alpha$/Fe]&-- \\
$T_{\rm eff}$&Effective temperature&K\\
err\_$T_{\rm eff}$&Uncertainty of effective temperature&K\\
log\,$g$&Surface gravity&--\\
err\_${\rm log} g$&Uncertainty of surface gravity&dex\\
dist&Distance&pc\\
err\_dist&Uncertainty of distance&pc\\
flg\_dist&Flag to indicate the method used to derive distance, which takes the values ``parallax", ``CMF", and ``NO"&--\\
\text{age}&Stellar age&Gyr\\
\text{err\_age}&Uncertainty of stellar age&Gyr\\
rv&Radial velocity&km s$^{-1}$\\
err\_rv&Uncertainty of radial velocity&km s$^{-1}$\\
flg\_rv&Flag to indicate the source of radial velocity, which takes the values ``GALAH", ``APOGEE'', ``Gaia",&--\\
& ``RAVE", ``LAMOST", ``SEGUE"&--\\
parallax&Parallax from {\it Gaia} EDR3&mas\\
err\_parallax&Uncertainty of parallax from {\it Gaia} EDR3&mas\\
pmra&Proper motion in Right Ascension direction from  {\it Gaia} EDR3&mas yr$^{-1}$\\
err\_pmra&Uncertainty of proper motion in Right Ascension direction from  {\it Gaia} EDR3&mas yr$^{-1}$\\
pmdec&Proper motion in Declination direction  from  {\it Gaia} EDR3&mas yr$^{-1}$\\
err\_pmdec&Uncertainty of proper motion in Declination direction from  {\it Gaia} EDR3&mas yr$^{-1}$\\
ruwe&Renormalised unit weight error from  {\it Gaia} EDR3&--\\
type&Flag to indicate classifications of stars, which takes the values ``dwarf" and ``giant" &--\\
subtype&Flag to indicate further sub-classifications of dwarf stars, which takes the values  ``TO", ``MS" and ``Binary"&--\\
\hline
\end{tabular}
\end{table*}

 \begin{table*}
\centering
\caption{The Applicable Range and Typical Uncertainty of Derived Parameters}
\begin{tabular}{cccc}
\hline
Parameter & Luminosity classification& Applicable Range & Typical Uncertainty\\
\hline
\hline
\multirow{2}{*}{$T_{\rm eff}$ } & Dwarf stars& [3800, 8000]\,K & 100\,K\\
& Giant stars& [3800, 6500]\,K & 100\,K\\
\multirow{2}{*}{[Fe/H]} & Dwarf stars& [$-4.0$, +1.0] & 0.1\,dex for [Fe/H]\,$> -2.0$ and 0.15-0.25\,dex for [Fe/H]\,$< -2.5$\\
& Giant stars& [$-4.0$, +1.0]&  0.1-0.2\,dex for [Fe/H]\,$> -2.0$ and 0.2-0.4\,dex for [Fe/H]\,$< -1.0$\\
\multirow{2}{*}{[C/Fe]} & Dwarf stars& [$-1.5$, +4.0]&0.1-0.2\,dex\\
& Giant stars& [$-1.5$, +4.0]&  0.1-0.2\,dex \\
\multirow{2}{*}{[Mg/Fe]} & Dwarf stars& [$-0.3$, +0.6]&0.1-0.2\,dex\\
& Giant stars& [$-0.3$, +0.6]&  0.1-0.2\,dex \\
\multirow{2}{*}{[$\alpha$/Fe]} & Dwarf stars& [$-0.2$, +0.5]&0.03-0.06\,dex\\
& Giant stars& [$-0.2$, +0.5]&  0.02-0.05\,dex \\
Age & -- & Turn-off main-sequence and sub-giant stars & 20\% \\
log$g$&--&[0.0, 5.0]&0.1-0.2\,dex\\
\hline 
\hline
\end{tabular}
\end{table*}

\vskip 1cm
\section{Summary and Future Prospects}
In this paper, we determine stellar parameters (including effective temperature, 
surface gravity, [Fe/H], and age) and the important elemental-abundance ratios ([C/Fe], [Mg/Fe] and [$\alpha$/Fe]) for over five million stars (4.5 million dwarf stars and 0.5 million giant stars) using 13 colors from a combination of narrow- and medium-band filter photometry from J-PLUS DR3 and ultra wide-band photometry from {\it Gaia} EDR3.
To obtain estimates of metallicity and the elemental-abundance ratios, we have constructed a largetraining set consisting of millions of spectroscopically targeted stars. 
The scales for the metallicity and elemental-abundance ratios are carefully calibrated to previous results from high-resolution spectroscopic studies.
The typical uncertainty is 0.10--0.20\,dex for [Fe/H] and [C/Fe] and 0.05\,dex for [Mg/Fe] and [$\alpha$/Fe] over much of the range in metallicity.

Due to use of the narrow/medium-band filters employed by J-PLUS for both [Fe/H] ($J0395$) and [C/Fe] ($J0430$), the degeneracy between metallicity and carbonicity is successfully broken in this study. This is of particular importance for the VMP stars, where large fractions of carbon-enhanced stars are found, which have confounded metallicity estimates in previous photometric surveys (e.g., SAGES, SMSS, and Pristine).

Our photometric determination of [Fe/H] is well-estimated down to [Fe/H] $\sim -4.0$, with a precision of 0.40\,dex and 0.25\,dex for dwarf and giant stars, respectively, with no significant offsets.
This sample thus opens the window to studies of the changes in the MDF and the fractions of CEMP stars for various disk and halo stellar populations based on a large, relatively bias-free sample of stars. 
Similar to previous efforts in this series, effective temperatures from broad-band colors and photometric-metallicity estimates, distances from either Gaia parallaxes or metallicity-dependent color-absolute magnitude fiducials, and ages from isochrone comparisons are included in the final parameter catalog.

The J-PLUS effort is still underway, and will at least double the numbers of stars in the Northern sky for which we can determine precision metallicity and elemental-abundance estimates once it is completed in the next few years.  The Southern Photometric Local Universe Survey (S-PLUS; \citealt{Mendes2019}) is a parallel survey of the Southern sky (using an identical telescope and filter set as J-PLUS), for which we will report results from a similar analysis for the stars in its soon-to-be publicly released DR4 (Herpich et al., in prep.) in the next paper in this series (Huang et al., in prep.). We are also presently extending our techniques to include estimates of the [N/Fe] and [Ca/Fe] abundance ratios based on other narrow/medium-band filters employed by both J-PLUS and S-PLUS. 

We can expect tens of millions of stars with precise elemental-abundance estimates once both surveys are completed, including stars in the disk and halo populations, in the direction of the Galactic Bulge, for stars associated with stellar streams, and for nearby canonical dwarf spheroidal galaxies and ultra-faint dwarf galaxies.  One obvious application will be the construction of 
``blueprints" of Galactic stellar populations following the methods described in the series of papers by \citet{An2020, An2021a, an2021b} and \citet{An2023}. Other applications include analysis of the chemo-dynamical nature of stars in the disk and halo systems of the MW, such as the identification of dynamically and chemo-dynamically tagged groups, and their associations with recognized substructures (e.g., \citealt{Cabrera2024}, \citealt{Shank2023}, \citealt{Zepeda2023}, and references therein), and the identification of candidate very and extremely 
metal-poor stars in the disk system (e.g., \citealt{Hong2024}, and references therein).
Clearly, our catalogs will also prove useful for identifying stars of particular interest for medium- and high-resolution spectroscopic follow-up studies.

 \section*{Acknowledgements} 
We would like to thank the referee for helpful comments.
Y.H. thanks Prof. H.W. Zhang and H.L. Chen for useful discussions on the uncertainties of [Fe/H] and element-abundance ratios in this work.
Based on observations made with the JAST/T80 telescope at the Observatorio Astrof{\'i}sico de Javalambre (OAJ), in Teruel, owned, managed, and operated by the Centro de Estudios de F{\'i}sica del Cosmos de Arag{\'o}n. We acknowledge the OAJ Data Processing and Archiving Unit (UPAD) for reducing and calibrating the OAJ data used in this work. Funding for the J-PLUS Project has been provided by the Governments of Spain and Arag\'on through the Fondo de Inversiones de Teruel; the Aragonese Government through the Research Groups E96, E103, E16\_17R, E16\_20R, and E16\_23R; the Spanish Ministry of Science and Innovation (MCIN/AEI/10.13039/501100011033 y FEDER, Una manera de hacer Europa) with grants PID2021-124918NB-C41, PID2021-124918NB-C42, PID2021-124918NA-C43, and PID2021-124918NB-C44; the Spanish Ministry of Science, Innovation and Universities (MCIU/AEI/FEDER, UE) with grants PGC2018-097585-B-C21 and PGC2018-097585-B-C22; the Spanish Ministry of Economy and Competitiveness (MINECO) under AYA2015-66211-C2-1-P, AYA2015-66211-C2-2, AYA2012-30789, and ICTS-2009-14; and European FEDER funding (FCDD10-4E-867, FCDD13-4E-2685). The Brazilian agencies FINEP, FAPESP, and the National Observatory of Brazil have also contributed to this project.

The Guoshoujing Telescope (the Large Sky Area Multi-Object Fiber Spectroscopic Telescope, LAMOST) is a National Major Scientific Project built by the Chinese Academy of Sciences. Funding for the project has been provided by the National Development and Reform Commission. LAMOST is operated and managed by the National Astronomical Observatories, Chinese Academy of Sciences.

This work is supported by the National Key R \& D Program of China, grant No.
2019YFA0405500, and the National Natural Science Foundation of China, grant Nos. 11903027, 11973001, 11833006, and 12222301. T.C.B. and J.H. acknowledge partial support from grant PHY 14-30152; Physics Frontier Center/JINA Center for the Evolution of the Elements (JINA-CEE), and from OISE-1927130: The International Research Network for Nuclear Astrophysics (IReNA), awarded by the US National Science Foundation. 
Y.S.L. acknowledges support from the National Research Foundation (NRF) of Korea grant funded by the Ministry of Science and ICT (NRF-2021R1A2C1008679). Y.S.L. also gratefully acknowledges partial support for his visit to the University of Notre Dame from OISE-1927130: The International Research Network for Nuclear Astrophysics (IReNA), awarded by the US National Science Foundation.  
P.C. (Paul Coelho) acknowledges support from Conselho Nacional de Desenvolvimento Cient\'ifico e Tecnol\'ogico (CNPq) under grant 310555/2021-3 and from Funda\c{c}\~{a}o de Amparo\`{a} Pesquisa do Estado de S\~{a}o Paulo (FAPESP) process number 2021/08813-7.
S.D. acknowledges CNPq/MCTI for grant 306859/2022-0. 
F.J.E acknowledge financial support from MCIN/AEI/10.13039/501100011033/ through grant PID2020-112949GB-I00.
P.C. (Patricia Cruz) and F.J.E acknowledge financial support from MCIN/AEI/10.13039/501100011033/ through grant PID2020-112949GB-I00.
C.H.-M. acknowledges the support from the Spanish Ministry of Science and Education via project PID2021-126616NB-I00.
A.E. acknowledges the financial support from the Spanish Ministry of Science and Innovation and the European Union - NextGenerationEU through the Recovery and Resilience Facility project ICTS-MRR-2021-03-CEFCA.
LSJ acknowledges the support from CNPq (308994/2021-3)  and FAPESP (2011/51680-6).

\appendix
\setcounter{table}{0}   
\setcounter{figure}{0}
\renewcommand{\thetable}{A\arabic{table}}
\renewcommand{\thefigure}{A\arabic{figure}}
\section{Calibrations}
The APOGEE DR17, GALAH+ DR3 and SDSS/SEGUE DR12 are cross-matched with the collected HRS sample (PASTEL+SAGA), and the stars in common are used to examine the metallicity scales of these spectroscopic surveys.  The results are shown in Fig.\,A1.
Generally, the metallicity of the three surveys are consistent with that of the HRS sample, but deviate significantly toward the metal-poor region.
To correct for these systematics, second- and third-order polynomial functions are applied.
In Fig.\,A2, we adjust the LAMOST DR9 [Fe/H] scale to that of APOGEE DR17, correcting for small systematic trends with $T_{\rm eff}$, log\,$g$, and [Fe/H].
Finally, the comparisons in Fig.\,A3 show that the metallicity scale of the LAMOST/SEGUE VMP samples is quite consistent with that of HRS.
A summary of the calibrations is presented in Table\,A1.

For [C/Fe], [Mg/Fe], and [$\alpha$/Fe], the scales of APOGEE DR17 are adopted as the reference ones. The elemental-abundance ratios derived from GALAH DR3 and LAMOST/SEGUE VMP samples are examined with APOGEE DR17 (see Figs.\,A4 to A5). The results are summarized in Tables\,A2 to A4. We note that no correlations are found for [C/Fe] between APOGEE DR17 and GALAH DR3. Therefore, no calibrations are performed for GALAH DR3.

\begin{figure*}
\begin{center}
\includegraphics[scale=0.425,angle=0]{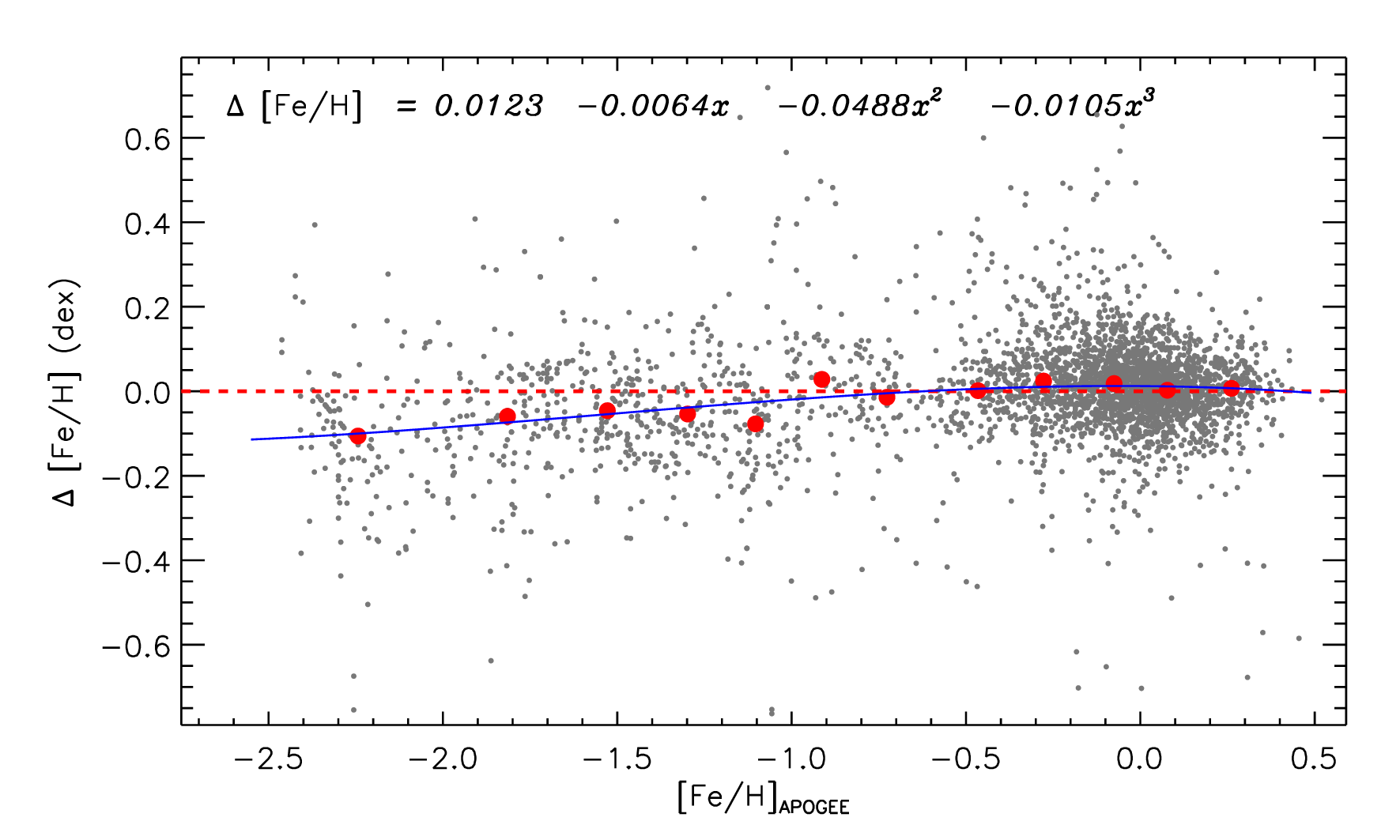}
\includegraphics[scale=0.425,angle=0]{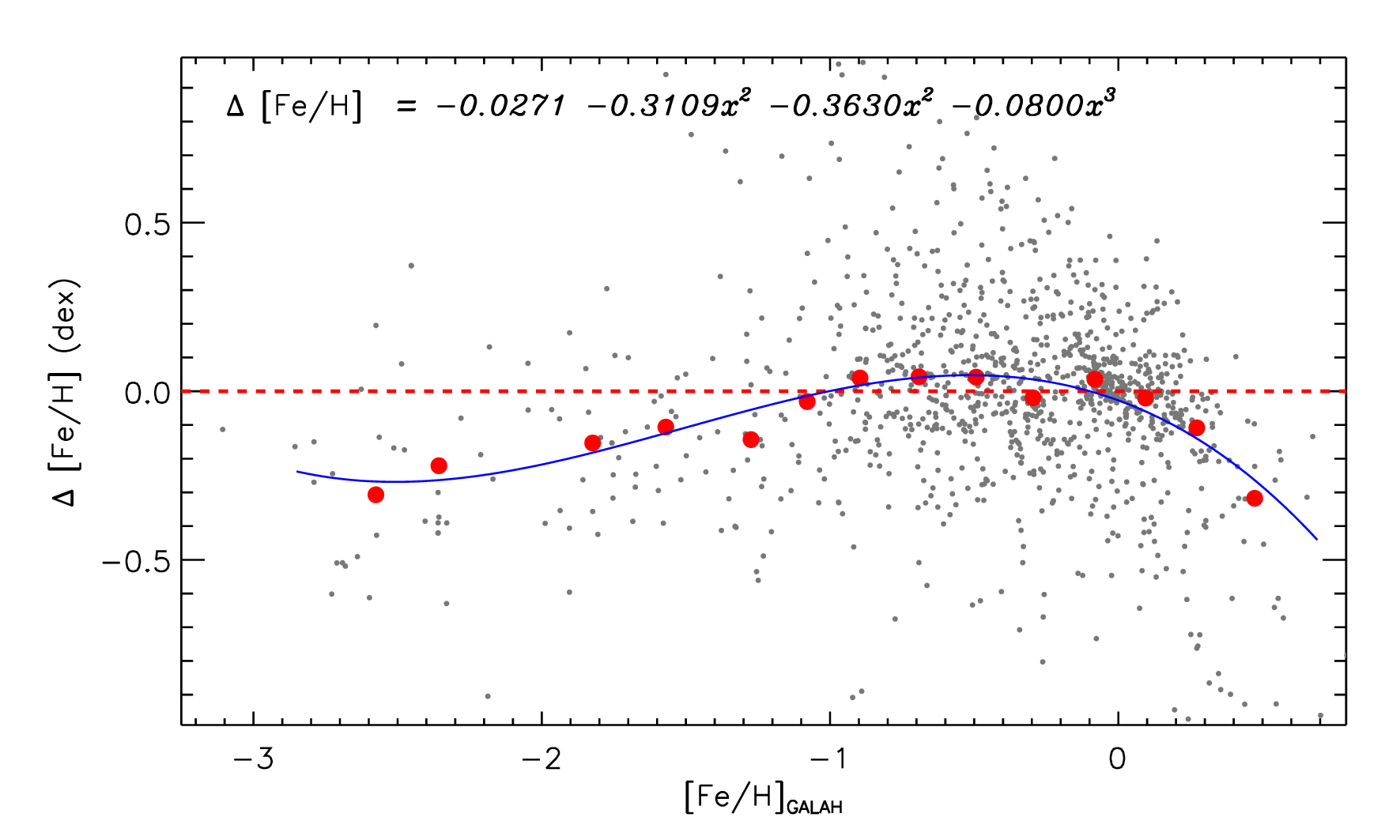}
\includegraphics[scale=0.425,angle=0]{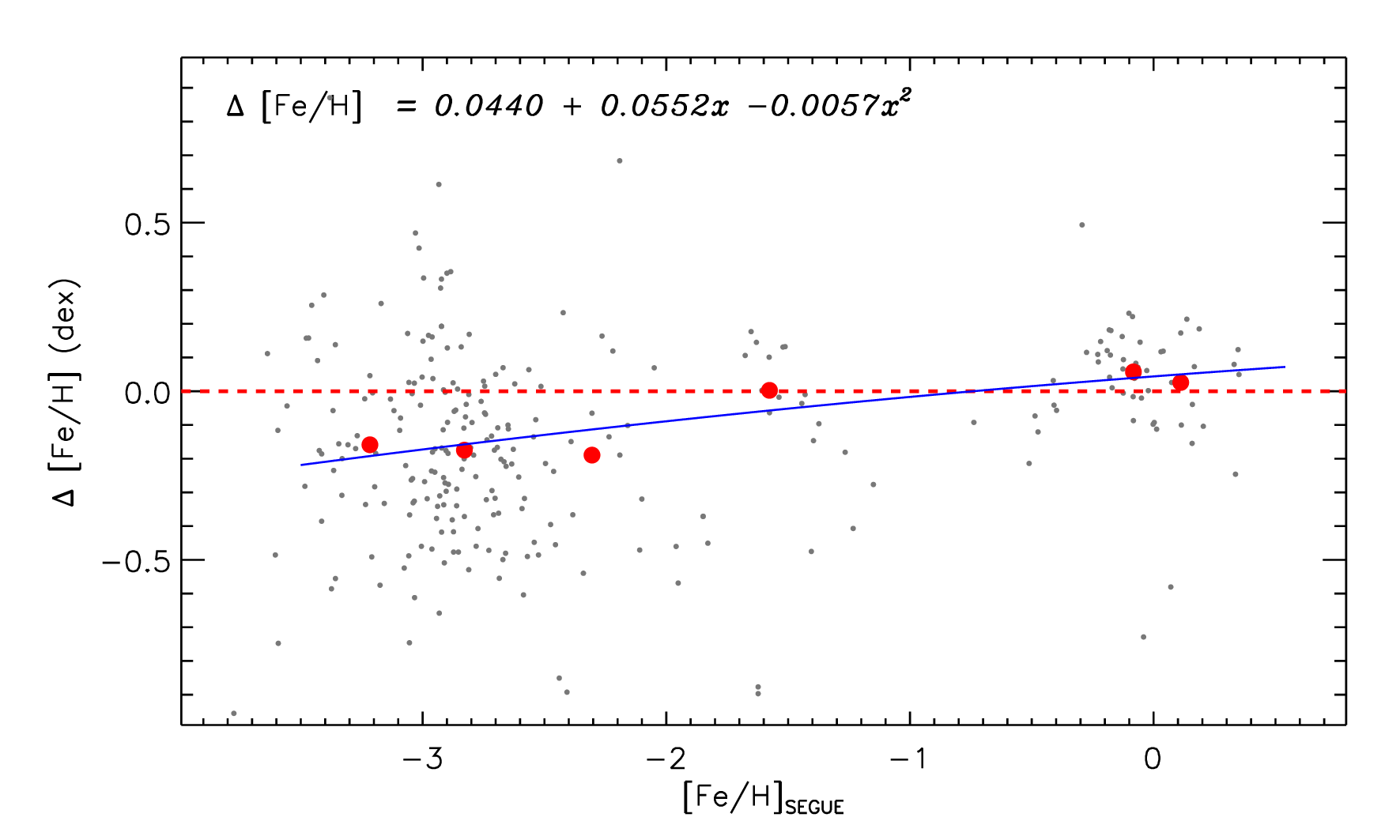}
\caption{Metallicity differences (HRS minus APOGEE/GALAH/SDSS) between the stars in common between APOGEE (top panel), GALAH (middle panel), SDSS/SEGUE (bottom panel), and the HRS, as a function of [Fe/H]. The red dots in each panel represent the median of the metallicity differences in the individual metallicity bins. Blue lines (with the functions marked in the top-left corner; here $x$ is the [Fe/H] of each spectroscopic survey) show second- to third-order polynomial fits to the blue data points. The red-dashed lines indicate zero residuals in each panel.}
\end{center}
\end{figure*}

\begin{figure*}
\begin{center}
\includegraphics[scale=0.425,angle=0]{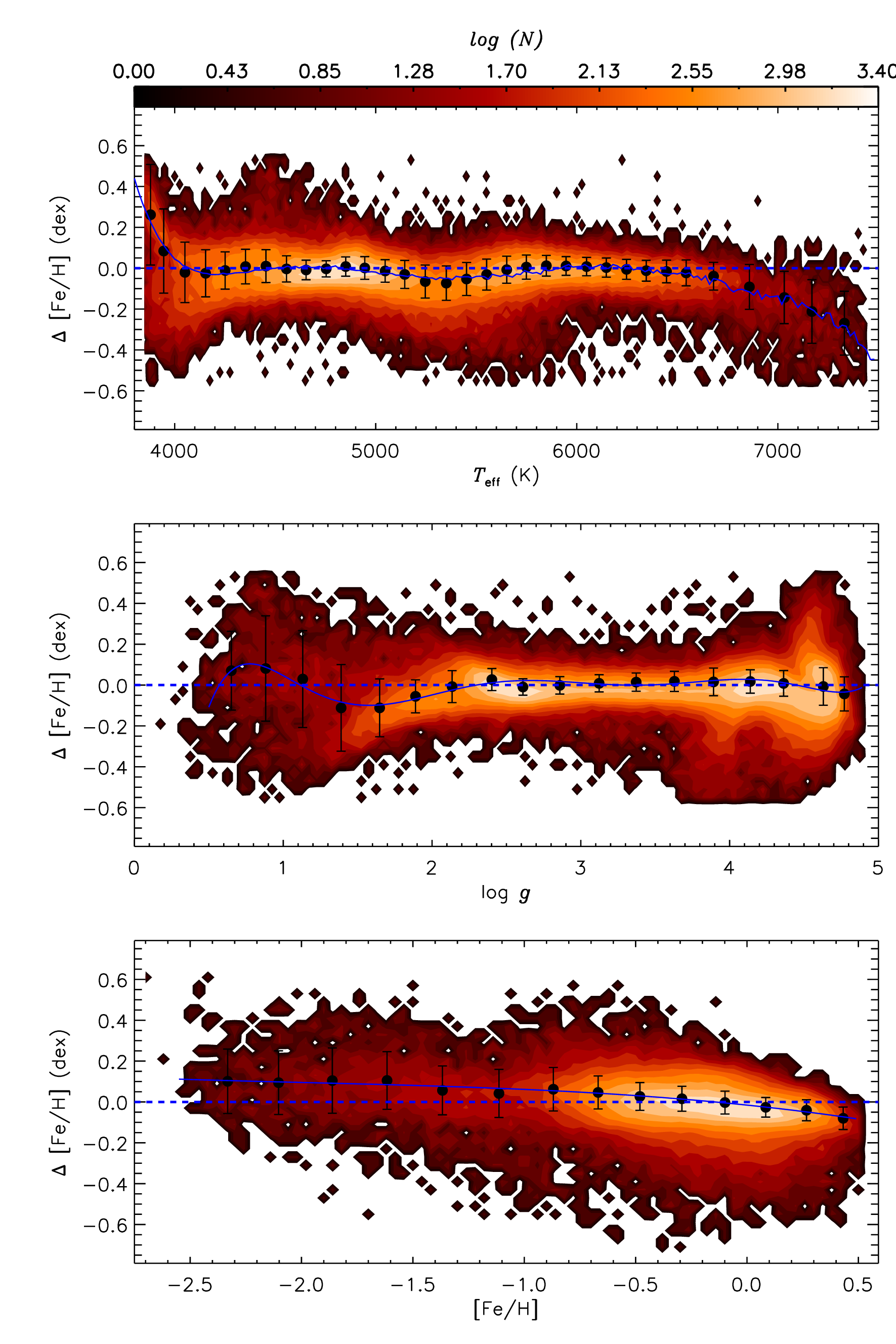}
\caption{Density distributions of the metallicity differences (APOGEE minus LAMOST) as a function of APOGEE effective temperature (top panel), surface gravity (middle panel), and [Fe/H] (bottom panel). The black dots and error bars in each panel represent the median and dispersion of the metallicity differences in the individual parameter bins.
Blue lines show third- and seventh-order polynomial fits to the black data points.
The function in the top panel is $\Delta {\rm [Fe/H] = 5.89680\times10^3 - 7.50396\times10^0T_{\rm eff}} + 4.05925\times10^{-3}T_{\rm eff}^2 - 1.21006\times10^{-6}T_{\rm eff}^3 + 2.14707\times10^{-10}T_{\rm eff}^4 - 2.26794\times10^{-14}T_{\rm eff}^5 + 1.32079\times10^{-18}T_{\rm eff}^6 - 3.27244\times10^{-23}T_{\rm eff}^{7}$.
The function in the middle panel is $\Delta {\rm [Fe/H]} = -3.49234\times10^0 + 1.43439\times10^1{\rm log}g - 2.18911\times10^1{\rm log}g^2 + 1.65358\times10^1{\rm log}g^3 - 6.85637\times10^0{\rm log}g^4 + 1.59063\times10^0{\rm log}g^5 - 1.93735\times10^{-1}{\rm log}g^6 + 9.64794\times10^{-3}{\rm log}g^7$.
The function in the bottom panel is $\Delta {\rm [Fe/H]} = -1.54691\times10^{-2} - 1.12912\times10^{-1}{\rm [Fe/H]} - 4.32802\times10^{-2}{\rm [Fe/H]}^2 - 7.26999\times10^{-3}{\rm [Fe/H]}^3$.
A color bar representing the numbers of stars is provided above the top panel. The blue-dashed lines indicate zero residuals in each panel.}
\end{center}
\end{figure*}

\begin{figure*}
\begin{center}
\includegraphics[scale=0.425,angle=0]{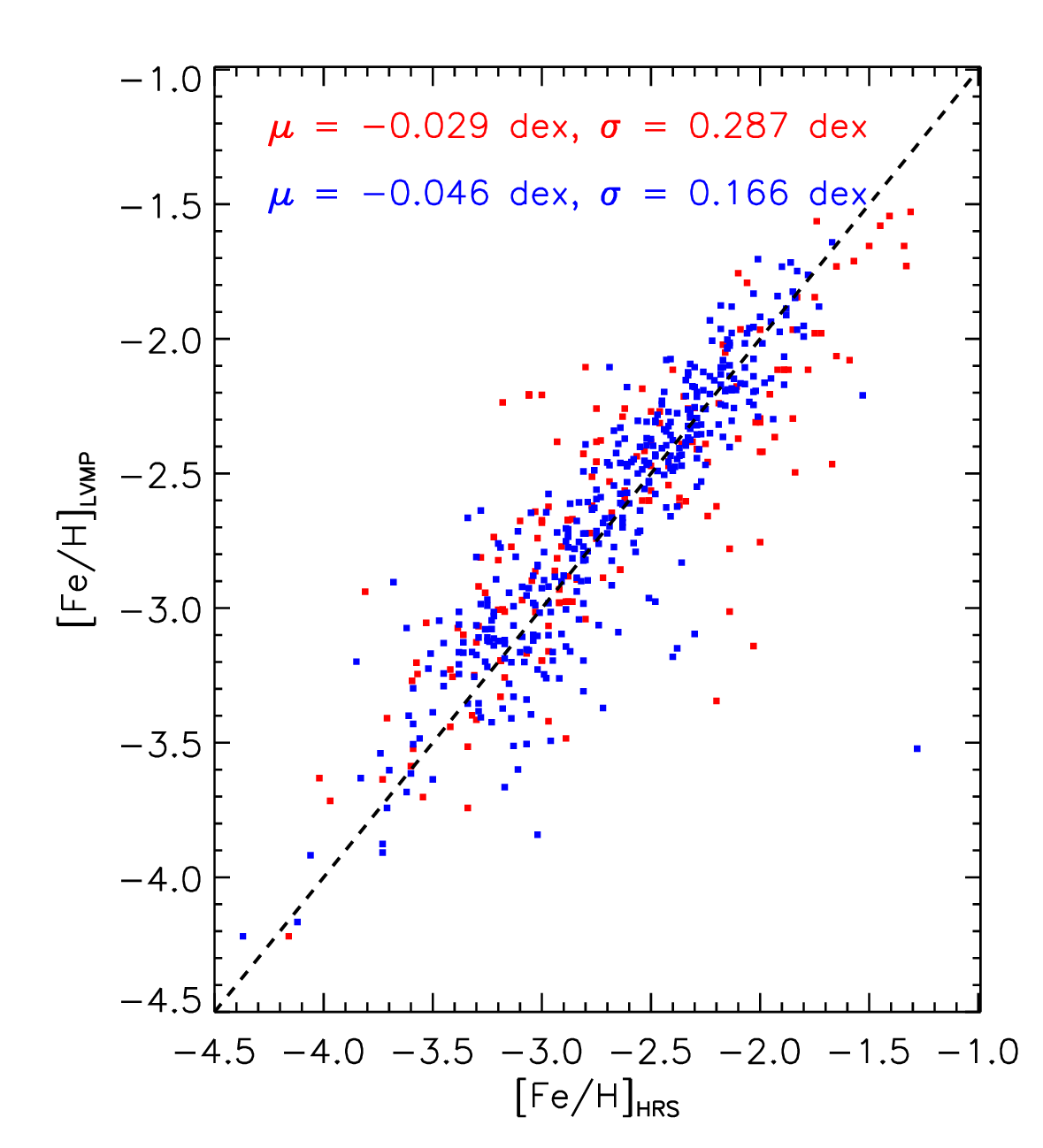}
\includegraphics[scale=0.425,angle=0]{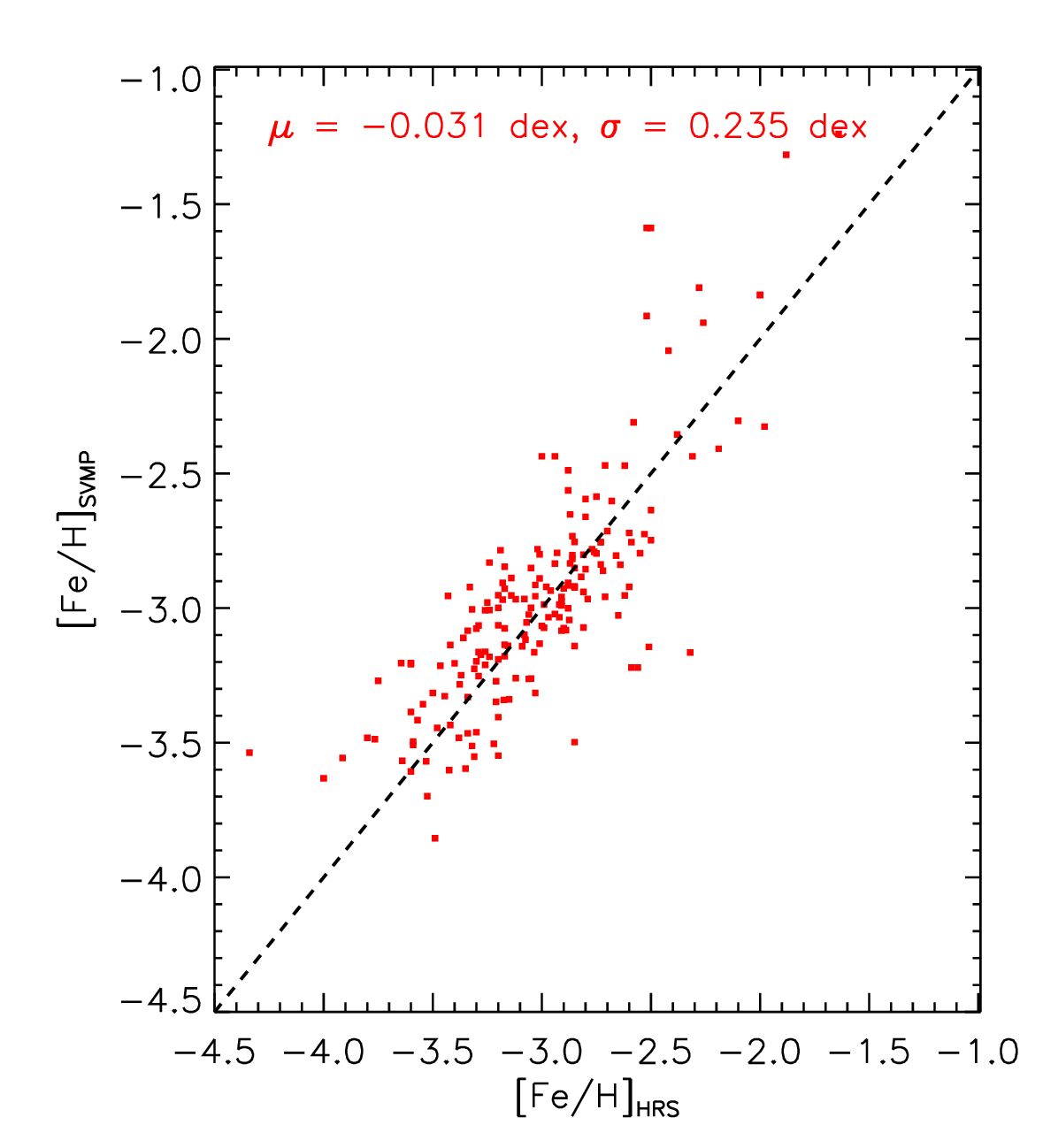}
\caption{Left panel: Comparisons between the LAMOST VMP (LVMP for short) sample (derived with the LSSPP; see Section\,3.1 for details) and the HRS sample compiled from PASTEL+SAGA (red squares), and  the LAMOST-Subaru HRS sample (blue squares) from \citet{2022ApJ...931..147L}. The overall median offset and standard
deviation are marked in the top-left corner.  Right panel: Comparison between SDSS/SEGUE VMP (SVMP for short) sample (derived from the SSPP) and the HRS sample compiled from PASTEL+SAGA. The black-dashed lines are the one-to-one-lines.}
\end{center}
\end{figure*}

%\begin{figure}
%\begin{center}
%\includegraphics[scale=0.425,angle=0]{fig/calib_galah_cfe.eps}
%\caption{Comparison between [C/Fe] from GALAH DR3 and that from APOGEE DR17 for stars (requiring spectral SNR greater than 50 per pixel in each survey) in common.  The blue-dashed line is the one-to-one line.}
%\end{center}
%\end{figure}

\begin{figure*}
\begin{center}
\includegraphics[scale=0.425,angle=0]{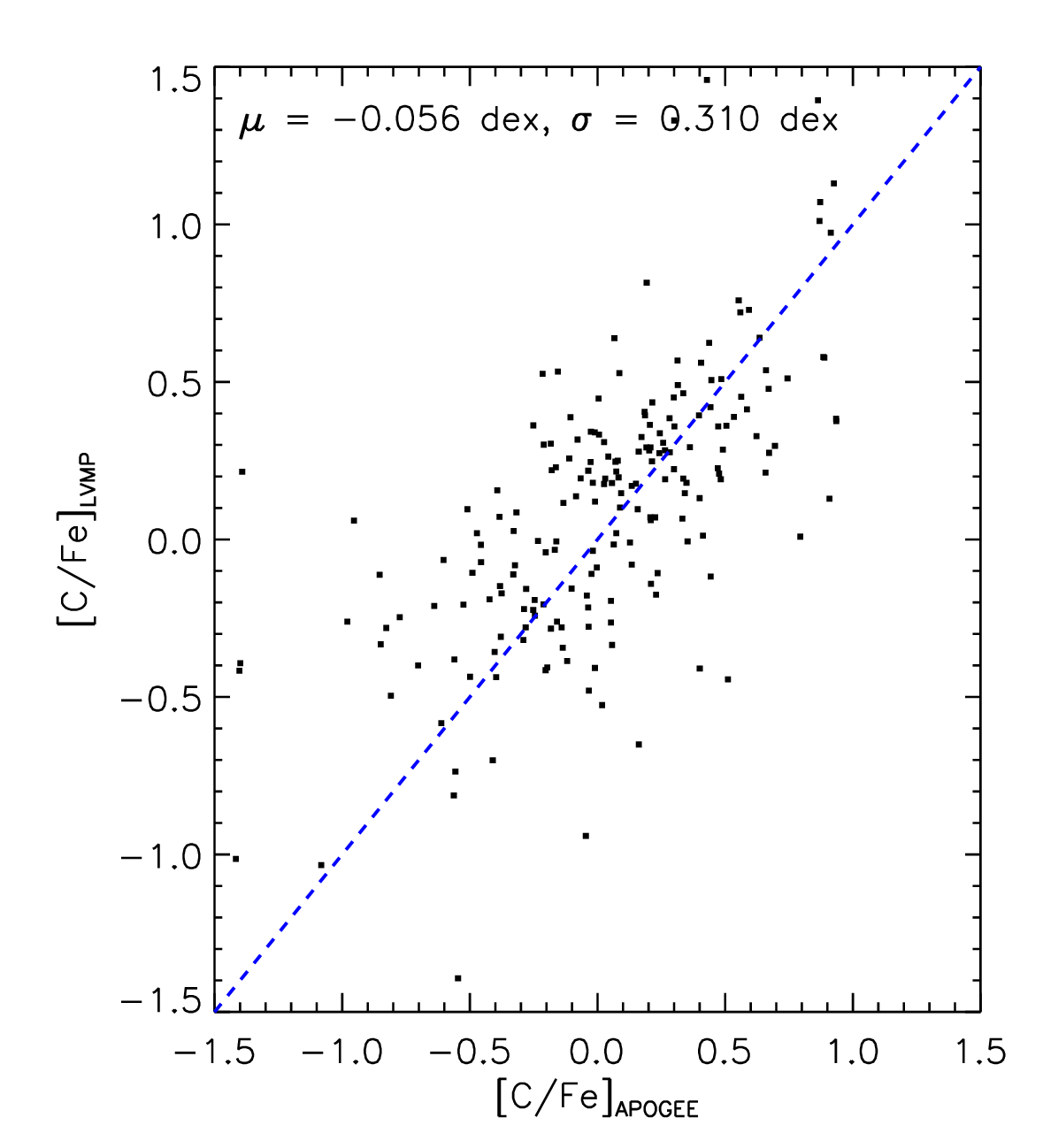}
\includegraphics[scale=0.425,angle=0]{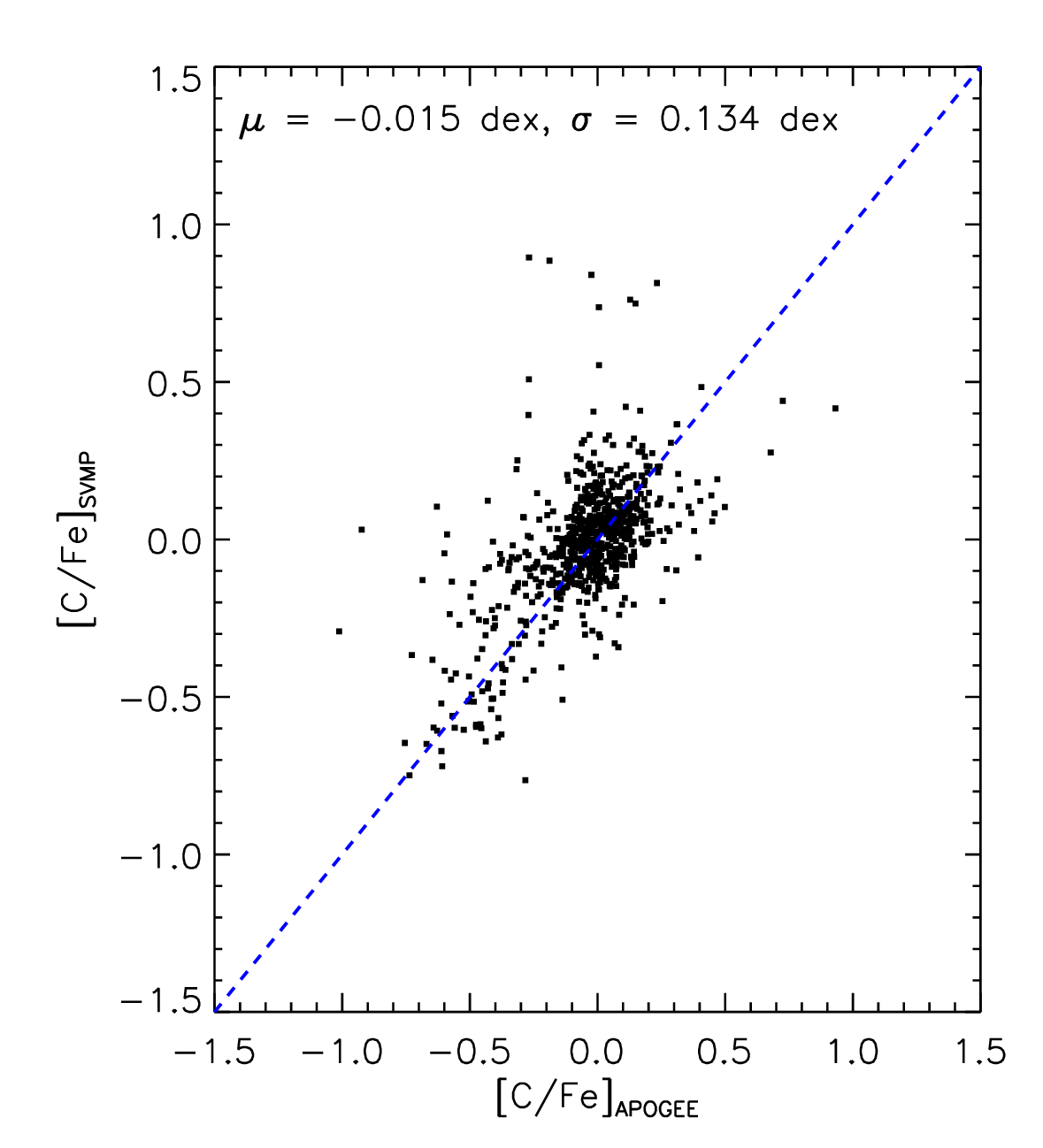}
\caption{Comparisons between [C/Fe] from the LAMOST VMP (LVMP for short) sample (left panel) and the SDSS/SEGUE VMP (SVMP for short) sample (right panel) and that of APOGEE DR17 for stars (requiring spectral SNR greater than 50 per piwel in each survey) in common. The overall median offset and standard deviation are marked in the top-left corner of each panel.}
\end{center}
\end{figure*}

\begin{figure*}
\begin{center}
\includegraphics[scale=0.425,angle=0]{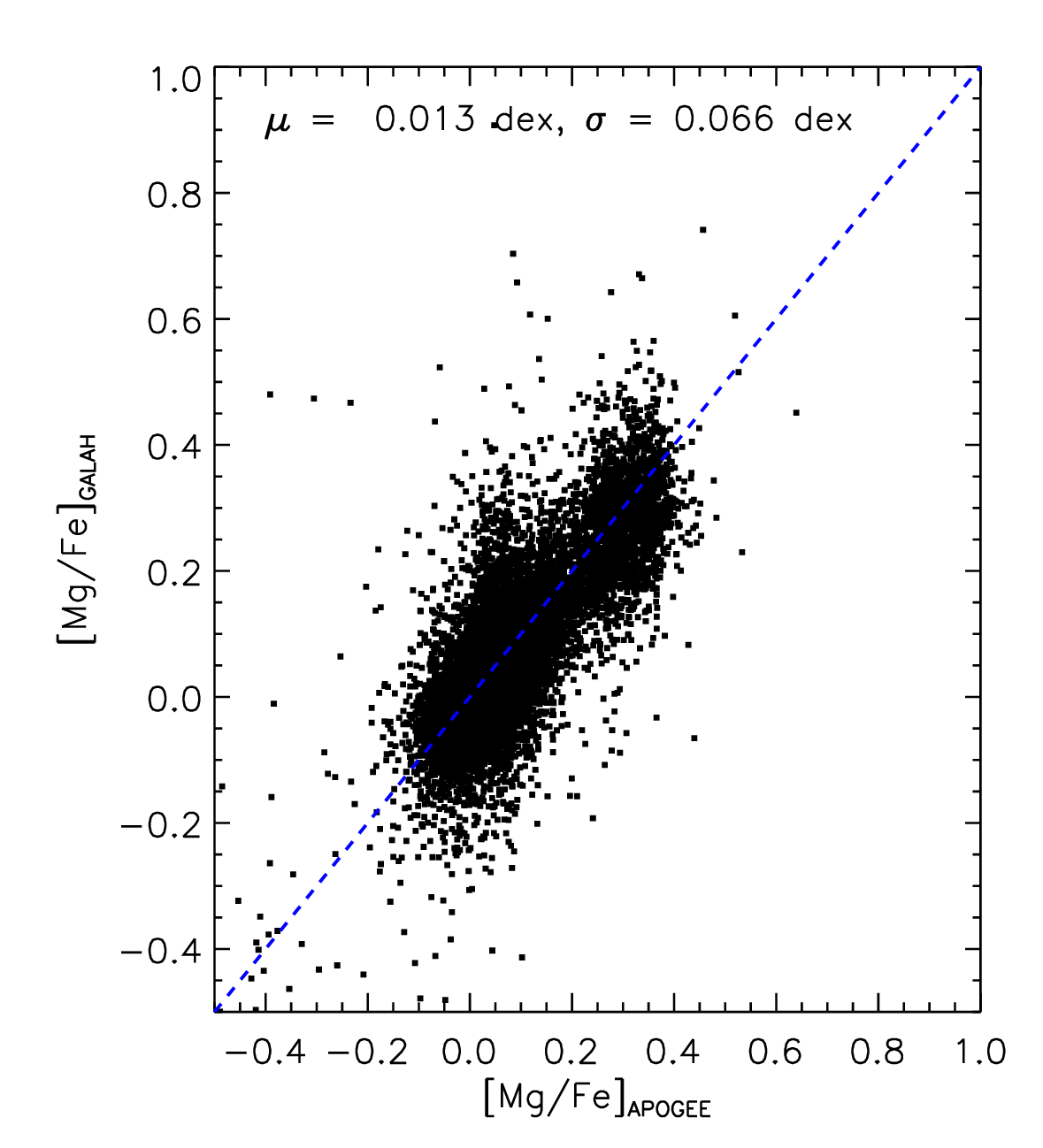}
\includegraphics[scale=0.425,angle=0]{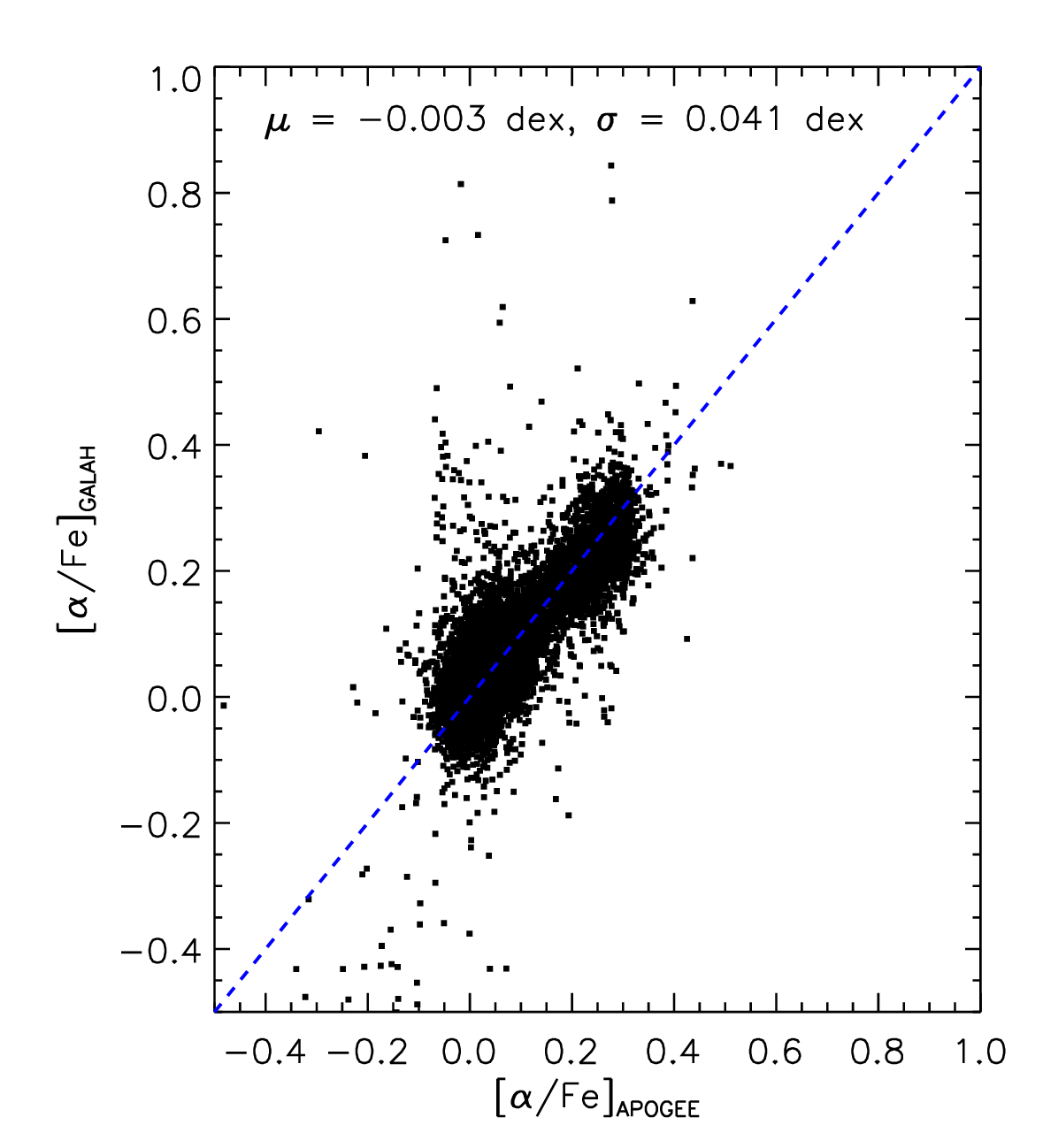}
\caption{Comparisons between [Mg/Fe] (left panel) and [$\alpha$/Fe] (right panel) of GALAH DR3 and those of APOGEE DR17. The overall median offset and standard deviation are marked in the top-left corner or each panel. Note that the spectral SNR of the stars in common are required to be greater than 50 per pixel in each survey.}
\end{center}
\end{figure*}

\begin{table*}
\centering
\caption{Summary of the Training and Testing Samples for Metallicity ([Fe/H])}
\begin{threeparttable}
\begin{tabular}{cccccc}
\hline
Catalog& $N$\tnote{a} & Metallicity Range & $\sigma_{\rm 1}$\tnote{b}& $\sigma_{\rm 2}$\tnote{c}&Calibration Note \\
&&&(dex)&(dex)&\\
\hline
\hline
PASTEL+SAGA&24,160&[$-5.70$, +1.00]&\dots&\dots&Reference scale\\
APOGEE DR17 & 642,616&[$-2.47$, +0.70]&0.075&0.073&Calibrated to the reference scale\\
GALAH DR3 & 438,397&[$-4.53$, +1.00]&0.172&0.162&Calibrated to the reference scale\\
LAMOST DR9& 4,755,823&[$-2.50$, +1.00]&0.053&0.047&Calibrated to the scale of APOGEE DR17\tnote{d}\\
SDSS DR12& 385,326&[$-4.50$, +0.75]&0.258&0.213&Calibrated to the reference scale\\
LAMOST VMP&42,221&[$-4.78$, $-1.80$]&0.287&\dots&No corrections\\
SDSS VMP&163,525&[$-4.41$, $-0.80$]&0.237&\dots&No corrections\\
\hline 
\hline
\end{tabular}
\begin{tablenotes}
\item[a]Here $N$ is the number of unique stars in the catalog with spectral SNR greater than 10 per pixel.
\item[b]$\sigma_1$ represents the standard deviation of the metallicity difference between the specific catalog and the reference scale.
\item[c]$\sigma_2$ represents the standard deviation of the metallicity difference between the specific catalog with calibrations and the reference scale.
\item[d]After calibration with APOGEE DR17, the metallicity scale of LAMOST DR9 can be further tied to the reference scale using the relations found for APOGEE DR17.
\end{tablenotes}
\end{threeparttable}
\end{table*}

 \begin{table*}
\centering
\caption{Summary of the Training and Testing Samples for [C/Fe]}
\begin{threeparttable}
\begin{tabular}{cccccc}
\hline
Catalog& $N$\tnote{a} &[C/Fe] Range & $\mu$& $\sigma$&Calibration Note \\
&&&(dex)&(dex)&\\
\hline
\hline
APOGEE DR17 & 642,616&[$-2.05$, +1.30]&\dots&\dots&Reference scale\\
%GALAH DR3 & 229,781&[$-1.41$, +3.00]&$-0.026$&0.098&No corrections\\
LAMOST VMP&37,716&[$-5.77$, $+4.41$]&$-0.056$&0.310&No corrections\\
SDSS VMP&152,504&[$-2.23$, $+4.14$]&$-0.015$&0.134&No corrections\\
\hline 
\hline
\end{tabular}
\begin{tablenotes}
\item[a]Here $N$ is the number of unique stars with [C/Fe]  measured in the catalog with spectral SNR ratio greater than 10 per pixel.
\end{tablenotes}
\end{threeparttable}
\end{table*}

\begin{table*}
\centering
\caption{Summary of the Training Samples for [Mg/Fe]}
\begin{threeparttable}
\begin{tabular}{cccccc}
\hline
Catalog& $N$\tnote{a} &[Mg/Fe] Range & $\mu$& $\sigma$&Calibration Note \\
&&&(dex)&(dex)&\\
\hline
\hline
APOGEE DR17 & 642,616&[$-1.71$, +1.87]&\dots&\dots&Reference scale\\
GALAH DR3 &425,203&[$-1.46$, +1.50]&$-0.003$&0.041&No corrections\\
LAMOST VMP\tnote{b}&32,485&[$-1.52$, $+2.63$]&$-0.008$&0.171&No corrections\\
SDSS VMP&101,770&[$-1.01$, $+2.66$]&$-0.144$&0.096&Corrected\\
\hline 
\hline
\end{tabular}
\begin{tablenotes}
\item[a]Here $N$ is the number of unique stars with [Mg/Fe]  measured in the catalog with spectral SNR greater than 10 per pixel.
\item[b]Here [$\alpha$/Fe] measurements are used since there are no [Mg/Fe] measurements for the 
LAMOST VMP sample.
\end{tablenotes}
\end{threeparttable}
\end{table*}

  \begin{table*}
\centering
\caption{Summary of the Training Samples for [$\alpha$/Fe]}
\begin{threeparttable}
\begin{tabular}{cccccc}
\hline
Catalog& $N$\tnote{a} &[$\alpha$/Fe] Range & $\mu$& $\sigma$&Calibration Note \\
&&&(dex)&(dex)&\\
\hline
\hline
APOGEE DR17 & 642,616&[$-1.68$, +1.70]&\dots&\dots&Reference scale\\
GALAH DR3 &425,203&[$-1.26$, +2.81]&$-0.003$&0.041&No corrections\\
LAMOST VMP&32,485&[$-1.52$, $+2.63$]&$-0.051$&0.174&Corrected\\
SDSS VMP&101,770&[$-1.01$, $+2.61$]&$-0.135$&0.077&Corrected\\
\hline 
\hline
\end{tabular}
\begin{tablenotes}
\item[a]Here $N$ is the number of unique stars with [$\alpha$/Fe] measured in the catalog with spectral SNR greater than 10 per pixel.
\end{tablenotes}
\end{threeparttable}
\end{table*}

\vfill\eject
\bibliography{sppara_calib}{}
\bibliographystyle{aasjournal}
\end{document}